\documentclass{JHEP3} 

\usepackage{amsfonts}
\usepackage{amsmath}
\usepackage{amssymb}
\usepackage{epsfig,multicol,bbm}
\usepackage{fancyhdr}
\usepackage{float}
\usepackage{graphicx}
\usepackage{mathrsfs}
\usepackage{subfigure}
\usepackage{caption} 
\newcommand\fverb{\setbox\fverbbox=\hbox\bgroup\verb}
\newcommand\fverbdo{\egroup\medskip\noindent%
			\fbox{\unhbox\fverbbox}\ }
\newcommand\fverbit{\egroup\item[\fbox{\unhbox\fverbbox}]}
\newbox\fverbbox

\newcommand{\eqnlab}[1]{\label{eqn:#1}}
\newcommand{\eqnref}[1]{(\ref{eqn:#1})}

\input{colordvi.tex}

\title{M5-branes, toric diagrams and gauge theory duality}

\preprint{HU-Mathematik: 2011 - 23
\\
HU-EP-11/58
\\
YITP-11-105
\\
SISSA 66/2011/EP}

\author{Ling Bao$^{a}$\footnote{Email: ling.bao@chalmers.se}
$\,$, Elli Pomoni$^{b}$\footnote{Email: pomoni@mathematik.hu-berlin.de}
$\,$, Masato Taki$^{c}$\footnote{Email: taki@yukawa.kyoto-u.ac.jp}
$\,$  and Futoshi Yagi$^{d}$\footnote{Email: fyagi@sissa.it} 
\\
\\
\it $^a$ Fundamental Physics, Chalmers University of Technology, 41296 G\"oteborg, Sweden
\\
\it $^b$ Institut f\"ur Mathematik und Institut f\"ur Physik, 
Humboldt-Universit\"at zu Berlin\\
Johann von Neumann-Haus, Rudower Chaussee 25, 12489 Berlin, Germany
\\
\it $^c$ Yukawa Institute for Theoretical Physics, Kyoto University,
Kitashirakawa Oiwake-Cho, Sakyo-Ku, Kyoto, Japan
\\
\it $^d$ International School of Advanced Studies (SISSA)
via Bonomea 265, 34136 Trieste, Italy and INFN, Sezione di Trieste
}
 	 
\abstract{

\bigskip

In this article we explore the duality between the low energy effective theory of five-dimensional $\mathcal{N}=1$ $SU(N)^{M-1}$ and $SU(M)^{N-1}$ linear quiver gauge theories compactified on $S^1$.
The  theories we study are the five-dimensional uplifts of four-dimensional superconformal linear quivers.
We study this duality by comparing the Seiberg-Witten curves and the Nekrasov partition functions of the two dual theories. 
The Seiberg-Witten curves are obtained by minimizing the worldvolume of an M5-brane with nontrivial geometry.
Nekrasov partition functions are computed using topological string theory.
The  result of our study is a map between the gauge theory parameters, i.e., Coulomb moduli, masses and  UV  coupling constants, of the two dual theories. 
Apart from the obvious physical interest, this duality also leads to compelling mathematical identities.
Through the AGTW conjecture these five-dimentional gauge theories are related to  $q$-deformed Liouville and Toda SCFTs in two-dimensions.
The duality we study implies the relations between Liouville and Toda correlation functions through the map we derive.
}


\begin{document}

\newpage
\section{Introduction}

$\mathcal N=2$ gauge theories have been of great interest in the past twenty-five years. While $\mathcal N=4$ SYM has trivial non-perturbative physics 
the more realistic $\mathcal N=1$ gauge theories are yet to be solved.
 $\mathcal N=2$ gauge theories exhibit many interesting phenomena, such as confinement and monopole condensation. Moreover, their topological sector gives access to their non-perturbative regime.

Seiberg and Witten derived the Wilsonian low energy effective action of the $\mathcal N=2$ $SU(2)$ gauge theory by encoding the problem into a two-dimensional (2D) holomorphic curve \cite{Seiberg:1994rs}. Their work was soon after generalized to other gauge groups and matter contents \cite{Argyres:1994xh,Klemm:1994qs,Seiberg:1994aj,Argyres:1995wt}. Although for the paradigmatic $SU(2)$ case the Seiberg-Witten (SW) curve was derived from first principles \cite{Seiberg:1994rs}, its construction becomes difficult for generic quiver gauge theories. 
Therefore, other methods have been employed, e.g., integrability \cite{Donagi:1995cf}, geometric engineering \cite{Katz:1996fh,Katz:1997eq} and the type IIA/M-theory brane constructions \cite{Witten:1997sc,Kol:1997fv,Brandhuber:1997ua}. The SW curve was initially introduced as an auxiliary space \cite{Seiberg:1994rs}, however, it was later understood that it is part of the M-theory target space \cite{Witten:1997sc}. Using string theory, $\mathcal{N}=2$ gauge theories can be realized as world volume theories on D4-branes, which are suspended between NS5-branes. Uplifting this brane setup to M-theory, all the branes can be seen as one single M5-brane with a non-trivial topology. The geometry of this M5-brane is encoded in the SW curve. Therefore, the SW curve can also be derived by studying the minimal surface of the M5-brane \cite{Witten:1997sc}.

An alternative way to derive the Seiberg-Witten results was discovered by Nekrasov \cite{Nekrasov:2002qd}. He succeeded in finding the instanton partition functions of the $\mathcal N=2$ gauge theories by introducing a special deformation called the $\Omega$ background. The deformed theory should in fact be interpreted as a five-dimensional (5D) $\mathcal N=1$ gauge theory defined on the space $\mathcal{M}_4 \times S^1$. This class of 5D gauge theories was first studied by Seiberg \cite{Seiberg:1996bd} and their relation to the four-dimensional (4D) $\mathcal N=2$ gauge theories on $\mathcal{M}_4$ was explored in \cite{Nekrasov:1996cz}. Further, it was found that the 5D $\mathcal N=1$ gauge theories can be realized using D5- and NS5-branes \cite{Aharony:1997ju,Aharony:1997bh}. This D5/NS5 brane construction is T-dual to the D4/NS5 system discussed above \cite{Witten:1997sc} as well as the original D3/NS5 Hanany-Witten set-up \cite{Hanany:1996ie}. The 5D extension of the SW curve has been studied in \cite{Kol:1997fv,Brandhuber:1997ua}. The curve was obtained by compactifying one of the directions along which the NS5-branes extend in the D4/NS5 setup. After T-duality along the compactified direction, D4-branes turn into D5-branes, whose world volume theory is a 5D $\mathcal{N}=1$ gauge theory.

An intriguing relation between the gauge theory partition function and topological string theory was conjectured by Nekrasov \cite{Nekrasov:2002qd}. String theory compactified on Calabi-Yau threefold (CY$_3$) yields $\mathcal{N}=2$ gauge theory on the 4D transverse space. The partition function of this gauge theory is equivalent to the field theory limit of the topological string partition function. This relation has been tested and verified by several authors \cite{Iqbal:2003ix,Iqbal:2003zz,Eguchi:2003sj}. The topological string theory computation leads to a special case of $\Omega$ deformed gauge theories. The generic $\Omega$ deformation of gauge theories is obtained by considering an extension called refined topological string partition function \cite{Iqbal:2007ii,Awata:2008ed,Taki:2007dh}.
Topological strings without field theory limit gives the generating function of the BPS states 
coming from M2-branes wrapped on two-cycles inside CY$_3$.
This means that the topological string theory describes the holomorphic sector of  M-theory on CY$_3$.  
The topological string partition function is then equivalent to the Nekrasov partition function  
for 5D gauge theory via M-theory lift of the geometric engineering.

\begin{figure}[t]
 \begin{center}
  \includegraphics[width=150mm,clip]{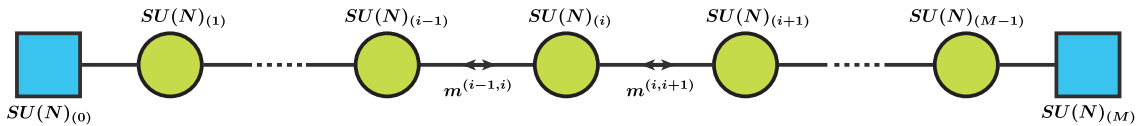}
 \end{center}
 \caption{The circle $SU(N)_{(i)}$ corresponds to the $i$-th gauge group,
the segments between two circles are bifundamental hypermultiplets,
and the flavor symmetries are illustrated by the two blue boxes $SU(N)_{(0)}$ and $SU(N)_{(M)}$ at the ends of the quiver.}
 \label{quiver}
\end{figure}

In the present article, we consider  the 5D $\mathcal{N}=1$ $SU(N)^{M-1}$ liner quivers depicted in Figure~\ref{quiver}. Their type IIA string theory description involves $N$ D4-branes and $M$ NS5-branes. In this set-up, the NS5-branes wrap a coordinate circle  $S^1$.  
From the M-theory point of view, there is, in addition, an M-theory circle around which the M5-branes that lead to D4-branes wrap. 
We have thus two compact circles, whose roles can be exchanged. In other words, all we have is an M5-brane with non-trivial topology, which yields the SW curve of either $SU(N)^{M-1}$ theory or $SU(M)^{N-1}$ theory. In this sense, $SU(N)^{M-1}$ gauge theory is dual to $SU(M)^{N-1}$ gauge theory. Although the conceptual understanding of this duality has been discussed previously \cite{Katz:1997eq,Aharony:1997bh}\footnote{For related work see also \cite{Muneyuki:2011qu}.}, the explicit duality map was not known.

We take a first step toward understanding this duality in detail. The strategy we adopt is to compare the low energy effective theories of 5D $SU(N)^{M-1}$ and $SU(M)^{N-1}$ gauge theories (Figure~\ref{quiver}). This is achieved by independently using both the Seiberg-Witten formalism and Nekrasov's partition function. We derive the map between the ultraviolet (UV) parameters of the two gauge theories, through which they are dual to each other.  
The Seiberg-Witten curves are obtained by minimizing the worldvolume of an M5-brane with nontrivial geometry.
Nekrasov partition functions are computed using topological string theory.
Both in the M-theory and the topological string theory descriptions the duality is geometrically realized simply as a rotation of the M5-brane and  toric diagram respectively.
We would  also like to mention that there is another duality for 4D   $\mathcal{N}=1$ gauge theories that is based on performing (non-trivial in this case) operations on the toric diagrams. The  $\mathcal{N}=1$  toric duality  is studied in \cite{Feng:2000mi}.

\bigskip

This article is organized as follows. In Section \ref{sec:Review}, we review well known tools and notions that will be used for our study of the duality. In particular, we will describe the Seiberg-Witten framework adopted for the 5D gauge theories, as well as the derivation of Nekrasov's partition function using topological string theory. In Section \ref{sec:MtheoryDeriv}, we compute the duality map for the gauge theory parameters based on analysis of the SW curve. The same map will then be re-derived independently in Section \ref{sec:TopStringDeriv}, where we calculate Nekrasov's partition function via the topological string partition function for toric CY$_3$. Starting from the toric diagram, one notices that the duality is manifest. The consequences of the duality for 2D CFTs through AGTW are discussed in  Section \ref{sec:GaugeToCFT} together with the simplest extension to  the generic $\Omega$ background. Section \ref{sec:Discussion} is devoted to discussions of our results and possible future applications.

\section{Background material}
\label{sec:Review}

\subsection{Seiberg-Witten formalism}
\label{sec:SWReview}

We begin by summarizing the Seiberg-Witten solution for 4D $\mathcal{N}=2$ gauge theories. Nice reviews of this topic can be found in \cite{Bilal:1995hc,Lerche:1996xu,Klemm:1997gg,Peskin:1997qi}. The complete description of the low energy effective theory (up to two derivatives and quartic fermion terms) is encoded in the prepotential $\mathcal{F}(a)$ according to
\begin{equation}
S_{eff} = \int d^4x d^4 \theta \mathcal{F} (a) + \int d^4x d^4 \bar{\theta}  \bar{\mathcal{F}} (\bar{a}) \, .
\end{equation}
The prepotential is a holomorphic function of the vacuum expectation values (${a}_i$) of the scalar fields in the $\mathcal{N} = 2$ vector multiplet. The holomorphic gauge couplings are obtained as
\begin{equation}
\tau_{ij} =\frac{ \partial^2\mathcal{F}(a)}{\partial {a}_i \partial {a}_j} \, ,
\end{equation}
while the expectation values of the scalar fields in the dual (magnetic) theory are given by
\begin{equation}
{a_D}^i =  \frac{\partial \mathcal{F}(a)}{\partial {a}_i} \, .
\eqnlab{PrepotentialDeriv}
\end{equation}
The electromagnetic duality acts on the Coulomb moduli as the modular transformation
\begin{equation}
   \left(   \begin{array}{c}
      {a_D}^i \\
       {a}_i \\
   \end{array}
   \right) \rightarrow
   \left(   \begin{array}{cc}
      a & b \\
      c & d \\
   \end{array}
   \right)
   \left(   \begin{array}{c}
      {a_D}^i \\
       {a}_i \\
   \end{array}
   \right)
   \quad \mbox{with} \quad   \left(   \begin{array}{cc}
      a & b \\
      c & d \\
   \end{array}
   \right) \in SL(2, \mathbb{Z}) \, .
\end{equation}
The prepotential is determined using an auxiliary curve called the SW curve
\begin{equation}
F_{4D}(t,v) = 0
\end{equation}
together with a meromorphic differential $\lambda_{SW}$. The derivatives of the meromorphic one-form with respect to the  moduli of the SW curve\footnote{The  moduli of the SW curve $u$  for the SU(2)  example is the gauge invariant Coulomb moduli $u=\langle \text{tr}\phi^2 \rangle + \dots$.} are the holomorphic differentials of the curve. The Coulomb moduli are then computed according to
\begin{equation}
a_i = \oint_{A_i} \lambda_{SW} \quad \mbox{and} \quad {a_D}^i = \oint_{B^i} \lambda_{SW} \, ,
\label{a_aD}
\end{equation}
where $A_i$ and $B_i$ are the basic cycles of the algebraic curve with intersection number $A_i \cdot B^j = \delta_i^j$. The prepotential itself can be found by integrating \eqnref{PrepotentialDeriv}. Moreover, contour integrals of the meromorphic differential $\lambda_{SW}$ around its poles give linear combinations of the bare quark masses ($m_i$).

The SW curve and one-form can also be derived from M-theory \cite{Witten:1997sc}. To do this we consider the brane setup in Table \ref{config}, where $N$ D4-branes are suspended between $M$ NS5-branes. We introduce also $2N$ flavor branes attached to the two outermost NS5-branes and extended to infinity. The theory described by this setup is 4D 
$\mathcal{N}=2$ $SU(N)^{M-1}$ gauge theory, which is asymptotically conformal. The rotation of $x^{4}$ and $x^{5}$ coordinates corresponds to $U(1)_R$ symmetry, while rotation of $x^{7}$, $x^{8}$, and $x^{9}$ corresponds to 
$SU(2)_R$ symmetry.
\begin{table}
\centering
\begin{tabular}{|c|c|c|c|c|c|c|c|c|c|c|c|}
\hline
&$x^0$ & $x^1$ &$x^2$ &$x^3$ &$x^4$ &$x^5$ &$x^6$ &$x^7$ &$x^8$ &$x^9$ &($x^{10}$) \\
\hline
$M$ NS5-branes &$-$&$-$&$-$&$-$&$-$&$-$&.&.&.&.&.\\
\hline
$N$ D4-branes &$-$&$-$&$-$&$-$&.&.&$-$&.&.&.&$-$\\
\hline
\end{tabular}
\caption{Brane configuration in type IIA string theory}
\label{config}
\end{table}
Table \ref{config} is a classical configuration from the gauge theory point of view. Taking the tension of the branes into account, the configuration has to be modified to include the quantum effects. Uplifting to M-theory and minimizing the world volume of the corresponding M5-brane under fixed boundary condition yields the SW curve. This curve describes a 2D subsurface inside the space spanned by the coordinates $\{x^4, x^5, x^6, x^{10}\}$, where $x^{10}$ is the direction of the M-theory circle.

To obtain 5D $\mathcal{N}=1$ gauge theory we compactify the $x^{5}$ coordinate. After T-duality along $x_5$, the system becomes an D5/NS5 brane system in type IIB string theory with a 5D $\mathcal N=1$ gauge theory living on the D5-branes (Table \ref{configIIB}). This is the 5D $\mathcal N=1$ gauge theory for which we are constructing the SW curve. The spacetime of this gauge theory is $\mathcal{M}_4 \times S^1$ with the circumference of the IIB circle being
\begin{equation}
\beta =\frac{  2 \pi   \alpha'}{R_5} = \frac{  2 \pi   \ell^3_{p}}{R_5 R_{10}} \, ,
\end{equation}
where $\alpha' = \ell_{s}^2 = \frac{\ell_{\text{p}}^3}{R_{10}}$.  Going back to the type IIA description, we define the complex coordinates $v$ and $s$ according to
\begin{equation}
  v \equiv x^4 + i x^5 \quad \text{and} \quad s \equiv x^6 + i x^{10} \, .
\end{equation}
Due to the periodic nature of $x^5$ and $x^{10}$ it is natural to introduce another pair of complex coordinates
\begin{equation}
  w \equiv e^{-\frac{v}{R_5}} \quad \text{and} \quad t \equiv e^{-\frac{s}{R_{10}}} \, .
\label{def_tw}
\end{equation}
The radius of the $x^{5}$ circle is denoted as $R_5$ and that of the M-theory circle as $R_{10}$.
\begin{table}
\centering
\begin{tabular}{|c|c|c|c|c|c|c|c|c|c|c|c|}
\hline
&$x^0$ & $x^1$ &$x^2$ &$x^3$ &$x^4$ &$x^5$ &$x^6$ &$x^7$ &$x^8$ &$x^9$ &($x^{10}$) \\
\hline
$M$ NS5-branes &$-$&$-$&$-$&$-$&$-$&$-$&.&.&.&.&.\\
\hline
$N$ D5-branes &$-$&$-$&$-$&$-$&.&$-$&$-$&.&.&.&$-$\\
\hline
\end{tabular}
\caption{Brane configuration in type IIB string theory}
\label{configIIB}
\end{table}

The SW curve of the 5D $SU(N)^{M-1}$ theory is now written as a polynomial of degree $N$ in $w$ and degree $M$ in $t$ as
\begin{align}
F (t,w) \equiv \sum_{i=0}^N \sum_{j=0}^M C_{p,q} w^p t^q \, .
\label{SW_curve}
\end{align}
The periodic boundary condition along the $x^5$ coordinate makes the curve invariant under a shift of the positions of the color branes ($a'$) and flavor branes ($m'$) by $2 \pi R_5$. Therefore, the coefficients of the curve $C_{p,q}$ depend only on the gauge coupling $q$ and
\begin{equation}
\label{tildema}
\begin{split}
\tilde{m} &\equiv e^{-m' / R_5} = e^{- \beta m} \, , \\
\tilde{a} &\equiv e^{-a' / R_5} = e^{- \beta a} \, ,
\end{split}
\end{equation}
in which periodicity is manifest. Note that quantities that have dimension of mass are related to the ones with dimension of length (primed) as
\begin{equation}
a = \frac{a'}{2\pi \ell^2_s} \quad \text{and} \quad
m = \frac{m'}{2\pi \ell^2_s} \, .
\end{equation}
The coefficients $C_{p,q}$ will be determined explicitly in Section \ref{sec:MtheoryDeriv}.

\bigskip

The M-theory derivation of the SW one-form can be found in \cite{Fayyazuddin:1997by, Henningson:1997hy, Mikhailov:1997jv}. We summarize it for pure $SU(2)$ theory here. The extension to generic quiver theories is straightforward. The idea is to relate two different expressions of the masses of BPS states. On one hand, the mass of a BPS particle is given by
\begin{align}
{m_{\text{BPS}}}^2 = |n_e a + n_m a_D|^2 \, ,
\label{BPS}
\end{align}
where $n_e$ and $n_m$ are the electric and magnetic charges of the BPS state respectively. This formula can be rewritten using the SW one-form as
\begin{align}
{m_{\text{BPS}}}^2 = \left| \int_{n_e A + n_m B} \lambda_{\rm SW} \right| ^2 \, .
\label{mass1}
\end{align}
On the other hand, a BPS state is interpreted as an open M2-brane attached to an M5-brane whose volume is minimized. The boundary of a such minimal M2-brane with charge $(n_e,n_m)$ is the cycle $n_e A + n_m B$. Finally, the mass of this BPS state is calculated using the volume-form of the M2-brane
\begin{align}
\omega = ds \wedge dv = d \left[ \log t \, (d \log w) \right]
\end{align}
and reads
\begin{align}
{m_{\text{BPS}}}^2 = \left| \frac{1}{(2 \pi)^2 \ell_p{}^3} \int_{M_2} \omega \right| ^2 \, ,
\label{mass2}
\end{align}
where $1/ (2 \pi)^2 \ell_p{}^3 $ is the tension of the M2-brane. Comparing (\ref{mass1}) with (\ref{mass2}), we find that the SW one-form takes the form
\begin{align}
\lambda_{\rm SW} = - \frac{i}{(2 \pi)^2 \ell_p{}^3} \log t \, (d \log w) \, .
\label{SW_1-form}
\end{align}

\subsection{Partition function and topological vertex}

The microscopic way to obtain the prepotential is via  Nekrasov's partition function  \cite{Nekrasov:2002qd,Nekrasov:2003rj}
\begin{equation}
\label{ZF}
Z(a;\epsilon_1 ,\epsilon_2) = e^{ \frac{\mathcal{F} (a)}{\epsilon_1 \epsilon_2}  + \cdots} \, ,
\end{equation}
which contains the full low energy effective description of $\mathcal{N} = 2$ gauge theories in a deformed background. More details can be found in \cite{Nekrasov:2005wg,Bruzzo:2002xf,Marino:2004cn,Tachikawa:2004,Shadchin:2005mx}. The starting point of Nekrasov's derivation is 5D $\mathcal{N} = 1$ gauge theory on $\mathcal{M}_4 \times S^1$. This theory depends on two deformation parameters $(\epsilon_1, \epsilon_2)$ and the circumference of the circle $\beta$. Taking the limit $\beta \rightarrow 0$ leads to the so called $\Omega$ deformed 4D $\mathcal{N}=2$ gauge theory. The deformation parameterized by $\epsilon_1$ and $\epsilon_2$ breaks the $SO(4)$ Lorentz symmetry down to $SO(2){\times}SO(2)$. In this way the path integral is localized to one point on $\mathcal{M}_4$ and the computation of the partition function is simplified to supersymmetric quantum mechanics along $S^1$.

Nekrasov's partition function $Z(a,m,q;\epsilon_1,\epsilon_2)$ of 4D $\mathcal{N}=2$ gauge theory is a function of the set of moduli $a$ parameterizing the Coulomb branch, the masses $m$ of all the flavor and bifundamental fields, the coupling constants $q=e^{2\pi i \tau}$ and the two parameters $\epsilon_1$ and $\epsilon_2$. It can be factorized as
\begin{equation}
\label{Partition}
Z = Z_{\rm pert}\,Z_{\rm inst} \, ,
\end{equation}
where $Z_{\rm pert}$ is the perturbative part containing tree-level and one-loop contributions, while $Z_{\rm inst}$ is the contribution from the instantons. The instanton part can be expanded with respect to the instanton number $k$
\begin{equation}
\label{Instantons}
Z_{\rm inst} = \sum_{k} q^{k} Z_{k} \, .
\end{equation}

As discussed previously, one way to realize 4D $\mathcal{N}=2$ gauge theories is the Hanany-Witten setup in Table \ref{config}.
Another way is to consider CY$_3$ compactification of type IIA string theory. 
These two different points of view are connected by a series of duality transformations \cite{Karch:1998yv}. Starting from the Hanany-Witten setup, the transformations consist of a T-duality along the $x^6$ coordinate, followed by an S-duality involving $x^6$ and $x^{10}$ and lastly another T-duality along the new $x^6$ coordinate. The resulting theory is IIA string theory on non-compact CY$_3$ without any branes.
The gauge symmetry of the 4D theory is geometrically realized by the vanishing cycles inside CY$_3$. A special class of CY$_3$ which yields $\mathcal{N}=2$ gauge theories is the toric type \cite{Aganagic:2003db}. Its generic configuration is a fibration of special Lagrangian $T^2 \times \mathbb{R}$ over the base $\mathbb{R}^3$. For $SU(N)$ gauge symmetry it is further required that the CY$_3$ manifold is a non-trivial fibration of $A_{N-1}$ singularity over the space $\mathbb{P}^1$ \cite{Iqbal:2003zz}.

Already in \cite{Nekrasov:2002qd} Nekrasov suggested that the partition function of $\mathcal{N}=2$ gauge theories is the field theory limit of the topological string partition function on toric CY$_3$. 
 For toric CY$_3$ the topological string partition function can be computed graphically using the so called toric diagram\footnote{In this article we use the word toric diagram for the dual graph of the toric data. This is also called web-diagram  \cite{Aharony:1997bh}.}, which characterizes the toric Calabi-Yau manifold. The toric diagram consists of a collection of trivalent vertices, which are joined together by oriented straight lines.

\begin{figure}[t]
\center
\includegraphics[scale=0.6]{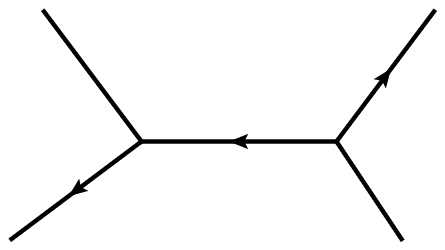}
\qquad   \qquad   
\includegraphics[scale=0.6]{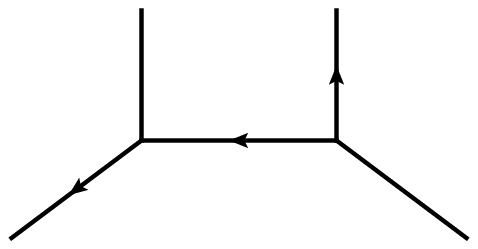}
\qquad   \qquad  \quad  
\includegraphics[scale=0.6]{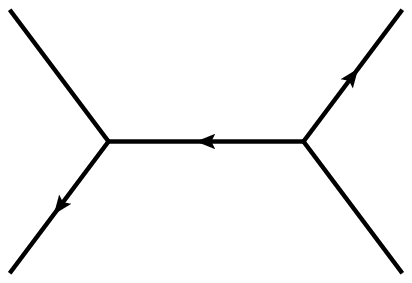}   
    \put(-336,8){\footnotesize$ \vec{v}_{\text{in}}$}   
       \put(-258,32){\footnotesize$ \vec{v}_{\text{out}}$}
              \put(-296,21){\footnotesize$R$} 
     \put(-296,10){\footnotesize$m$} 
       \put(-180,8){\footnotesize$m=1$}       
  \put(-48,10){\footnotesize$m=0$} 
\caption{In the definition of the framing factor we have $m=\det \left(\vec{v}_{\text{in}} \cdot  \vec{v}_{\text{out}}\right)$. We graphically clarify its definition and give two examples.}
\label{mfigure}
\end{figure}

Writing down  the topological string partition function is simple using the topological vertex formalism. The procedure is very similar to computing Feynman diagrams for usual field theory, where the internal momentum integrals are replaced by sums over the Young diagrams $R$
\begin{equation}
\int d p  \quad  \longrightarrow  \quad \sum_{R}  \nonumber \, .
\end{equation}
Schematically, it takes the form
\begin{equation}
  \mathcal{Z} = \sum_{R} \, (\text{three-vertices}) \times (\text{oriented lines}) \, .
  \eqnlab{PartitionFuncSum}
\end{equation}
The topological three-vertex describes the open string amplitude on a local $\mathbb{C}^3$ coordinate patch. In the case $\hbar = \epsilon_1 = - \epsilon_2$, the contribution from the topological vertex is given by \cite{Aganagic:2003db}
\begin{equation}
  C_{R_1 R_2 R_3}(\mathfrak{q}) = \mathfrak{q}^{\frac{\kappa_{R_3}}{2}}S_{R_2}(\mathfrak{q}^\rho) \sum_\eta S_{R_1/\eta}(\mathfrak{q}^{R_2^T+\rho})\, S_{R_3^T/\eta}(\mathfrak{q}^{R_2+\rho}) \, ,
  \label{def_topv}
\end{equation}
where $S_{\alpha}$ and $S_{\alpha / \eta}$ are the Schur and skew-Schur functions, respectively. 
We also introduce the symbol $\mathfrak{q}=e^{ - \beta \hbar}$ for the exponentiated $\Omega$ background.
The three free indices represent the three straight lines going out from the vertex. Each line is labeled by an infinite set of all possible Young tableaux associated with the group $U(\infty)$. 
In the Feynman diagram analogy, the vertex
\begin{figure}[h!]
\qquad    \qquad   \qquad  \quad
\includegraphics[scale=0.6]{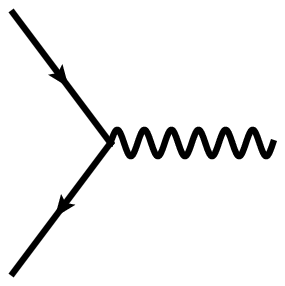}
\qquad   \qquad    
\put(17,21){$\longrightarrow$}
\qquad   \qquad  
\includegraphics[scale=0.6]{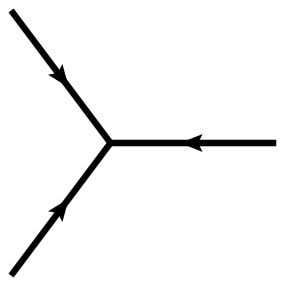}   
\qquad   \qquad   
\put(13,21){$= \quad C_{R_1 R_2 R_3}$}
\put(-19,17){\tiny$R_1$}
\put(-50,32){\tiny$R_3$}
\put(-50,13){\tiny$R_2$}
\end{figure}
is replaced by the topological vertex (\ref{def_topv}), \\ while for the propagator
\begin{equation}
G(p)=\frac{1}{p^2}  \quad  \longrightarrow  \quad (-Q)^{|R|} (-1)^m \mathfrak{q}^{-\frac{m}{2}\kappa_{R}} \, ,    \nonumber
\end{equation}
where $Q=e^{-t}$ is the exponentiated K\"ahler moduli (size) of the two-cycle represented by the segment. The framing factor ($(-1)^m \mathfrak{q}^{-\frac{m}{2}\kappa_{R}}$) of the ``propagator'' contains the second Casimir $\kappa_R$ of the representation $R$, which is defined as $\kappa_R = \sum_j R_j (R_j-2j-1)=-\kappa_{R^T}$ where $R_j$ is the number of boxes in the $j$-th line of the tableu and $R^T$ is the transposed Young tableu. The  integer $(-m-1)$ is the self-intersection number of the two-cycle and is illustrated in Figure \ref{mfigure} together with two examples. 

The closed string amplitude on the full CY$_3$ is obtained by gluing together the local $\mathbb{C}^3$ patches as in \eqnref{PartitionFuncSum}. The sum in \eqnref{PartitionFuncSum} is taken over all the Young tableaux sets attached to the internal lines of the toric diagram. After carrying out the summation explicitly, it is straightforward to compare with the gauge theory partition function given by Nekrasov. The topological vertex formalism gives thus an alternative derivation for
 Nekrasov's partition function based on the geometric shape of the corresponding toric diagram.

When $\epsilon_1 \neq - \epsilon_2$, the topological vertex function above should be replaced by the \textit{refined} topological vertex function \cite{Iqbal:2007ii,Awata:2008ed}
\begin{align}
\begin{split}
C_{R_1 R_2 R_3} (\mathfrak{t},\mathfrak{q}) = & {\left( {\frac{\mathfrak{q}}{\mathfrak{t}}} \right)}^{\frac{{\left\| R_1  \right\|^2  + \left\| R_2 \right\|^2 }}{2}} \mathfrak{t}^{\frac{{{\kappa}_{R_1}}}{2}} 
P_{{R_2}^T } (\mathfrak{t}^{ - \rho } ;\mathfrak{q},\mathfrak{t}) \, \times \\
& \sum_\eta {{\left( {\frac{\mathfrak{q}}{\mathfrak{t}}} \right)}^{\frac{{\left| \eta  \right| + \left| R_3 \right| - \left| R_1  \right|}}{2}} 
S_{R_1 /\eta } (\mathfrak{t}^{ - {R_2}^T } \mathfrak{q}^{ - \rho } )}
S_{{R_3}^T /\eta } (\mathfrak{t}^{ - \rho } \mathfrak{q}^{ - R_2 } ) \, ,
\end{split}
\end{align}
where 
$\mathfrak{q}=e^{-\beta \epsilon_1},\, \mathfrak{t}=e^{\beta \epsilon_2}$,
and $P_{R} (\mathop \mathfrak{t}\nolimits^{ - \rho } ;\mathfrak{q},\mathfrak{t})$ is the principal specialization of the Macdonald function
\begin{align}
P_{R^T } (\mathfrak{t}^{ - \rho } ;\mathfrak{q},\mathfrak{t})=\mathfrak{t}^{\frac{1}{2}||R||^2} \prod_{(i,j)\in R}(1-\mathfrak{t}^{{R_j}^T-i+1} \mathfrak{q}^{R_i-j})^{-1} \, .
\end{align}
The refined topological vertex function is a generalization which reduces to the ordinary vertex function when choosing $\epsilon_1 = - \epsilon_2$. It has slightly different properties compared to the ordinary topological vertex. E.g., instead of being entirely cyclic symmetric, one of its legs indicates a preferred direction. Slicing invariance is a conjecture claiming that the full partition function should be invariant under a change of the choice of the preferred direction.

\subsection{Introducing the duality}
\label{subsec:review_duality}

The first hint toward a duality between the 5D  gauge theories with gauge groups $SU(N)^{M-1}$ and $SU(M)^{N-1}$ is given by counting the physical parameters of these two theories. Indeed, we find that the number of parameters matches exactly. For this counting we can ignore the infinite tower of Kaluza-Klein modes and count only the zero modes\footnote{To include all the Kaluza-Klein modes only the circumference of the circle $\beta$ is added as an extra parameter. However, the circumference $\beta$ always appears multiplied with the gauge parameters whose mass dimension is $1$, so it does not actually introduce any new degree of freedom.}. The zero modes coincide with the parameters of the corresponding 4D gauge theories on $\mathcal{M}_4$, we will therefore use 4D terminology in the rest of this section.

The infrared (IR) physics of $SU(N)^{M-1}$ and $SU(M)^{N-1}$ gauge theories at generic points on the Coulomb branch are both described by the $U(1)^{(N-1)(M-1)}$ theory. They are thus described by $(N-1)(M-1)$ IR effective coupling constants
\begin{equation}
\tau_{IR}^i = \tau_{IR}^i\left(\tau_{UV} , m_{\text{f}} , m_{\text{bif}} , a \right) \, ,
\end{equation}
which depend holomorphically on the gauge theory parameters. $\tau_{UV}$ are the UV coupling constants, $m_{\text{f}}$ are the mass parameters of the flavor hypermultiplets, $m_{\text{bif}}$ are the mass parameters of the bifundamental hypermultiplets and $a$ are the Coulomb moduli parameters. The counting of the parameters for asymptotically superconformal $SU(N)^{M-1}$ and $SU(M)^{N-1}$ gauge theories is summarized in Table~\ref{tab:CountParam}. Summing all the parameters shows that there are in total $[(N+1)(M+1)-3]$ parameters in both theories, allowing the possibility to derive a map between them.
\begin{table}[h]
	 \centering
    \begin{tabular}{| c || c | c |}
    \hline
    & $SU(N)^{M-1}$ & $SU(M)^{N-1}$ \\
    \hline
	 \hline
	 $\tau_{UV}$ & $M-1$ & $N-1$ \\
	 \hline
	 $m_{\text{f}}$ & $2N$ & $2M$ \\
	 \hline
	 $m_{\text{bif}}$ & $M-2$ & $N-2$ \\
	 \hline
	 $a$ & $(N-1)(M-1)$ & $(M-1)(N-1)$ \\
	 \hline
	 \hline
	 Total & $(N+1)(M+1)-3$ & $(M+1)(N+1)-3$ \\
	 \hline
    \end{tabular}
	 \caption{Counting of the gauge theory parameters}
	 \label{tab:CountParam}
\end{table}

One of approaches we use is to match the coefficients of the SW curves and the SW one-form of the two dual theories. Before attempting that, we first count the degrees of freedom that are encoded in the SW curve. The SW curve of the 5D  $SU(N)^{M-1}$ gauge theory is a polynomial of degree $M$ in the variable $t$ and $N$ in the variable $w$. We have therefore $[(M+1)(N+1)-1]$ complex coefficients, where one has been subtracted to allow an overall coefficient. Moreover, there is the freedom to set the origins of the coordinates $s$ and $v$. Removing two more coefficients we find $[(M+1)(N+1)-3]$ degrees of freedom in total. Thus, the number of coefficients in the SW curve is always the same as the number of physical parameters.

If we exchange the role of the variables $t$ and $w$,  the SW curve (\ref{SW_curve}) of the original $SU(N)^{M-1}$ theory 
can be read as the SW curve of the dual $SU(M)^{N-1}$ theory.
The coefficients $C_{p,q}$ in the original curve 
get reinterpreted as the coefficients in the curve of the dual theory  $(C_{q,p})_d$.
In addition, the SW one-form  (\ref{SW_1-form})  also remains the same  up to a minus sign (\ref{equalSWoneform}).
Using (\ref{a_aD}) the IR effective coupling constant is given by
\begin{equation} 
 \tau_{IR} 
  =  \frac{   \frac{\partial a_D}{\partial u}   }{  \frac{\partial a}{\partial u}   }   
  =    \frac{   \int_B   \omega    }{    \int_A   \omega    }  \, ,
\end{equation}
where $\omega$ is the holomorphic differential. Since the  holomorphic differential does not distinguish\footnote{This is true because $\omega$ has no poles as opposed to the  meromorphic $\lambda_{SW}$ that does have poles.}
the cycle $A$ (or $B$) of the original theory from $A_d$ (or $B_d$) of the dual theory, 
we get that the dual  IR effective coupling constant is equal to the original one.
Therefore, once the relation between the gauge theory parameters and the coefficients $C_{p,q}$ in the SW curve is established, it is straightforward to find the duality map. 
The map is obtained by equating the coefficients $C_{p,q}$, written in terms of the gauge theory parameters of the original $SU(N)^{M-1}$ theory, with the coefficients $(C_{q,p})_d$, written in terms of the parameters of the dual $SU(M)^{N-1}$ theory.

The interpretation of this duality in the context of the brane setup in IIA/M and IIB theories \cite{Aharony:1997bh} is the following. Consider M-theory compactified on $T^2$. The cycles of the torus correspond to the two phases of the variables $t$ and $w$. Exchanging $t$ and $w$ is the holomorphic extension of a particular $SL(2,\mathbb{Z})$ modular transformation on the compactification torus, where the M-theory circle is exchanged with the $x^5$ circle. This modular transformation is equivalent to S-duality in IIB theory compactified on $S^1$ via T-duality along the $x^5$ circle. 
The modular transformation  in the IIA theory limit exchanges D4-branes with  NS5-branes compactified along the $x^5$ circle, while S-duality in IIB theory  exchanges D5-branes with NS5-branes.
Rigorously, the 90 degree rotation  of the brane configuration ($w\rightarrow t$ and $t\rightarrow w^{-1}$) corresponds to this S-duality, but in the main body of this paper we study the $t\leftrightarrow w$ reflection that it is technically simpler. The 90 degree rotation case is presented in detail in the appendix \ref{app:90Rotation}.
Note that this S-duality is different from the one which appears as the 
electric-magnetic duality in the 4D Seiberg-Witten theory. 
As we will see in the following sections, it acts on the gauge theory parameters 
in a totally different manner.
The difference of these two types of S-dualities is due to the difference of the brane setup.
It is known that the Montonen-Olive duality,
which is the extension of the electric-magnetic duality for 4D $\mathcal{N}=4$ theory,
 is obtained by compactifying the M5-branes on a torus \cite{Vafa:1997mh, Tachikawa:2011ch}.
In the brane setup of Table~\ref{config} the $x^6$ direction has to be compactified instead of $x^5$ (as in our case). 

The duality described here was originally found in the context of geometric engineering \cite{Katz:1997eq}. On the IIB string theory side, the SW curve is embedded in the CY manifold and the duality can be seen in a similar way as the M-theory analysis above. In the mirror IIA theory, on the other hand, the duality is most clear from the toric diagram. Indeed, the toric diagram for the $SU(N)^{M-1}$ theory is exactly the same as the one for the $SU(M)^{N-1}$ theory, up to a simple reflection or 90 degree rotation. This duality is therefore manifest at the level of the topological string partition function. Depending on which sums are carried out explicitly in \eqnref{PartitionFuncSum}, we obtain  Nekrasov's partition function of $SU(N)^{M-1}$ theory or $SU(M)^{N-1}$ theory. The topological vertex formalism provides the extension of the duality for the non-zero self-dual $\Omega$ background.

\section{M-theory derivation}
\label{sec:MtheoryDeriv}

In this section we present the first derivation of the duality map using the Seiberg-Witten formalism reviewed in Section \ref{sec:SWReview}. Another, independent derivation based on the topological vertex formalism is given in \ref{sec:TopStringDeriv}. The map between the gauge theory parameters of the 5D $\mathcal{N}=1$ $SU(N)^{M-1}$ and $SU(M)^{N-1}$ liner quiver gauge theories compactified on $S^1$ is obtained by comparing their Seiberg-Witten curves. The SW curves are derived  using the M-theory approach \cite{Witten:1997sc}. We, firstly, warm up with the  self-dual case of $SU(2)$ gauge theory with four flavors and then turn to the generic duality between $SU(N)^{M-1}$ and $SU(M)^{N-1}$. The special  case ($M=2$) between $SU(N)$ and $SU(2)^{N-1}$ is given at the end of this section.

\subsection{$SU(2)$ gauge theory with four flavors}
\label{subsec:Msu2}

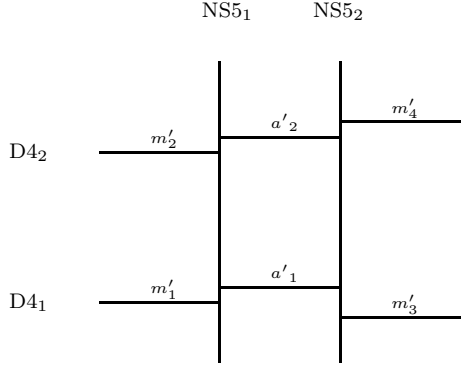
\begin{figure}[htb]
\centering
\tiny
\setlength{\unitlength}{4cm}
\vspace{10mm}
\begin{picture}(1,1)(0,0)
\linethickness{0.25mm}
\scriptsize
\put(0.5,0){\line(0,1){1}}
\put(0.44,1.15){NS5$_1$}
\put(0.9,0){\line(0,1){1}}
\put(0.81,1.15){NS5$_{2}$}
\tiny

\put(0.1,0.2){\line(1,0){0.4}}
\put(0.27,0.23){$m'_{1}$}
\put(0.1,0.7){\line(1,0){0.4}}
\put(0.27,0.73){$m'_{2}$}

\scriptsize
\put(-0.2,0.18){D4$_1$}
\put(-0.2,0.68){D4$_2$}
\tiny

\put(0.5,0.25){\line(1,0){0.4}}
\put(0.67,0.28){$a'{}_{1}$}
\put(0.5,0.75){\line(1,0){0.4}}
\put(0.67,0.78){$a'{}_{2}$}

\put(0.9,0.15){\line(1,0){0.4}}
\put(1.07,0.18){$m'_{3}$}
\put(0.9,0.8){\line(1,0){0.4}}
\put(1.07,0.83){$m'_{4}$}
\end{picture}
\normalsize
\caption{Brane configuration for $SU(2)$ gauge theory.}
\label{su2branesetup}
\end{figure}

We begin by deriving the SW curve for the simplest case; the compactified 5D $SU(2)$ gauge theory with four flavors.\footnote{
An alternative derivation of the SW curve is given in \cite{Minahan:1997ch} using a different point of view that exploits  the enhancement of the global symmetry to $E_5 = SO(10)$ \cite{Seiberg:1996bd}.} 
The brane setup is described in Table \ref{config} together with Figure \ref{su2branesetup} and includes $M=2$ NS5-branes with $N=2$ D4-branes suspended between them.

The asymptotic behavior of the NS5-branes is determined by the holomorphic extension of the equations of motion $\nabla^2 s=0$, which minimizes its worldvolume. If the $x^5$ direction is not compactified the asymptotic behavior of an NS5-brane at large $|v|$ is given by 
\begin{align}
\frac{s}{R_{10}} = \sum_{i=1}^2 \log (v-a'_i)
- \sum_{i=1}^{2} \log (v-b'_i)
+ {\rm const} \, ,
\end{align}
where $a'_i$ and $b'_i$ are the classical positions on the $v$-plane of the D4-branes attached to the NS5-brane from the left and the right respectively. Compactifying the $x^5$ direction is equivalent to periodically attaching D4-branes on a non-compact $x^5$ coordinate. Firstly, we concentrate on the first NS5-brane. The two flavor D4-branes that are attached to it at $v=m'_1$ and $v=m'_2$ can be reinterpreted as infinitely many D4-branes attached at $v = m'_1 + 2 \pi i R_5 n$ and $v=m'_2 + 2 \pi i R_5 n$, with $n=\cdots, -1,0,1,\cdots$. Similarly, two color D4-branes are attached at $v = a'_1 + 2 \pi i R_5 n$ and $v=a'_2 + 2 \pi i R_5 n$ from the other side. The asymptotic behavior of the first NS5-brane is, therefore,  given by
\begin{equation}
\begin{split}
\frac{s_{(1)}}{R_{10}} 
=& \sum_{n=-\infty}^{\infty} \left( \log (v - m'_1 - 2 \pi i R_5 n)
 + \log(v - m'_2 - 2 \pi i R_5 n) \right) - \\
& \sum_{n=-\infty}^{\infty} \left( \log (v - a'_1 - 2 \pi i R_5 n)
 + \log(v - a'_2 - 2 \pi i R_5 n) \right)
+ {\rm const} \, .
\end{split}
\nonumber
\end{equation}
Using the definitions of the periodic coordinates (\ref{def_tw}) and gauge theory parameters (\ref{tildema}) we can write the position of the first NS5-brane as
\begin{align}
t_{(1)} = & \, C \, 
\frac{\sinh\left( \frac{v-a'_1}{2R_5} \right)\sinh\left( \frac{v-a'_2}{2R_5} \right)}
{\sinh\left( \frac{v-m'_1}{2R_5} \right)\sinh\left( \frac{v-m'_2}{2R_5} \right)} \quad \longrightarrow \quad C 
\left\{
\begin{array}{l}
\sqrt{
\frac{ \tilde{m}_1 \, \tilde{m}_2
}{
\tilde{a}_1\,  \tilde{a}_2
}
}
\quad (w \to \infty) \\
\sqrt{ \frac{
\tilde{a}_1 \,  \tilde{a}_2
}
{
\tilde{m}_1  \, \tilde{m}_2
}
}
\quad (w \to 0)
\end{array}
\right. \, ,
\label{asym_t1}
\end{align}
where the expressions after the arrow are the asymptotic behaviors in the  $w \to \infty$ and $w \to 0$ regions.
Similarly, for the second NS5-brane we have
\begin{align}
t_{(2)}
= & C' \frac{\sinh\left( \frac{v-m'_3}{2R_5} \right)\sinh\left( \frac{v-m'_4}{2R_5} \right)}
{\sinh\left( \frac{v-a'_1}{2R_5} \right)\sinh\left( \frac{v-a'_2}{2R_5} \right)} \quad \longrightarrow \quad C' \left\{
\begin{array}{l}
\sqrt{ \frac{
\tilde{a}_1  \,  \tilde{a}_2
}
{
\tilde{m}_3  \,   \tilde{m}_4
} }
\quad (w \to \infty) \\
\sqrt{ \frac{
\tilde{m}_3 \, \tilde{m}_4
}{
\tilde{a}_1  \,  \tilde{a}_2
}
}
\quad (w \to 0)
\end{array}
\right. \, .
\end{align}
Following \cite{Witten:1997sc}, the distance between the two NS5-branes should give the 4D bare coupling constant $q \equiv \exp \left( 2 \pi i \tau_{\rm bare} \right)$ in the limit $R_5 \to \infty$. However, since we are studying the compactified 5D case, 
\begin{align}
\frac{t_{(2)}}{t_{(1)}}
= & \exp \left( \frac{s_{(1)}-s_{(2)}}{R_{10}} \right) 
= \frac{C'}{C} 
\frac{\prod_{\mathfrak{i}=1}^4 \sinh\left( \frac{v-m'_\mathfrak{i}}{2R_5} \right)}%
{\sinh^2 \left( \frac{v-a'_1}{2R_5} \right) \sinh^2 \left( \frac{v-a'_2}{2R_5} \right)} \, 
\label{1lp}
\end{align}
and the asymptotic distance between the NS5-branes at $w \to 0$ is different from the distance at $w \to \infty$ by a factor\footnote{Note that the index $i=1,2$ counts the color, while the index $\mathfrak{i}=1,\dots,4$ counts the flavor.} $\prod_\mathfrak{i} \tilde{m}_\mathfrak{i} \prod_i \tilde{a}_i^{-2}$. Thus, relating the constants $C$ and $C'$ to the 4D gauge theory parameters is subtle. In the rest of this section, we assume that these constants do not depend on the radius $R_5$ and that
\begin{align}
\frac{C'}{C} = q
\label{CCq}
\end{align}
is an exact relation for arbitrary $R_5$. 
This assumption indicates that the bare coupling constant is identified as the average of the
 two asymptotic distances, which is one of the most natural possibilities.
Indeed, as discussed in section \ref{subsec:topsu2}, this identification is justified by 
comparing the topological string partition function with Nekrasov partition function.

We continue by writting the 5D SW curve as a polynomial of degree two in $w$
\begin{align}
q_1(t) w^2 + q_2(t) w + q_3(t) =0 \, .
\label{WPolynomial}
\end{align}
In the $w \to \infty$ region, having two NS5-branes at $t=C \left(\frac{ \tilde{m}_1 \, \tilde{m}_2 }{ \tilde{a}_1\, \tilde{a}_2 }\right)^{\frac{1}{2}}$ 
and $t=Cq \left( \frac{ \tilde{a}_1 \, \tilde{a}_2 }{ \tilde{m}_3 \, \tilde{m}_4 } \right)^{\frac{1}{2}}$ leads to
\begin{align}
q_1(t) = \left(t-C \left(\frac{ \tilde{m}_1 \, \tilde{m}_2 }{ \tilde{a}_1\,  \tilde{a}_2 }\right)^{\frac{1}{2}}
\right)
\left(
t-Cq 
\left( \frac{ \tilde{a}_1 \, \tilde{a}_2 }{ \tilde{m}_3 \, \tilde{m}_4 }   \right)^{\frac{1}{2}}
\right) \, .
\end{align}
Similarly, in the $w \to 0$ region, we have the two NS5-branes at $t=C \left(\frac{ \tilde{a}_1\, \tilde{a}_2 }{ \tilde{m}_1 \, \tilde{m}_2 }\right)^{\frac{1}{2}}$ and $t=Cq \left( \frac{  \tilde{m}_3 \, \tilde{m}_4 }{ \tilde{a}_1 \, \tilde{a}_2 } \right)^{\frac{1}{2}}$, so  we obtain
\begin{align}
q_3(t) = d' \left(t-C \left(\frac{ \tilde{a}_1\, \tilde{a}_2 }{ \tilde{m}_1 \, \tilde{m}_2 }\right)^{\frac{1}{2}}
\right)
\left(
t-Cq 
\left( \frac{ \tilde{m}_3 \, \tilde{m}_4 }{ \tilde{a}_1 \, \tilde{a}_2 }   \right)^{\frac{1}{2}}
\right) \, ,
\end{align}
where $d'$ is a temporarily undetermined constant.
If  we, now, write the 5D SW curve as a polynomial of degree two in $t$ and consider the asymptotic behavior of the flavor D4-branes we can determine some more coefficients. In the $t \to \infty$ ($s \to -\infty$) region there are two flavor D4-branes at $w=\tilde{m}_1$ and $w=\tilde{m}_2$ and in the $t \to 0$ ($s \to \infty$) region two flavor D4-branes at $w=\tilde{m}_3$ and $w=\tilde{m}_4$. These boundary conditions constrain the SW curve to be of the form
\begin{align}
(w-\tilde{m}_1)(w-\tilde{m}_2) t^2 + P_2(w) t + d (w-\tilde{m}_3)(w-\tilde{m}_4) = 0 \, ,
\label{TPolynomial}
\end{align}
where $d$ is another undetermined constant that we will now fix.
The two forms (\ref{WPolynomial}) and (\ref{TPolynomial}) of the SW curve are simultaneously satisfied if we write
\footnotesize
\begin{equation}
\begin{split}
& (w-\tilde{m}_1)(w-\tilde{m}_2) t^2 \\
& - \left( \left[
C \left(\frac{ \tilde{m}_1 \, \tilde{m}_2 }{ \tilde{a}_1\, \tilde{a}_2 }\right)^{\frac{1}{2}}
+ Cq
\left( \frac{ \tilde{a}_1 \, \tilde{a}_2 }{ \tilde{m}_3 \, \tilde{m}_4 }   \right)^{\frac{1}{2}}
\right]w^2
- b \, w 
\right. \\
& \quad \left.
+ \, \tilde{m}_1 \, \tilde{m}_2 \left[
C \left(\frac{ \tilde{a}_1\, \tilde{a}_2 }{ \tilde{m}_1 \, \tilde{m}_2 }\right)^{\frac{1}{2}}
+ Cq
\left( \frac{  \tilde{m}_3 \, \tilde{m}_4 }{ \tilde{a}_1 \, \tilde{a}_2 }   \right)^{\frac{1}{2}}
\right]
\right) t \\
& + C^2 q \left( \frac{ \tilde{m}_1 \, \tilde{m}_2 }{ \tilde{m}_3 \,   \tilde{m}_4 } \right)^{\frac{1}{2}} (w-\tilde{m}_3)(w-\tilde{m}_4) \, =0 \, ,
\label{curveC}
\end{split}
\end{equation}
\normalsize
or equivalently
\footnotesize
\begin{equation}
\begin{split}
&\left(t-C \left(\frac{ \tilde{m}_1 \, \tilde{m}_2 }{ \tilde{a}_1\, \tilde{a}_2 }\right)^{\frac{1}{2}}
\right)
\left(
t-Cq
\left( \frac{ \tilde{a}_1 \, \tilde{a}_2 }{ \tilde{m}_3 \, \tilde{m}_4 }   \right)^{\frac{1}{2}}
\right) w^2 \\
& + \left( - \left( \tilde{m}_1 + \tilde{m}_2 \right) t^2
+ b \, t
- \, C^2q \left( \frac{ \tilde{m}_1 \, \tilde{m}_2 }{ \tilde{m}_3 \,   \tilde{m}_4 } \right)^{\frac{1}{2}}
\left( \tilde{m}_3 + \tilde{m}_4 \right)
\right) w \\
& + \,
\tilde{m}_1 \, \tilde{m}_2
\left(t-C \left(\frac{ \tilde{a}_1\, \tilde{a}_2 }{ \tilde{m}_1 \, \tilde{m}_2 }\right)^{\frac{1}{2}}
\right)
\left(
t-Cq
\left( \frac{ \tilde{m}_3 \, \tilde{m}_4 }{ \tilde{a}_1 \, \tilde{a}_2 }   \right)^{\frac{1}{2}}
\right)
\, =0 \, .
\label{curveCD}
\end{split}
\end{equation}
\normalsize
We have, thus, determined all the coefficients in the curve except for $b$, which is related to the Coulomb moduli parameter.

A comment on the weak coupling limit  ($q \equiv C'/C \ll 1$) of the obtained curve is in order. In this limit, the curve (\ref{curveC}) reduces to
\small
\begin{equation}
\begin{split}
& (w-\tilde{m}_1)(w-\tilde{m}_2) t^2 
- \left( 
C \left(\frac{ \tilde{m}_1 \, \tilde{m}_2 }{ \tilde{a}_1\, \tilde{a}_2 }\right)^{\frac{1}{2}} w^2
- b \, w 
+ \,  
C \left( \tilde{m}_1 \, \tilde{m}_2 \tilde{a}_1\, \tilde{a}_2 \right)^{\frac{1}{2}}
\right) t \\
& + C^2 q \left( \frac{ \tilde{m}_1 \, \tilde{m}_2 }{ \tilde{m}_3 \, \tilde{m}_4 } \right)^{\frac{1}{2}} (w-\tilde{m}_3)(w-\tilde{m}_4) = 0 \, .
\label{reduced_C}
\end{split}
\end{equation}
\normalsize
If we choose $C=1$ and assume that $b=\tilde{a}_1+\tilde{a}_2$ with $\tilde{a}_1 = \tilde{a}_2^{-1}$ the curve (\ref{reduced_C}) coincides with the one previously given in \cite{Nekrasov:1996cz,Brandhuber:1997cc,Eguchi:2000fv}. However, we want to emphasize that this expression is valid only under the weak coupling approximation. 
Moreover, we want to briefly comment on the 4D limit ($R_5 \to \infty$) of our 5D curve (\ref{curveC}). Details are provided in Appendix \ref{app:4DLimit}. In the 4D limit the curve (\ref{curveC}) reduces to the one obtained in \cite{Eguchi:2009gf}. This is an additional check of our result.

\bigskip

We are now ready to derive the duality map that corresponds to the exchange of the coordinates
\begin{align}
t_d = w  \, , \qquad w_d = t \, ,
\end{align}
where $d$ stands for dual.
Without any loss of generality we pick $|\tilde{m}_1| \geq |\tilde{m}_2|$,  $|\tilde{m}_3| \geq |\tilde{m}_4|$, $|(\tilde{m}_1)_d| \geq |(\tilde{m}_2)_d|$ and  $|(\tilde{m}_3)_d| \geq |(\tilde{m}_4)_d|$. Then, simply by comparing the two expressions (\ref{curveC}) and (\ref{curveCD}) of the SW curve we obtain the duality transformation
\begin{equation}
\begin{split}
(\tilde{m}_1)_d = C \left(\frac{ \tilde{m}_1 \, \tilde{m}_2 }{ \tilde{a}_1 \,  \tilde{a}_2 }\right)^{\frac{1}{2}} \, ,&
\qquad
(\tilde{m}_2)_d = \, C q
\left( \frac{ \tilde{a}_1 \, \tilde{a}_2 }{ \tilde{m}_3 \, \tilde{m}_4 }   \right)^{\frac{1}{2}}
\, , 
\\
(\tilde{m}_3)_d = C \left(\frac{ \tilde{a}_1 \, \tilde{a}_2 }{ \tilde{m}_1 \, \tilde{m}_2 }\right)^{\frac{1}{2}} \, ,&
\qquad
(\tilde{m}_4)_d = \, C q
\left( \frac{ \tilde{m}_3 \, \tilde{m}_4 }{ \tilde{a}_1 \, \tilde{a}_2 }   \right)^{\frac{1}{2}} \, ,
\\
b_d = b \, ,& \qquad
q_d = \left( \frac{\tilde{m}_2\tilde{m}_4}{\tilde{m}_1 \tilde{m}_3} \right)^{\frac{1}{2}} \, ,
\\
C_d = \tilde{m}_1^{\frac{1}{2}} \tilde{m}_3^{\frac{1}{2}}  
\, ,&
\qquad
(\tilde{a}_1)_d{} (\tilde{a}_2)_d{} = \, C^2 q 
\left( \frac{\tilde{m}_2\tilde{m}_3}{\tilde{m}_1 \tilde{m}_4} \right)^{\frac{1}{2}}
\, .
\label{Map_SU(2)_C}
\end{split}
\end{equation}

\smallskip

So far we have not specified the  $v$; a natural choice is to set it at the center of mass of the two D4-branes, where
\begin{align}
a_1 = - a_2 \quad \Rightarrow \quad \tilde{a}_1 = \tilde{a}_2^{-1} \, .
\label{U(1)}
\end{align}
Similarly, we pick the  $s=v_d$ so that 
\begin{align}
(\tilde{a}_1)_d = (\tilde{a}_2)_d^{-1}
\label{U(1)D}
\end{align}
is realized. This condition is satisfied when the constant $C$ is
\begin{align}
C = \left( \frac{\tilde{m}_1 \tilde{m}_4}{\tilde{m}_2\tilde{m}_3} \right)^{\frac{1}{4}}
q^{-\frac{1}{2}} 
\, = \, (\tilde{m}_1)_d^{\frac{1}{2}} (\tilde{m}_3)_d^{\frac{1}{2}} \, .
\end{align}


\begin{figure}[t]
 \begin{center}
  \includegraphics[width=6cm,clip]{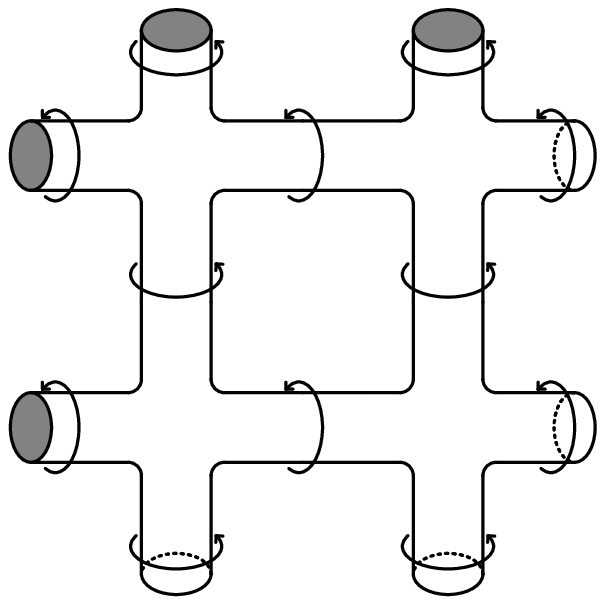}
 \put(-210,-20){\vector(0,1){190}}
 \put(-220,170){\vector(0,-1){190}}
  \put(-210,-20){\vector(1,0){240}}
    \put(30,-30){\vector(-1,0){240}}
   \put(50,-25){$s$}
      \put(50,-35){$t$}
 \put(-223,180){$w$}
  \put(-213,180){$v$}
    \put(-200,50){$w=\tilde{m}_1$}
        \put(-200,120){$w=\tilde{m}_2$}
           \put(-5,50){$w=\tilde{m}_3$}
        \put(-5,120){$w=\tilde{m}_4$}
  \put(-153,170){\footnotesize$t=\sqrt[4]{ \frac{ \tilde{m}_4}{ \tilde{m}_1 \tilde{m}_2^3 \tilde{m}_3  q^{2}} } $}
  \put(-73,170){\footnotesize$t=\sqrt[4]{ \frac{\tilde{m}_1 \tilde{m}_3  \tilde{m}_4^3 q^{2}}{  \tilde{m}_2 } } $}
    \put(-150,-5){\footnotesize$t=\sqrt[4]{ \frac{\tilde{m}_1^3 \tilde{m}_2  \tilde{m}_4}{ \tilde{m}_3 \, q^{2}} } $}
  \put(-73,-5){\footnotesize$t=\sqrt[4]{ \frac{\tilde{m}_1 \, q^{2}}{  \tilde{m}_2 \tilde{m}_3^3  \tilde{m}_4} } $}
          \put(-90,30){\small$A_1$}
        \put(-90,100){\small$A_2$}
           \put(-150,30){\small$M_1$}
        \put(-150,100){\small$M_2$}
               \put(-25,30){\small$M_3$}
        \put(-25,100){\small$M_4$}
            \put(-32,145){\small$T_4$}
           \put(-102,145){\small$T_3$}
              \put(-32,20){\small$T_2$}
           \put(-102,20){\small$T_1$}
               \put(-37,80){\footnotesize$\left(A_2\right)_d$}
           \put(-107,80){\footnotesize$\left(A_1\right)_d$}
 \end{center}
 \caption{In this figure the configuration of the M5-brane that leads to 5D $SU(2)$ four flavor is depicted.}
  \label{web}
\end{figure}

The transformation rule $b_d=b$ contains the duality transformation of the Coulomb moduli implicitly. We could in principle calculate $b$ explicitly in terms of the Coulomb moduli, as in \cite{Eguchi:2009gf}. However, for our purpose it is enough to consider the contour integrals of the SW one-form around the $A$ and the $A_d$ cycles. We depict all the cycles on the M5-brane in Figure \ref{web}. For the four junctions in Figure \ref{web} we find the following topological relations
\begin{equation}
\begin{split}
A_1 - T_1 - M_1 + \left(A_1\right)_d = 0 \, , \qquad
& - \left(A_1\right)_d - M_2 + T_3 + A_2 = 0 \, , \\
-A_2 + T_4 + M_4 - \left(A_2\right)_d = 0 \, , \qquad
& \left(A_2\right)_d+ M_3 - T_2 - A_1 = 0
\end{split}
\label{cycle_cons}
\end{equation}
among the cycles. In our conventions the integrals are positive when we go around $w = \infty$ and $t=\infty$  clock-wise. The relations (\ref{cycle_cons})  among the cycles can also be read as relations among cycle integrals by replacing $A \to \int_{A} \lambda_{SW}$. Using the expression of the SW one-form given in (\ref{SW_1-form})
\begin{align}
\lambda_{SW} 
= \frac{i}{(2\pi)^2 \ell_p{}^3} v ds
= \frac{i R_5 R_{10}}{(2\pi)^2 \ell_p{}^3} \log w \frac{dt}{t} \, ,
\end{align}
where the factor $1/(2\pi)^2 l_s{}^3$ is the tension of the M2-brane, we can calculate the cycle integrals. The integral around $M_1$ is obtained by considering the limit $t \to \infty$ and regarding the coordinate $w$ as a multivalued function of $t$. The curve around this region is approximately given by
\begin{equation}
(w(t) - \tilde{m}_1)(w(t) - \tilde{m}_2) t^2 \approx 0 
\qquad \Rightarrow \qquad  
w(t) = \left\{
\begin{array}{l}
\tilde{m}_1 + {\cal O}(t^{-1}) \\
\tilde{m}_2 + {\cal O}(t^{-1})
\end{array}
\right. \, .
\end{equation}
The first branch contributes to the integral around the cycle $M_1$. The contour integral can be carried out as
\begin{equation}
\oint_{M_1} \log w(t) \frac{dt}{t} 
= - \oint_{t=\infty} \left( \log \tilde{m}_1 + {\cal O}(t^{-1}) \right) \frac{dt}{t} = 2 \pi i \log \tilde{m}_1 \, .
\end{equation}
Similarly, we obtain the rest of the integrals around the cycles $M_\mathfrak{i}$ and $T_\mathfrak{i}$
\begin{equation}
\begin{split}
\oint_{M_\mathfrak{i}} \lambda_{SW} &= - \frac{R_5 R_{10}}{2 \pi \ell_p{}^3} \log\tilde{m_\mathfrak{i}} = m_\mathfrak{i} \, ,
\\
\oint_{T_\mathfrak{i}} \lambda_{SW} &= \frac{R_5 R_{10}}{2 \pi \ell_p{}^3} \log (\tilde{m_\mathfrak{i}})_d  = \left( m_\mathfrak{i} \right)_d \, ,
\end{split}
\label{cint}
\end{equation}
where $\mathfrak{i}=1,\dots,4$ is the flavor index. What is more, the contour integrals around the cycles $A_i$ and $\left(A_i\right)_d$, by definition, give the Coulomb moduli:
\begin{equation}
\begin{split}
\oint_{A_1} \lambda_{SW} 
&= - \oint_{A_2} \lambda_{SW}
= - \frac{R_5 R_{10}}{2 \pi \ell_p{}^3} \log \tilde{a} = a \, ,
\\
\oint_{A'_1} \lambda_{SW} 
&= - \oint_{A'_2} \lambda_{SW} 
= \frac{R_5 R_{10}}{2 \pi \ell_p{}^3} \log \tilde{a}_d = - a_d \, ,
\end{split}
\label{aint}
\end{equation} 
where $i=1,2$ is the color index. The first equality in each line is ensured by (\ref{cycle_cons}). The sign of $a_d$ in the second line is inverted because
\begin{align}
\label{equalSWoneform}
(\lambda_{SW})_d 
= \frac{i (R_5)_d (R_{10})_d}{(2\pi)^2 \ell_p{}^3} \log (w)_d \frac{d(t)_d}{(t)_d}
= \frac{i R_{10} R_5}{(2\pi)^2 l_p{}^3} \log t \frac{dw}{w}
= - \lambda_{SW} \, .
\end{align}
The four conditions in (\ref{cycle_cons}) consistently lead to the duality relation
\begin{align}
\tilde{a}_d = \left( \frac{\tilde{m}_2 \tilde{m}_4}
{\tilde{m}_1\tilde{m}_3} \right)^{\frac{1}{4}} q^{-\frac{1}{2}} \tilde{a} \, .
\label{dual_a}
\end{align}
The positions $a_1$ and $a_2$ of the color D4-branes were originally defined ``classically'' in the D4/NS5 brane setup and are not necessarily {equal} to $a$ defined by the cycle integral (\ref{aint}). However, the first equality of each line in (\ref{aint}) indicates that the classical conditions (\ref{U(1)}) and (\ref{U(1)D}) are satisfied even when we include the quantum effects. 

Summarizing, the duality map for the self-dual $SU(2)$ case is
\begin{equation}
\begin{split}
\label{map_reflection_su2}
(\tilde{m}_1)_d 
= \tilde{m}_1^{\frac{3}{4}} \tilde{m}_2^{\frac{1}{4}}
\tilde{m}_3^{-\frac{1}{4}} \tilde{m}_4^{\frac{1}{4}} q^{-\frac{1}{2}} \, ,&
\qquad
(\tilde{m}_2)_d 
= \tilde{m}_1^{\frac{1}{4}} \tilde{m}_2^{-\frac{1}{4}}
\tilde{m}_3^{-\frac{3}{4}} \tilde{m}_4^{-\frac{1}{4}} q^{\frac{1}{2}} \, ,
\\
(\tilde{m}_3)_d 
= \tilde{m}_1^{-\frac{1}{4}} \tilde{m}_2^{-\frac{3}{4}}
\tilde{m}_3^{-\frac{1}{4}} \tilde{m}_4^{\frac{1}{4}} q^{-\frac{1}{2}} \, ,&
\qquad
(\tilde{m}_4)_d 
= \tilde{m}_1^{\frac{1}{4}} \tilde{m}_2^{-\frac{1}{4}}
\tilde{m}_3^{\frac{1}{4}} \tilde{m}_4^{\frac{3}{4}} q^{\frac{1}{2}} \, ,
\\
q_d = \left( \frac{\tilde{m}_2\tilde{m}_4}{\tilde{m}_1 \tilde{m}_3} \right)^{\frac{1}{2}} \, ,&
\qquad
\tilde{a}_d = \left( \frac{\tilde{m}_2 \tilde{m}_4}%
{\tilde{m}_1\tilde{m}_3} \right)^{\frac{1}{4}} q^{-\frac{1}{2}} \tilde{a} \,
\end{split}
\end{equation}
and can, alternatively, be reorganized as
\begin{equation}
\begin{split}
(\tilde{m}_1)_d (\tilde{m}_2)_d (\tilde{m}_3)_d (\tilde{m}_4)_d 
= \frac{\tilde{m}_1 \tilde{m}_4}{\tilde{m}_2 \tilde{m}_3} \, ,&
\qquad
\frac{(\tilde{m}_1)_d (\tilde{m}_4)_d}{(\tilde{m}_2)_d (\tilde{m}_3)_d}
= \tilde{m}_1 \tilde{m}_2 \tilde{m}_3 \tilde{m}_4 \, ,
\\
\frac{(\tilde{m}_2)_d (\tilde{m}_4)_d}{(\tilde{m}_1)_d (\tilde{m}_3)_d}
= q^2 \, ,&
\qquad
{q_d}^2
= \frac{\tilde{m}_2 \tilde{m}_4}{\tilde{m}_1 \tilde{m}_3} \, ,
\\
\frac{(\tilde{m}_1)_d (\tilde{m}_2)_d}{(\tilde{m}_3)_d (\tilde{m}_4)_d}
= \frac{\tilde{m}_1 \tilde{m}_2}{\tilde{m}_3 \tilde{m}_4} \, ,&
\qquad
q_d^{-\frac{1}{2}} \tilde{a}_d = q^{-\frac{1}{2}} \tilde{a} \, .
\end{split}
\label{Map_5DC}
\end{equation}

\bigskip

\subsection{$SU(N)^{M-1} \leftrightarrow SU(M)^{N-1}$ duality}

\begin{figure}[t]
\centering
\tiny
\setlength{\unitlength}{4cm}
\vspace{10mm}
\begin{picture}(1,1)(0,0)
\linethickness{0.25mm}
\scriptsize
\put(-0.3,0){\line(0,1){1}}
\put(-0.36,1.15){NS5$_1$}
\put(0.1,0){\line(0,1){1}}
\put(0.06,1.15){NS5$_2$}
\put(0.5,0){\line(0,1){1}}
\put(0.44,1.15){NS5$_i$}
\put(0.9,0){\line(0,1){1}}
\put(0.81,1.15){NS5$_{i+1}$}
\put(1.3,0){\line(0,1){1}}
\put(1.19,1.15){NS5$_{M-1}$}
\put(1.7,0){\line(0,1){1}}
\put(1.63,1.15){NS5$_M$}
\tiny

\put(-0.7,0.1){\line(1,0){0.4}}
\put(-0.65,0.13){$m'_{1} = a'{}_{1}^{(0)}$}
\put(-0.7,0.25){\line(1,0){0.4}}
\put(-0.65,0.28){$m'_{2} = a'{}_{2}^{(0)}$}
\put(-0.5,0.45){\circle*{0.01}}
\put(-0.5,0.55){\circle*{0.01}}
\put(-0.5,0.65){\circle*{0.01}}
\put(-0.7,0.75){\line(1,0){0.4}}
\put(-0.78,0.78){$m'_{N-1} = a'{}_{N-1}^{(0)}$}
\put(-0.7,0.9){\line(1,0){0.4}}
\put(-0.66,0.93){$m'_{N} = a'{}_{N}^{(0)}$}

\scriptsize
\put(-1.1,0.08){D4$_1$}
\put(-1.1,0.23){D4$_2$}
\put(-1.1,0.73){D4$_{N-1}$}
\put(-1.1,0.88){D4$_{N}$}
\tiny

\put(-0.3,0.15){\line(1,0){0.4}}
\put(-0.17,0.18){$a'{}_{1}^{(1)}$}
\put(-0.3,0.3){\line(1,0){0.4}}
\put(-0.17,0.33){$a'{}_{2}^{(1)}$}
\put(-0.1,0.47){\circle*{0.01}}
\put(-0.1,0.53){\circle*{0.01}}
\put(-0.1,0.59){\circle*{0.01}}
\put(-0.3,0.7){\line(1,0){0.4}}
\put(-0.19,0.73){$a'{}_{N-1}^{(1)}$}
\put(-0.3,0.85){\line(1,0){0.4}}
\put(-0.17,0.88){$a'{}_{N}^{(1)}$}

\put(0.22,0.5){\circle*{0.01}}
\put(0.3,0.5){\circle*{0.01}}
\put(0.38,0.5){\circle*{0.01}}

\put(0.5,0.1){\line(1,0){0.4}}
\put(0.63,0.13){$a'{}_{1}^{(i)}$}
\put(0.7,0.27){\circle*{0.01}}
\put(0.7,0.34){\circle*{0.01}}
\put(0.7,0.41){\circle*{0.01}}
\put(0.5,0.5){\line(1,0){0.4}}
\put(0.63,0.53){$a'{}_{\alpha}^{(i)}$}
\put(0.7,0.67){\circle*{0.01}}
\put(0.7,0.74){\circle*{0.01}}
\put(0.7,0.81){\circle*{0.01}}
\put(0.5,0.9){\line(1,0){0.4}}
\put(0.63,0.93){$a'{}_{N}^{(i)}$}

\put(1.02,0.5){\circle*{0.01}}
\put(1.1,0.5){\circle*{0.01}}
\put(1.18,0.5){\circle*{0.01}}

\put(1.3,0.1){\line(1,0){0.4}}
\put(1.39,0.13){$a'{}_{1}^{(M-1)}$}
\put(1.3,0.3){\line(1,0){0.4}}
\put(1.39,0.33){$a'{}_{2}^{(M-1)}$}
\put(1.5,0.48){\circle*{0.01}}
\put(1.5,0.54){\circle*{0.01}}
\put(1.5,0.6){\circle*{0.01}}
\put(1.3,0.7){\line(1,0){0.4}}
\put(1.39,0.73){$a'{}_{N-1}^{(M-1)}$}
\put(1.3,0.9){\line(1,0){0.4}}
\put(1.39,0.93){$a'{}_{N}^{(M-1)}$}

\put(1.7,0.05){\line(1,0){0.4}}
\put(1.73,0.08){$m'_{N+1} = a'{}_{1}^{(M)}$}
\put(1.7,0.2){\line(1,0){0.4}}
\put(1.73,0.23){$m'_{N+2} = a'{}_{2}^{(M)}$}
\put(1.9,0.41){\circle*{0.01}}
\put(1.9,0.51){\circle*{0.01}}
\put(1.9,0.61){\circle*{0.01}}
\put(1.7,0.75){\line(1,0){0.4}}
\put(1.73,0.78){$m'_{2N-1} = a'{}_{N-1}^{(M)}$}
\put(1.7,0.87){\line(1,0){0.4}}
\put(1.73,0.9){$m'_{2N} = a'{}_{N}^{(M)}$}
\end{picture}
\vspace{5mm}
\normalsize
\caption{Brane setup for $SU(N)^{M-1}$ gauge theory, with vertical lines being D4-branes and horizontal ones being NS5-branes.
Without the loss of generality, we assume that $|\tilde{m}_1| \ge |\tilde{m}_2| \ge \cdots \ge |\tilde{m}_N|$ and $|\tilde{m}_{N+1}| \ge |\tilde{m}_{N+2}| \ge \cdots \ge |\tilde{m}_{2N}|$.
}
\label{branesetup}
\end{figure}
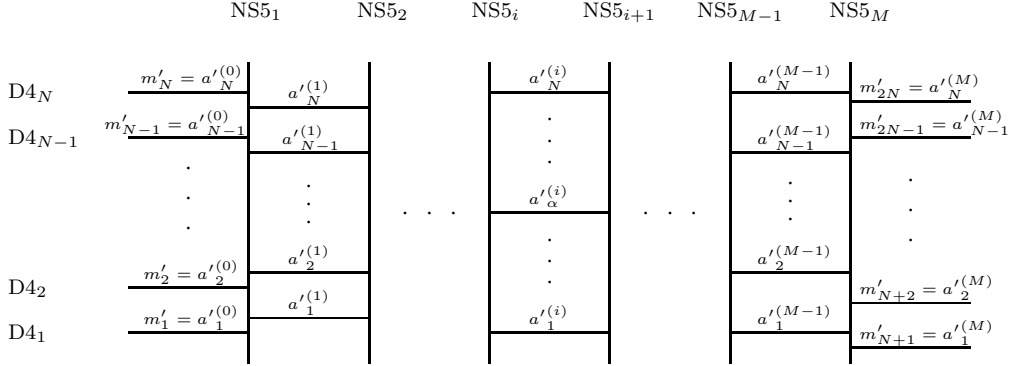

The extension of the analysis in the previous subsection to the generic linear quiver gauge theory is straightforward. The asymptotics of the NS5-branes and the D4-branes constrain the form of the SW curve of $SU(N)^{M-1}$ gauge theory to
\begin{align}
\prod^{N}_{\alpha=1} (w-\tilde{m}_\alpha)  t^M + \cdots + d \prod^{2N}_{\alpha=N+1} (w-\tilde{m}_\alpha)
=0
\label{NMw}
\end{align}
and
\footnotesize
\begin{align}
\prod^{M}_{i=1}\left[t-C_{(i)}\left(  
\frac{ \prod_{\alpha=1}^N \tilde{a}_\alpha^{(i-1)} }{ \prod_{\beta=1}^N \tilde{a}^{(i)}_\beta}
\right)^{1/2}
\right] w^N + \cdots
+ d' \prod^{M}_{i=1}\left[ t-C_{(i)}\left(
\frac{ \prod_{\alpha=1}^N \tilde{a}^{(i)}_\alpha}{ \prod_{\beta=1}^N \tilde{a}_\beta^{(i-1)} }
\right)^{1/2}
\right]
=0 \, ,
\label{NMt}
\end{align}
\normalsize
where we have defined
\begin{align}
\label{mass-a}
\tilde{a}_{\alpha}^{(0)} \equiv \tilde{m}_\alpha
\quad \text{and} \quad
\tilde{a}_{\alpha}^{(M)} \equiv \tilde{m}_{N+\alpha} \, .
\end{align}
The index $\alpha=1,\dots,N$ is used to count colors inside each single $SU(N)$ factor, whereas the index $i=1,\dots,M$ counts hypermultiplets along the quiver gauge group. This notation is further clarified in Figure \ref{branesetup}.
Similarly, the curve for $SU(M)^{N-1}$ can be written in two forms:
\begin{align}
\prod^{M}_{i=1} (w-\tilde{m}_i) t^N + \cdots
+ D \prod^{2M}_{i=M+1} (w-\tilde{m}_i)
= 0
\label{MNw}
\end{align}
and
\footnotesize
\begin{align}
\prod^{N}_{\alpha=1}\left[t - C_{(\alpha)}\left(
\frac{ \prod_{i=1}^M \tilde{a}_i^{(\alpha-1)} }{ \prod_{j=1}^M \tilde{a}^{(\alpha)}_j}
\right)^{1/2}
\right] w^M + \cdots
+ D' \prod^{N}_{\alpha=1}\left[t-C_{(\alpha)}\left(  
\frac{ \prod_{i=1}^M \tilde{a}^{(\alpha)}_i}{ \prod_{j=1}^M \tilde{a}_j^{(\alpha-1)} }
\right)^{1/2}
\right] = 0 \, ,
\label{MNt}
\end{align}
\normalsize
where, now, the index $i=1,\dots,M$  is used to count colors inside a single $SU(M)$ factor and the index $\alpha=1,\dots,N$ counts hypermultiplets along the product gauge group. 
Also, we define
\begin{align}
\tilde{a}_{i}^{(0)} \equiv \tilde{m}_i
\quad \text{and} \quad
\tilde{a}_{i}^{(N)} \equiv \tilde{m}_{M+i} \, .
\end{align}
As in the previous subsection, we now have to express the constants $C_{(i)}$ of the $SU(N)^{M-1}$ SW curve in terms of the gauge coupling constants $q^{(i)}$. The educated assumption 
\begin{align}
q^{(i)} = \frac{C_{(i+1)}}{C_{(i)}}  
\quad  \Rightarrow \quad
C_{(i)}= C \prod_{k=1}^{i-1} q^{(k)}   \, ,
\label{CK}
\end{align}
turns out to be the correct one, where $C \equiv C_{(1)}$ is some common constant that corresponds to the ambiguity of the rescaling of the coordinate $t$. The same relation (\ref{CK}) holds for the constants $C_{(\alpha)}$ of the $SU(M)^{N-1}$ SW curve in terms of the gauge coupling constants $q^{(\alpha)}$.
We are now ready to derive the duality map for the exchange $w \leftrightarrow t$. By comparing the SW curves (\ref{NMw}) and (\ref{MNt}) we obtain\footnote{We use the notation $(A B)_d\equiv A_d B_d$.}
\small
\begin{align}
\tilde{m}_\alpha
= \left( C_{(\alpha)}\left(  
\frac{ \prod_{i=1}^M \tilde{a}_i^{(\alpha-1)} }{ \prod_{j=1}^M \tilde{a}^{(\alpha)}_j}
\right)^{1/2} \right)_d \, ,
\quad
\tilde{m}_{N+\alpha}
= \left( C_{(\alpha)}\left(  
\frac{ \prod_{i=1}^M \tilde{a}^{(\alpha)}_i}{ \prod_{j=1}^M \tilde{a}_j^{(\alpha-1)} }
\right)^{1/2} \right)_d \, .
\label{comp1}
\end{align}
\normalsize
Furthermore, by comparing (\ref{NMt}) with (\ref{MNw}), we obtain
\small
\begin{align}
C_{(i)}\left(
\frac{ \prod_{\alpha=1}^N \tilde{a}_\alpha^{(i-1)} }{ \prod_{\beta=1}^N\tilde{a}^{(i)}_\beta}
\right)^{1/2}
= ( \tilde{m}_i )_d \, ,
\quad
C_{(i)}\left(  
\frac{ \prod_{\alpha=1}^N \tilde{a}^{(i)}_\alpha}{ \prod_{\beta=1}^N \tilde{a}_\beta^{(i-1)} }
\right)^{1/2}
= ( \tilde{m}_{M+i} )_d \, .
\label{comp2}
\end{align}
\normalsize
\begin{figure}[t]
 \begin{center}
  \includegraphics[width=5cm,clip]{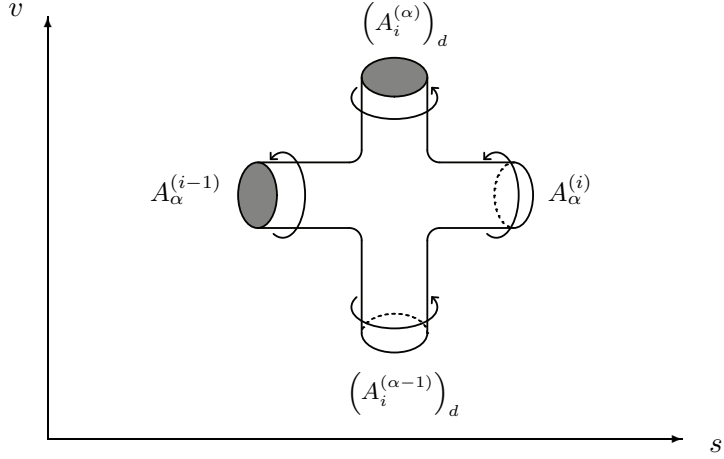}
  \put(-87,-5){\footnotesize$ \left( A^{(\alpha-1)}_i \right)_d$}
   \put(-11,70){\small $A^{(i)}_\alpha$}
  \put(-165,70){ \small$A^{(i-1)}_\alpha$}
 \put(-82,135){\footnotesize $\left( A^{(\alpha)}_i \right)_d$}
 \put(-200,-20){\vector(0,1){160}}
  \put(-200,-20){\vector(1,0){240}}
   \put(50,-25){$s$}
 \put(-215,140){$v$}
 \end{center}
 \caption{In this figure a ``junction'' of M5-branes is depicted. From this we read off  the rule for the ``conservation'' of the cycle integrals. }
\end{figure}

Again as in the previous subsection we have to impose  the ``conservation'' of the cycle integrals,
\begin{align}
\left( A^{(\alpha)}_i \right)_d 
- A^{(i)}_\alpha
- \left( A^{(\alpha-1)}_i \right)_d
+ A^{(i-1)}_\alpha = 0 \, ,
\label{cycle_cons_gen}
\end{align}
which leads to the map
\begin{align}
\label{aa_aa}
\frac{ (\tilde{a}^{(\alpha)}_i )_d}{(a^{(\alpha-1)}_i )_d}
= \frac{\tilde{a}^{(i)}_\alpha}{\tilde{a}^{(i-1)}_\alpha} \, .
\end{align}
Combining the equations above we get
\begin{equation}
\begin{split}
\left( \tilde{a}^{(\alpha)}_{i} \right)_d
&= C \left(  
\frac{\prod_{\gamma=1}^\alpha \tilde{a}^{(i)}_\gamma \prod_{\delta=\alpha+1}^N \tilde{a}^{(i-1)}_\delta}
{ \prod_{\gamma=1}^\alpha \tilde{a}_\gamma^{(i-1)} \prod_{\delta=\alpha+1}^N \tilde{a}_\delta^{(i)} }\right)^{1/2} \prod_{k=1}^{i-1} q^{(k)} \, , \\
\left( q^{(\alpha)} \right)_d
&= \left( \frac{\tilde{m}_{\alpha+1} \tilde{m}_{N+\alpha+1}}{\tilde{m}_{\alpha} \tilde{m}_{N+\alpha}} \right)^{1/2} \, .
\end{split}
\label{gen_map}
\end{equation}
The duality map as derived above still includes one unknown coefficient $C$ that reflects the freedom to rescale  the coordinate $t$. Moreover, the Coulomb moduli parameters are defined up to the choice of the  $v$.
A natural way to fix both is to impose 
\begin{align}
\prod_{\alpha=1}^N \tilde{a}^{(0)}_\alpha = 1
\quad \text{and} \quad
\prod_{i=1}^M (\tilde{a}^{(0)}_i)_d = 1 \, .
\label{prod_a}
\end{align}
The latter condition determines the constant $C$ to be
\begin{align}
C = 
{\prod_{\alpha=1}^N \left( \tilde{a}^{(M)}_{\alpha} \right)^{\frac{1}{2M}}}
\prod_{i=1}^{M-1} \left( q^{(i)} \right)^{- \frac{M-i}{M}} \, .
\label{Def_C}
\end{align}
At this point we have to stress that the constant $C$ depends on the choice of the origin (\ref{prod_a}) for the coordinates $v$ and $s$. Naively, one may think that the duality map depends on this choice. However, in terms of the {\it physical} gauge theory parameters the map is independent of this choice. A detailed description of the physical gauge theory parameters is given in Appendix \ref{phys-par}.

In terms of the physical gauge theory parameters the duality map is
\footnotesize
\begin{align}
& \left( \hat{a}_{i}^{(\alpha)} \right)_d
= \left( \tilde{m}_{\text{bif}}^{(i-1,i)} \right)^{\alpha-\frac{N}{2}}
\prod_{\gamma=1}^\alpha 
\left( 
\frac{\hat{a}_{\gamma}^{(i)} }{\hat{a}_{\gamma}^{(i-1)} } \right)
\left(
\frac{\hat{a}_\gamma^{(0)}}{\hat{a}_\gamma^{(M)}}
\right) ^{\frac{1}{M}}
 \prod_{k=1}^{M} 
\left( \tilde{m}_{\text{bif}}^{(k-1,k)} \right)^{\frac{N-2\alpha}{2M}}
\prod_{\ell=1}^{i-1} \left( q^{(\ell)} \right)^{\frac{i}{M}}
\prod_{\ell=i}^{M-1} \left( q^{(\ell)} \right) ^{- \frac{M-\ell}{M}} \, , \nonumber
\\
&\left( \tilde{m}^{(\alpha-1,\alpha)}_{\text{bif}} \right)_d
= \left(
\frac{\hat{a}_\alpha^{(M)}}{\hat{a}_\alpha^{(0)}}  
\prod_{k=1}^{M} \tilde{m}_{\text{bif}}^{(k-1,k)}
\right)^{\frac{1}{M}} \, ,
\\
& \left( q^{(\alpha)} \right)_d
= \left( \frac{\hat{a}^{(0)}_{\alpha+1} \hat{a}^{(M)}_{\alpha+1}}
{\hat{a}^{(0)}_{\alpha} \hat{a}^{(M)}_{\alpha}} \right)^{1/2} \, . \nonumber
\end{align}
\normalsize
%
%

\subsubsection*{$SU(M)$ $\leftrightarrow$ $SU(2)^{M-1}$ case}

Before ending this section we wish to consider the special case with $N=2$. This duality between $SU(M)$ and $SU(2)^{M-1}$ gauge theories is of particular interest due to its implications in 2D CFTs. Through the AGTW conjecture this duality relates four-point correlation functions of $q$-deformed $W_M$ Toda theories to $(M+2)$-correlation functions of $q$-deformed Liouville theories. This topic will be discussed in Section \ref{sec:GaugeToCFT}. For now we just give the map
\normalsize
\begin{equation}
\begin{split}
&(\tilde{m}_{1}^{\text{f}})_d 
= \left( \frac{(\tilde{m}^{\text{f}}_1)^{1+M} \tilde{m}^{\text{f}}_3}
{(\tilde{m}^{\text{f}}_2)^{1-M} \tilde{m}^{\text{f}}_4} \right) ^{\frac{1}{2M}} 
\prod_{k=1}^{M-1} \left( q^{(k)} \right)^{\frac{M-k}{M}} \, ,
\\
&(\tilde{m}_{i}^{\text{f}})_d 
= \left( \frac{\tilde{m}^{\text{f}}_1 \tilde{m}^{\text{f}}_3}%
{\tilde{m}^{\text{f}}_2 \tilde{m}^{\text{f}}_4} \right) ^{\frac{1}{2M}} 
\tilde{m}_{ \text{bif} }^{ (i-1,i)} 
\prod_{k=1}^{i-1} \left( q^{(k)} \right)^{-\frac{k}{M}}
\prod_{k=i}^{M-1} \left( q^{(k)} \right)^{\frac{M-k}{M}} \, ,
\\
&(\tilde{m}_{M}^{\text{f}})_d 
= \left( \frac{\tilde{m}^{\text{f}}_1 (\tilde{m}^{\text{f}}_3)^{1+M}}
{\tilde{m}^{\text{f}}_2 (\tilde{m}^{\text{f}}_4)^{1-M} } \right) ^{\frac{1}{2M}} 
\prod_{k=1}^{M-1} \left( q^{(k)} \right)^{-\frac{k}{M}} \, ,
\\
&(\tilde{m}_{M+1}^{\text{f}})_d 
= \left( \frac{(\tilde{m}_2^{\text{f}})^{1+M} \tilde{m}_4^{\text{f}}}
{(\tilde{m}_1^{\text{f}})^{1-M} \tilde{m}_3^{\text{f}}}
\right) ^{\frac{1}{2M}}
\prod_{k=1}^{M-1} \left( q^{(k)} \right)^{-\frac{M-k}{M}} \, ,
\\
&(\tilde{m}_{M+i}^{\text{f}})_d 
= \left( \frac{\tilde{m}_2^{\text{f}} \tilde{m}_4^{\text{f}}}
{\tilde{m}_1^{\text{f}} \tilde{m}_3^{\text{f}}}
\right) ^{\frac{1}{2M}}
\tilde{m}_{ \text{bif} }^{ (i-1,i)} 
\prod_{k=1}^{i-1} \left( q^{(k)} \right)^{\frac{k}{M}}
\prod_{k=i}^{M-1} \left( q^{(k)} \right)^{-\frac{M-k}{M}} \, ,
\\
&(\tilde{m}_{2M}^{\text{f}})_d 
= \left( \frac{\tilde{m}_2^{\text{f}} (\tilde{m}_4^{\text{f}})^{1+M}}
{\tilde{m}_1^{\text{f}} (\tilde{m}_3^{\text{f}})^{1-M}}
\right) ^{\frac{1}{2M}}
\prod_{k=1}^{M-1} \left( q^{(k)} \right)^{\frac{k}{M}} \, ,
\end{split}
\end{equation}
\normalsize
and
\normalsize
\begin{equation}
\begin{split}
&\left( \tilde{a}_{1}^{\text{f}} \right)_d
= \tilde{a}^{(1)}_{\text{f}} 
\left( 
\frac{(\tilde{m}_2^{\text{f}})^{1-M} \tilde{m}_4^{\text{f}}}
{(\tilde{m}_1^{\text{f}})^{1-M} \tilde{m}_3^{\text{f}}}
\right) ^{\frac{1}{2M}}
\prod_{i=1}^{M-1} \left( q^{(i)} \right)^{-\frac{M-k}{M}} \, ,
\\
&\left( \tilde{a}_{i}^{\text{f}} \right)_d
= \frac{\tilde{a}^{(i)}_{\text{f}} }{\tilde{a}^{(i-1)}_{\text{f}} }
\left( \frac{\tilde{m}_2^{\text{f}} \tilde{m}_4^{\text{f}}}
{\tilde{m}_1^{\text{f}} \tilde{m}_3^{\text{f}}}
\right) ^{\frac{1}{2M}}
\prod_{k=1}^{i-1} \left( q^{(k)} \right)^{\frac{k}{M}}
\prod_{k=i}^{M-1} \left( q^{(k)} \right)^{-\frac{M-k}{M}} \, ,
\\
&\left( \tilde{a}_{M}^{\text{f}} \right)_d
= \frac{1}{\tilde{a}^{(M-1)}_{\text{f}} }
\left( \frac{\tilde{m}_2^{\text{f}} \tilde{m}_4^{\text{f}}}
{\tilde{m}_1^{\text{f}} \tilde{m}_3^{\text{f}}}
\right) ^{\frac{1}{2M}}
\prod_{i=1}^{M-1} \left( q^{(i)} \right)^{\frac{k}{M}} \, ,
\\
&q_d
= \left( \frac{ \tilde{m}_{1}^{\text{f}} \tilde{m}_{4}^{\text{f}}}
{\tilde{m}_{2}^{\text{f}} \tilde{m}_{3}^{\text{f}}}
\right)^{1/2} \, .
\end{split}
\end{equation}
\normalsize
It is interesting to note that the mass parameters and the gauge coupling constant of the dual $SU(M)$ theory are completely independent of the Coulomb moduli parameters of the original $SU(2)^{M-1}$ theory.

\section{Topological string derivation}
\label{sec:TopStringDeriv}

In the previous section we presented a derivation of the duality map using the Seiberg-Witten formalism. Here, we will present an independent derivation (or check) using Nekrasov's partition function. We compute Nekrasov's partition functions for 5D $\mathcal{N}=1$ $SU(N)^{M-1}$ and $SU(M)^{N-1}$ linear quivers and show that they are equal when we relate their gauge theory parameters with the duality map (\ref{gen_map}).

The computation of Nekrasov's partition functions  is performed using topological string theory. As we reviewed in Section  \ref{sec:Review}, topological string theory offers an alternative derivation of the gauge theory partition functions and most importantly provides {\it a rewriting of the partition function} in a form in which the  duality is manifest.  
It is unlikely that gauge theory reasoning alone would lead to this rewriting. However, from the string theory point of view it is natural. Due to the fact that the partition function is read off from a toric diagram, symmetries that arise from the CY geometry (and are obscured otherwise) are manifest in this formalism.

In the previous section we used the type IIA D4/NS5 brane setup to realize the linear quiver gauge theories. As we discussed in Section \ref{sec:Review}, the D4/NS5 brane configuration is dual to type IIA string theory compactified on CY$_3$. We are interested in the special class of Calabi-Yau manifolds that satisfy the toric condition  and lead to $SU(N)$ gauge theory. Theses CY$_3$ are completely specified by their toric diagrams. In the case of linear quivers the toric diagram is essentially same as the brane diagram.

\begin{figure}[htbp]
 \begin{center}
  \includegraphics[width=130mm,clip]{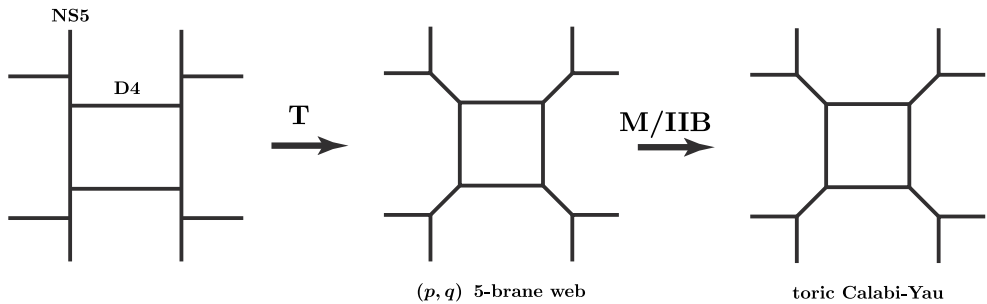}
 \end{center}
 \caption{The D4/NS5 system is T-dual to IIB $(p,q)$ 5-brane system.
The M/IIB duality relates it to M-theory on the corresponding toric CY.}
 \label{BraneToToric}
\end{figure}
Following \cite{Aharony:1997bh},  the D4/NS5  brane setup in IIA theory is T-dual to IIB $(p,q)$-brane web system (D5/NS5). When uplifting this system to M-theory via M/IIB duality, we obtain M-theory on $M^4\times \textrm{CY}_3\times S^1$ where CY is a toric three-fold whose $(p,q)$-cycle shrinks.
In this way the D4/NS5 system is equivalent to M-theory on toric CY,
or IIA on CY which is the usual geometric engineering setup. This connection is illustrated in Figure \ref{BraneToToric}.
 

Given the toric diagram we can use the topological vertex formalism to calculate Nekrasov's partition function of 4D $\mathcal{N}=2$ gauge theories. We should stress again that in this paper we study the Nekrasov partition function for the 5D uplift of the 4D gauge theory. The 5D Nekrasov partition function is precisely equal to the topological string partition function\footnote{To be precise, the obtained topological string partition function is the Nekrasov partition function for the $U(N)$ gauge theory whose Coulomb moduli parameters are constrained as $a_1=-a_2=a$. According to \cite{Alday:2009aq}, this constrained partition function is still not precisely $SU(N)$. The difference is the overall factor which in \cite{Alday:2009aq} is called the U(1) factor and is independent of the Coulomb moduli. This $U(1)$ factor does not affect the low-energy effective coupling constants which we studied in the previous section.}; of course after the appropriate identification of the gauge theory parameters with the string theory parameters.

Writing down  the topological string partition function is simple using the topological vertex formalism. The procedure was reviewed in Section~\ref{sec:Review}.
What is quite tedious is to bring the topological string partition function in the form given by Nekrasov. For that we have to perform the sums. Such calculations have previously been done by \cite{Aganagic:2002qg,Iqbal:2003zz,Eguchi:2003sj,Iqbal:2004ne,Taki:2007dh}. They involve summations over Young diagrams. The summand contains Schur and skew-Schur functions. The calculation is quite technical so we hide most of the details in Appendix \ref{app:NekrasovTop}. We first warm up with the  $SU(2)$ case and then present the general  $\mathcal{N}=2$ $SU(N)^{M-1}$ linear quiver in its full glory. Once we bring the topological string partition function in the form that was given by Nekrasov, we obtain the identification between the gauge theory and the string theory parameters. Using this identification we finally write down the duality map that is identical to the one found in Section \ref{sec:MtheoryDeriv}.

\subsection{$SU(2)$ gauge theory with four flavors}
\label{subsec:topsu2}

\begin{figure}[htbp]
 \begin{center}
  \includegraphics[width=40mm,clip]{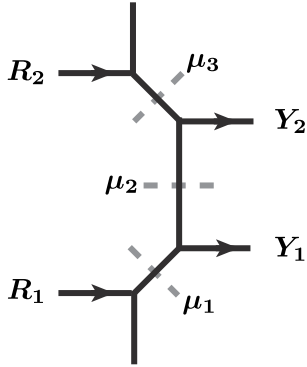}
 \end{center}
 \caption{The sub-diagram that engineers the bifundamental hypermultiplet of $SU(2)$ quiver gauge theories, where $R_i,Y_i,\mu_i$ denote Young diagrams. The parameters $Q_1,Q_2,Q_3$ are associated with the line labeled by the Young diagram $\mu_1, \mu_2, \mu_3$, respectively.}
 \label{fig:bifund}
\end{figure}

\begin{figure}[htbp]
 \begin{center}
  \includegraphics[width=70mm,clip]{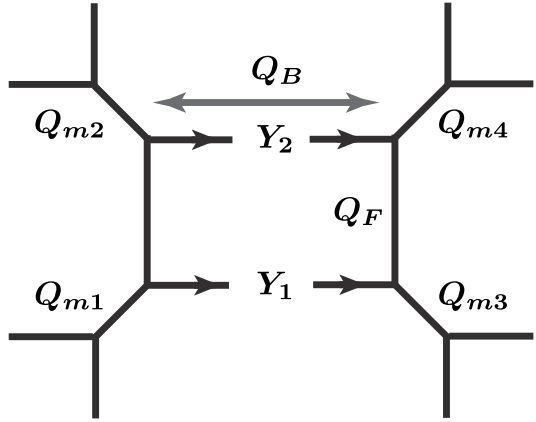}
 \end{center}
 \caption{The toric diagram that gives $SU(2)$ SQCD with four fundamental hypermultiplets. Since the eight external lines are semi-infinite half-lines, we assign the empty Young diagram $\emptyset$ to them.}
 \label{fig:4flavSQCD}
\end{figure}

In this subsection we compute the topological string partition function for $SU(2)$ SQCD with four flavors using the topological vertex formalism. The toric diagram from which we read off the partition functions is depicted in Figure~\ref{fig:4flavSQCD}. Due to the symmetry of the diagram, it is convenient to first consider the ``half-geometry'' of the corresponding toric CY shown in Figure~\ref{fig:bifund}.
This sub-diagram is dual to two horizontal D4-branes crossing a vertical NS5-brane. The vertical sequence of closed loops describes a combination of the two-cycles in CY$_3$ which give a vector multiplet and two fundamental hypermultiplets. As we will see, the K\"ahler parameters of the three two-cycles inside the CY geometry correspond to the Coulomb moduli of the $SU(2)$ gauge group and two of the hypermultiplet masses. After gluing the two ``half-geometries'' according to Figure~\ref{fig:4flavSQCD} we obtain a toric CY$_3$ with six two-cycles, which correspond to the Coulomb moduli parameter $a$, the four flavor masses $m_{\mathfrak{i}}$ ($\mathfrak{i}=1,\dots,4$) and the gauge coupling constant $q$.

First, we will focus on the computation of the contribution from this  ``half-geometry'' to the topological partition function. The Young diagram $R_i$ is kept to be arbitrary for as long as possible so that  this computation can be used also for more generic cases like $SU(2)^{M-1}$ gauge theory. For the $SU(2)$ gauge theory with four flavors we then set $R_1=R_2=\emptyset$ in order to get the partition function.

Using the topological vertex formalism we read off the following sub-amplitude for the local geometry depicted in Figure~\ref{fig:bifund} 
\footnotesize
\begin{equation}
\begin{split}
L^{\,R_1\,Y_1}_{\,R_2\,Y_2}
(Q_{1},\,Q_{2},\,Q_{3})
&\equiv
\sum_{\mu_1,\mu_2,\mu_3}
(-1)^{|\mu_2|}\mathfrak{q}^{\frac{1}{2}\kappa_{\mu_2} }\,
\prod_{i=1}^3(-Q_i)^{|\mu_i|}
\\
&\rule{0pt}{3ex}
\qquad \times 
C_{\emptyset R_1 \mu_1}(\mathfrak{q})\,
C_{\mu_2 Y_1^T \mu_1^T}(\mathfrak{q})\,
C_{\mu_3 Y_2^T\mu_2^T}(\mathfrak{q})\,
C_{\mu_3^T R_2 \emptyset}(\mathfrak{q})
\\
&\rule{0pt}{4ex}=
\,S_{R_1}(\mathfrak{q}^\rho)\,S_{R_2}(\mathfrak{q}^\rho)\,S_{Y_1^T}(\mathfrak{q}^\rho)\,S_{Y_2^T}(\mathfrak{q}^\rho)\\
&\rule{0pt}{4ex}
\qquad \times
\sum_{\mu_1,\mu_2,\mu_3}
\sum_{\zeta,\eta}
S_{\mu_1^T}(-Q_1\mathfrak{q}^{R_1+\rho})\,
S_{\mu_1/\zeta}(\mathfrak{q}^{Y_1^T+\rho})\,
S_{\mu_2/\zeta}(\mathfrak{q}^{Y_1+\rho})\\
&\qquad
\times
S_{\mu_2/\eta}(Q_2\mathfrak{q}^{Y_2^T+\rho})\,
S_{\mu_3/\eta}(Q_2^{-1}\mathfrak{q}^{Y_2+\rho})\,
S_{\mu_3^T}(-Q_2Q_3\mathfrak{q}^{R_2^T+\rho})\, .
\end{split}
\end{equation}
\normalsize
The second line of the equation is obtained by inserting the definition of the vertex function (\ref{def_topv}).  In order to get a closed form of the topological string amplitude we have to perform the summation explicitly. For that we employ the Cauchy formulas
\begin{equation}
\begin{split}
\sum_\eta
S_{\eta/R_1}(x)
S_{\eta/R_2}(y)
&=
\prod_{i,j}
(1-x_iy_j)^{-1}
\sum_\eta
S_{R_1/\eta}(y)
S_{R_2/\eta}(x) \, , \\
\sum_\eta
S_{\eta^T/R_1}(x)
S_{\eta/R_2}(y)
&=
\prod_{i,j}
(1+x_iy_j)
\sum_\eta
S_{R_1^T/\eta}(y)
S_{R_2^T/\eta^T}(x) \, .
\end{split}
\end{equation}
Notice that $S_{R/\emptyset}=S_{R}$ and $S_{\emptyset/R}=\delta_{R,\emptyset}$. By using these formulas repeatedly, we obtain the following closed form of the amplitude
\small
\begin{equation}
\begin{split}
L^{\,R_1\,Y_1}_{\,R_2\,Y_2}
(Q_{1},\,Q_{2},\,Q_{3})
&=
\,S_{R_1}(\mathfrak{q}^\rho)\,S_{R_2}(\mathfrak{q}^\rho)\,S_{Y_1^T}(\mathfrak{q}^\rho)\,S_{Y_2^T}(\mathfrak{q}^\rho)
\\
&\rule{0pt}{4ex}\times
\frac{
\left[R_1, Y_1^T \right]_{Q_1}
\left[Y_2, R_2^T \right]_{Q_3}
\left[R_1, Y_2^T \right]_{Q_1Q_2}
\left[Y_1, R_2^T, \right]_{Q_2Q_3}
}
{
\left[ Y_1, Y_2^T \right]_{Q_2}
\left[ R_1, R_2^T \right]_{Q_1Q_2Q_3}
} \, ,
\end{split}
\end{equation}
\normalsize
where $[*,*]_Q$ is defined as
\begin{align}
\left[Y_1, Y_2 \right]_{Q}
\equiv
\prod_{i,j=1}^\infty
(1-Q\mathfrak{q}^{Y_{1i}+Y_{2j}-i-j+1})
=\left[Y_2, Y_1 \right]_{Q} \, .
\end{align}
The instanton contribution of the gauge theory partition function is given by the normalized amplitude
\footnotesize
\begin{equation}
\begin{split}
&\tilde{L}^{\,R_1\,Y_1}_{\,R_2\,Y_2}
\equiv
\frac{L^{\,R_1\,Y_1}_{\,R_2\,Y_2}}{L^{\,\emptyset\,\emptyset}_{\,\emptyset\,\emptyset}}\\
&=\rule{0pt}{4ex}
2^{|R_1|+|R_2|+|Y_1|+|Y_2|}
\left(\sqrt{\frac{Q_1}{Q_3}}\right)^{|R_1|-|R_2|}
\left(\sqrt{{Q_1}{Q_3}}\right)^{|Y_1|+|Y_2|}
\mathfrak{q}^{-\frac{1}{4}(\kappa_{R_1}-\kappa_{R_2}-\kappa_{Y_1}+\kappa_{Y_2})}
\\
&\times\rule{0pt}{4ex}
\label{bifundamp}
S_{R_1}(\mathfrak{q}^\rho)\,S_{R_2}(\mathfrak{q}^\rho)\,S_{Y_1^T}(\mathfrak{q}^\rho)\,S_{Y_2^T}(\mathfrak{q}^\rho)
\frac{
P^{-1}_{Y_1R_1}(Q_1)
P^{-1}_{Y_2R_1}(Q_1Q_2)
P^{-1}_{R_2Y_1}(Q_2Q_3)
P^{-1}_{R_2Y_2}(Q_3)
}
{P^{-1}_{R_2R_1}(Q_1Q_2Q_3)P^{-1}_{Y_2Y_1}(Q_2)} \, ,
\end{split}
\end{equation}
\normalsize
where we define the function $P$ as follows \cite{Konishi:2003qq}:
\begin{equation}
\begin{split}
&\frac{1
}{P_{Y_1Y_2}(\mathfrak{q},Q)}
\equiv
\prod_{(i,j)\in Y_1}
\sinh \frac{\beta}{2}
\left(
a+\hbar(Y_{1\,i}+Y^T_{2\,j}-i-j+1)
\right)
\\
&\rule{0pt}{5ex}
\quad\qquad\qquad\quad
\times
\prod_{(i,j)\in Y_2}
\sinh \frac{\beta}{2}
\left(a+\hbar
(-Y^T_{1\,j}-Y_{2\,i}+i+j-1)\right)
\end{split}
\end{equation}
for $\mathfrak{q}=e^{-\beta \hbar}$ and $Q=e^{-\beta a}$. To get this expression we have used the formulas (\ref{NPrelation}) and (\ref{bracket_N}).

With this sub-amplitude at hand we move on to the computation of the full partition function of $SU(2)$ SQCD with four flavors. The associated toric diagram is depicted in Figure~\ref{fig:4flavSQCD}. The partition function for this toric diagram is obtained by gluing two sub-diagrams $\tilde{L}$ according to
\small
\begin{align}
Z_{\,\textrm{inst}}=
\sum_{Y_1,Y_2}
Q_B^{|Y_1|+|Y_2|}\,
\mathfrak{q}^{\frac{\kappa_{Y_1}}{2}-\frac{\kappa_{Y_2}}{2}}\,
\tilde{L}^{\,\emptyset\,Y_1}_{\,\emptyset\,Y_2}\,
(Q_{m1},Q_{F},Q_{m2})\,
\tilde{L}^{\,\emptyset\,Y_2^T}_{\,\emptyset\,Y_1^T}\,
(Q_{m4},Q_{F},Q_{m3}) \, .
\end{align}
\normalsize
This expression is written in terms of the string theory parameters used in geometric engineering. By comparing it with the Nekrasov partition function in \cite{Nekrasov:2002qd} we obtain the identifications
\small
\begin{equation}
\begin{split}
&Q_{m1}=e^{-\beta(m_1^{\text{f}} - a)}
=\frac{\tilde{m}_1}{\tilde{a}},\quad
Q_{m2}=e^{-\beta(-m_2^{\text{f}}-a)}
= \frac{1}{\tilde{m}_2\tilde{a}}
,\quad\cr
&Q_{m3}=e^{-\beta(m_3^{\text{f}}-a)}
=\frac{\tilde{m}_3}{\tilde{a}}
,\quad
Q_{m4}=e^{-\beta(-m_4^{\text{f}}-a)}
= \frac{1}{\tilde{m}_4 \tilde{a}}, \quad
Q_{F}=e^{-2a\beta}
=\tilde{a}^{2} \, ,
\label{Q_Def}
\end{split}
\end{equation}
\normalsize
where the second equality is written in the M-theoretical parametrization from Section \ref{sec:MtheoryDeriv}.

In particular, the ``numerator contribution" of the left sub-diagram
$
\tilde{L}^{\,\emptyset\,Y_1}_{\,\emptyset\,Y_2}\,
(Q_{m1},Q_{F},Q_{m2})
$
takes the form
\begin{equation}
\begin{split}
&P^{-1}_{Y_1\emptyset}(Q_{m1})
P^{-1}_{Y_2\emptyset}(Q_{m1}Q_F)
P^{-1}_{\emptyset Y_1}(Q_FQ_{m2})
P^{-1}_{\emptyset Y_2}(Q_{m2})
\cr
&\rule{0pt}{4ex} \qquad\qquad\qquad
= (-1)^{|Y_1|+|Y_2|}
\prod_{f=1,2}
Z_{\,\textrm{fund}}(\,a,\vec{Y},m_f,\hbar;\beta\,) \, ,
\end{split}
\end{equation}
where we have used (\ref{Pinverse}) together with Nekrasov's expresion (\ref{Nek5Dfund}). This is precisely the contribution from the two fundamental hypermultiplets with masses $m_1$ and $m_2$. The sub-diagram $\tilde{L}^{Y_2^T\emptyset}_{Y_1^T\emptyset}$ 
gives the contribution of the two fundamental hypermultiplets with masses $m_3$ and $m_4$ in a similar fashion. Moreover, the remaining part has the interpretation of contribution from the vector multiplet
\begin{align}
\frac{S_{Y_1}(\mathfrak{q}^\rho)\,S_{Y_2}(\mathfrak{q}^\rho)}{P^{-1}_{Y_1 Y_2}(Q_F)}
\frac{S_{Y_2^T}(\mathfrak{q}^\rho)\,S_{Y_1^T}(\mathfrak{q}^\rho)}{P^{-1}_{Y_2^T Y_1^T}(Q_F)}
 = (-4)^{-|Y_1|-|Y_2|} Z_{\textrm{vector}} (\,a,\vec{Y},\hbar;\beta\,)
\end{align}
where we have used (\ref{specializedSchur}), (\ref{Ptranspose}) and (\ref{Nek5Dvect2}). The details of the computation can be found in Appendix \ref{app:NekrasovTop}. We have thus exactly reproduced the Nekrasov partition function \cite{Nekrasov:2002qd}, where the instanton factor is given by
\begin{align}
q
=Q_B\sqrt{Q_{m1}Q_{m2}Q_{m3}Q_{m4}}
=Q_B\, \tilde{a}^{-2} \sqrt{\frac{\tilde{m}_1\tilde{m}_3}{\tilde{m}_2\tilde{m}_4}} \, .
\label{q_Def}
\end{align}
It is remarkable that the parametrization (\ref{q_Def}) does not depend on the $\Omega$ background parameter $\mathfrak{q}=e^{ - \beta \hbar}$.

We can interpret the identification of the parameters (\ref{Q_Def}) and (\ref{q_Def}) in the context of the brane setup. Taking into account that $\tilde{a}_{\alpha}$ and $\tilde{m}_\mathfrak{i}$, correspond to the positions of the color branes and the flavor branes respectively, the ratio of them corresponds to the distance between the corresponding branes as in (\ref{Q_Def}). In the similar way, by rewriting (\ref{q_Def}) as 
$$
q = \sqrt{(Q_{m1} Q_B Q_{m3}) \times (Q_{m2} Q_B Q_{m4})} 
$$
we see that $q$ can be interpreted as the average distance between the two NS5-branes in the $v \rightarrow \pm \infty$ asymptotic regions, as discussed in Section \ref{sec:MtheoryDeriv}. This observation justifies the identification (\ref{CCq}).

\subsubsection*{Reflection symmetry}

The topological string partition function $Z=\left(L_{\,\emptyset\emptyset}^{\,\emptyset\emptyset}\right)^2 Z_{\textrm{inst}}$ (without normalization) has the same symmetries as the toric diagram it is based on. The normalization factor $\left(L_{\,\emptyset\emptyset}^{\,\emptyset\emptyset}\right)^2$ gives the perturbative contribution of the Nekrasov partition function, while $Z$ is equivalent to the full Nekrasov partition function. Therefore, a graphical symmetry of the toric diagram is also a symmetry of the full quantum gauge theory, including perturbative and instanton corrections.

Typical examples are the reflection symmetries of Figure~\ref{fig:4flavSQCD}. The partition function is invariant under reflection along the diagonal axis when it is performed together with the transformation 
\begin{align}
Q_{m2}\leftrightarrow Q_{m3}\, , \quad
Q_{B}\leftrightarrow Q_{F} \, .
\end{align}
This reflection symmetry implies the following duality relations
\begin{equation}
\begin{split}
(Q_{m1})_d=Q_{m1}\, ,\quad
(Q_{m2})_d=Q_{m3}\, ,&\quad
(Q_{m3})_d=Q_{m2}\, ,\quad
(Q_{m4})_d=Q_{m4}\, ,\cr
(Q_{B})_d=Q_{F}\, ,&\quad
(Q_{F})_d=Q_{B}\, .
\end{split}
\end{equation}
In the M-theory language, it is an invariance of the Nekrasov partition function under the transformation
\small
\begin{equation}
\begin{split}
\frac{(\tilde{m}_1)_d}{(\tilde{a})_d}=
\frac{\tilde{m}_1}{\tilde{a}} \, , \quad 
\frac{1}{(\tilde{m}_2)_d (\tilde{a})_d}
=\frac{\tilde{m}_3}{\tilde{a}} \, , \quad
\rule{0pt}{4ex}
& \frac{(\tilde{m}_3)_d}{(\tilde{a})_d}
= \frac{1}{\tilde{m}_2\tilde{a}} \, , \quad 
\frac{1}{(\tilde{m}_4)_d (\tilde{a})_d} = \frac{1}{\tilde{m}_4 \tilde{a}} \, ,
\cr
\rule{0pt}{4ex}
q_d\,{(\tilde{a})_d^{2}}\sqrt{\frac{(\tilde{m}_2)_d(\tilde{m}_4)_d}{(\tilde{m}_1)_d(\tilde{m}_3)_d}}
=\tilde{a}^{2} \, , \quad 
\rule{0pt}{4ex}
& (\tilde{a})_d^{2}=q\,{\tilde{a}^{2}}\sqrt{\frac{\tilde{m}_2\tilde{m}_4}{\tilde{m}_1\tilde{m}_3}} \, .
\label{top_su2_map1}
\end{split}
\end{equation}
\normalsize
This is the self-duality of the holomorphic sector of the 5D gauge theory in the Coulomb branch.

Note that if we combine this duality map with a known symmetry of the Nekrasov partition function, we obtain another expression for this self-duality. In particular, we can combine with a simultaneous change of the sign of the Coulomb moduli and the masses discussed at the end of Appendix \ref{app:NekrasovTop}, which correspond to $\tilde{m}_{\mathfrak{i}} \to \tilde{m}_{\mathfrak{i}}^{-1}$, $\tilde{a} \to \tilde{a}^{-1}$. Acting this symmetry transformation on both the original and the dual theory in (\ref{top_su2_map1}) we obtain
\small
\begin{equation}
\begin{split}
\frac{(\tilde{a})_d}{(\tilde{m}_1)_d}=
\frac{\tilde{a}}{\tilde{m}_1} \, , \quad
(\tilde{m}_2)_d (\tilde{a})_d
=\frac{\tilde{a}}{\tilde{m}_3} \, ,\quad
\rule{0pt}{4ex}
& \frac{(\tilde{a})_d}{(\tilde{m}_3)_d}
=\tilde{m}_2\tilde{a} \, , \quad
(\tilde{m}_4)_d (\tilde{a})_d = \tilde{m}_4 \tilde{a} \, ,
\cr
\rule{0pt}{4ex}
q_d\,{(\tilde{a})_d^{-2}}\sqrt{\frac{(\tilde{m}_1)_d(\tilde{m}_3)_d}{(\tilde{m}_2)_d(\tilde{m}_4)_d}}
=\tilde{a}^{-2} \, , \quad
\rule{0pt}{4ex}
& (\tilde{a})_d^{-2}=q\,{\tilde{a}^{-2}}\sqrt{\frac{\tilde{m}_1\tilde{m}_3}{\tilde{m}_2\tilde{m}_4}} \, .
\label{top_su2_map2}
\end{split}
\end{equation}
\normalsize
It is now straightforward to see that (\ref{top_su2_map2}) is equivalent to the duality map (\ref{Map_5DC}) which was derived using the M5-brane construction in the previous section.\footnote{It is possible to obtain (\ref{top_su2_map2}) directly if we define the toric diagram in Figure~\ref{fig:4flavSQCD} with all the arrows reversed. In that case, the parametrization of the geometric engineering parameters in (\ref{Q_Def}) gets inverted. In this article, we use the standard parametrization from \cite{Iqbal:2004ne}.} The point here is that the self-dual $\Omega$-background deformation $\hbar$ maintains this duality, since we have shown that not only the Seiberg-Witten solution but also the Nekrasov partition function is invariant under the duality transformation. This result is due to the nontrivial fact that the duality map does not depend on the $\Omega$-background parameter $\hbar$. In the following we will see that this duality for high rank gauge theories also satisfy this non-trivial property.

{\subsection{$SU(N)^{M-1}\leftrightarrow SU(M)^{N-1}$ duality}}

\begin{figure}[htbp]
 \begin{center}
  \includegraphics[width=110mm,clip]{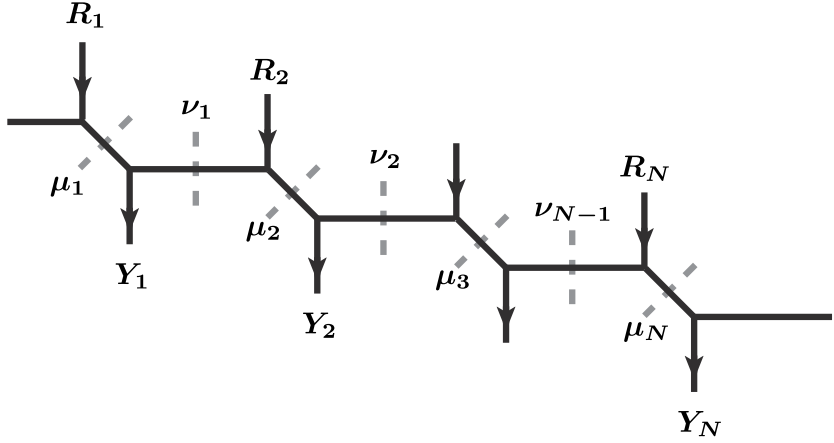}
 \end{center}
 \caption{The sub-diagram of the toric diagram for $SU(N)$ quiver gauge theories. The parameters $Q_{m\alpha}$ and $Q_{F\alpha}$ are associated with the lines labeled by the Young diagrams $\mu_{\alpha}$ and $\nu_{\alpha}$, respectively.}
 \label{fig:quiv2}
\end{figure}
We will now generalize to $SU(N)$ quiver gauge theories. As in the previous subsection, we divide the toric diagram into sub-diagrams along its symmetry lines. The sub-diagram of the generic ladder geometry we choose to compute is shown in Figure~\ref{fig:quiv2}. Using the topological vertex formalism, the contribution coming from this sub-diagram is
\small
\begin{equation}
\begin{split}
&H_{\,Y_1Y_2\cdots Y_N}^{\,R_1R_2\cdots R_N}\,(\,\mathfrak{q},Q_{m1},\cdots,Q_{mN},Q_{F1},\cdots,Q_{FN})
\\
&=\sum_{\mu_{1,\cdots ,N},\nu_{1,\cdots ,N}}
\prod_{\alpha=1}^N\,
(-Q_{m\alpha})^{|\mu_\alpha|}\,
\prod_{\alpha=1}^{N-1}\,
(-Q_{F\alpha})^{|\nu_\alpha|}\,
\\
&\times
C_{ \emptyset R_1\mu_1^T}\,
C_{ \nu_1^TY_1^T \mu_1 }\,
C_{ \nu_1 R_2 \mu_2^T}\,
C_{ \nu_2^T  Y_2^T \mu_2}\,
C_{  \nu_2 R_3\mu_3^T}\,
C_{ \nu_3^TY_3^T \mu_3}\,
\cdots
C_{ \nu_{N-1}R_N \mu_N^T}\,
C_{ \emptyset Y_N^T \mu_N} \, .
\end{split}
\end{equation}
\normalsize
By substituting the definition of the topological vertex, we obtain the following expression
\small
\begin{equation}
\begin{split}
&H_{\,Y_1Y_2\cdots Y_N}^{\,R_1R_2\cdots R_N}
\\
&=\rule{0pt}{4ex}
\prod_{\alpha=1}^N\,
S_{R_\alpha}(\mathfrak{q}^{\rho})\,
S_{Y_\alpha^T}(\mathfrak{q}^{\rho})\,
\sum_{\mu,\nu,\eta,\zeta}
\prod_{\alpha=1}^N\,
(-Q_{m\alpha})^{|\mu_\alpha|}\,
(-Q_{F\alpha})^{|\nu_\alpha|}
\\
&\quad\times
\prod_{\alpha=1}^N\,
S_{ \nu_{\alpha-1}/\zeta_{\alpha-1}}(\mathfrak{q}^{R_\alpha^T+\rho})\,
S_{ \mu_\alpha/\zeta_{\alpha-1}}(\mathfrak{q}^{R_\alpha+\rho})\,
S_{ \mu_\alpha^T/\eta_\alpha}(\mathfrak{q}^{Y_\alpha^T+\rho})\,
S_{ \nu_\alpha^T/\eta_\alpha}(\mathfrak{q}^{Y_\alpha+\rho}) \, .
\end{split}
\end{equation}
\normalsize
Note that the lines on the left and right edges are associated with a singlet or empty $\nu_0=\nu_N=\emptyset$ tableu. We can take the summation since all the $\kappa$-factors from the framing factors are canceled out in the partition function. This type of subdiagram is called ``the vertex on a strip geometry" and is studied extensively in \cite{Iqbal:2004ne}. By using the formula (B.1) from \cite{Taki:2007dh} we can compute it explicitly:
\begin{equation}
\begin{split}
H_{\,Y_1Y_2\cdots Y_N}^{\,R_1R_2\cdots R_N}
&=
\frac{\prod_{\alpha=1}^N\,
S_{R_\alpha}(\mathfrak{q}^{\rho})\,
S_{Y_\alpha^T}(\mathfrak{q}^{\rho})}
{\prod_{1\leq \alpha<\beta\leq N}\left[R_\alpha, R_\beta^T\right]_{Q_{\alpha\beta}}
\left[Y_\alpha,Y_\beta^T \right]_{Q_{m\alpha}^{-1}Q_{\alpha\beta}Q_{m\beta}}
}
\\
&\qquad\times\rule{0pt}{4ex}
\prod_{1\leq \alpha<\beta\leq N}
\left[Y_\alpha,R_\beta^T \right]_{Q_{m\alpha}^{-1}Q_{\alpha\beta}}
\prod_{1\leq \alpha\leq\beta\leq N}
\left[R_\alpha, Y_\beta^T\right]_{Q_{\alpha\beta}Q_{m\beta}} \, ,
\end{split}
\end{equation}
where $Q_{\alpha\beta} = \prod_{a=\alpha}^{\beta-1}Q_{m\,a}Q_{F\,a}$. Normalizing this sub-diagram by dividing with $H_{\emptyset\cdots}^{\emptyset\cdots}$ we obtain
\small
\begin{equation}
\begin{split}
\tilde{H}_{\,Y_1Y_2\cdots Y_N}^{\,R_1R_2\cdots R_N}
&=
\frac{\prod_{\alpha=1}^N\,
S_{R_\alpha}(\mathfrak{q}^{\rho})\,
S_{Y_\alpha^T}(\mathfrak{q}^{\rho})}
{\prod_{1\leq \alpha<\beta\leq N}
N_{R_\beta R_\alpha}(Q_{\alpha\beta})\,
N_{Y_\beta Y_\alpha}(Q_{m\alpha}^{-1}Q_{\alpha\beta}Q_{m\beta})
}
\\
&\qquad\times\rule{0pt}{4ex}
\prod_{1\leq \alpha<\beta\leq N}
N_{R_\beta Y_\alpha}(Q_{m\alpha}^{-1}Q_{\alpha\beta})
\prod_{1\leq \alpha\leq\beta\leq N}
N_{Y_\beta R_\alpha}(Q_{\alpha\beta}Q_{m\beta}) \, ,
\end{split}
\label{Htilde}
\end{equation}
\normalsize
where 
\begin{align}
N_{Y_1Y_2}(\mathfrak{q},Q)
\equiv \frac{\left[Y_1^T, Y_2 \right]_{Q}}
{\left[\emptyset, \emptyset \right]_{Q}}
=N_{Y_2^TY_1^T}(\mathfrak{q},Q) \, .
\end{align}

\begin{figure}[htbp]
 \begin{center}
  \includegraphics[width=100mm,clip]{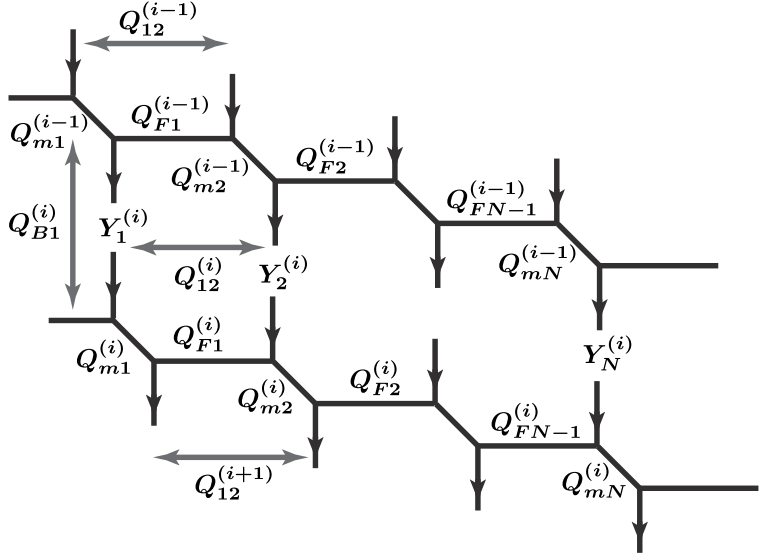}
 \end{center}
 \caption{The local structure of the toric diagrams for the $SU(N)$ linear quivers.}
 \label{fig:genQuiver}
\end{figure}

\begin{figure}[htbp]
 \begin{center}
  \includegraphics[width=140mm,clip]{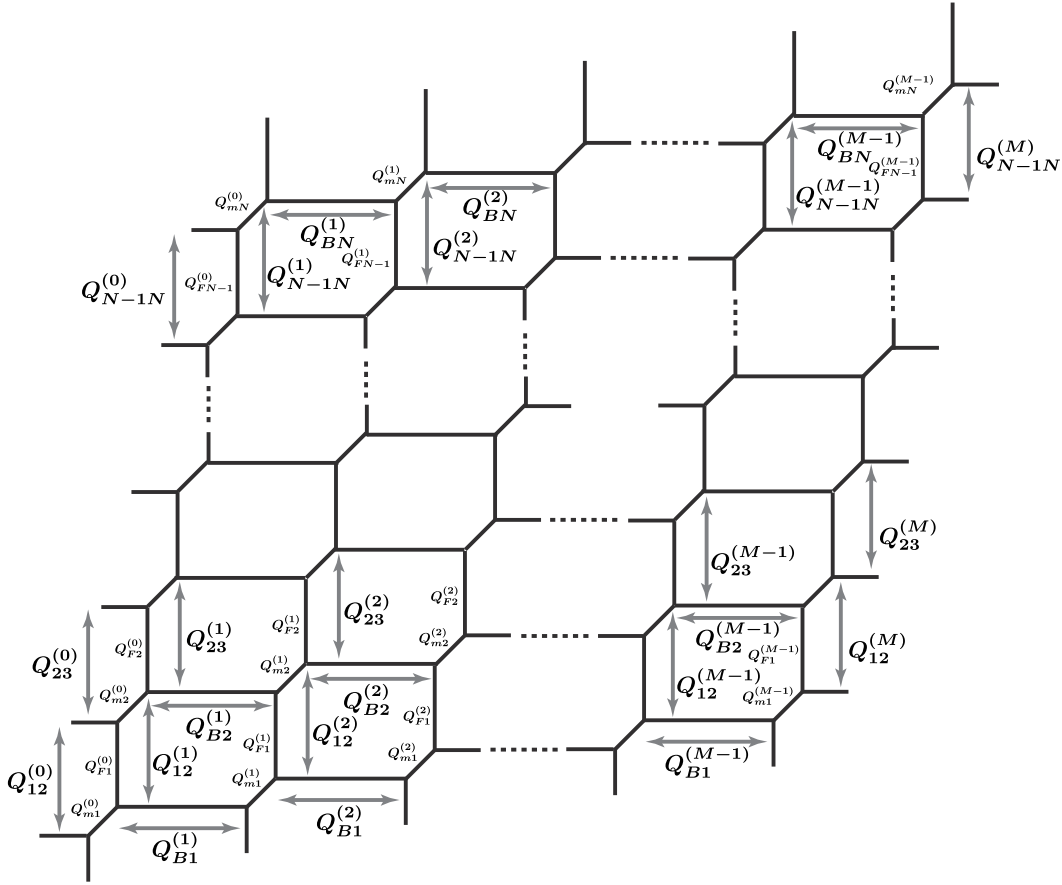}
 \end{center}
 \caption{The toric diagram for the linear $SU(N)$  quiver gauge theory. $Q_{B}^{(i)}$ is related to the gauge coupling constant $q^{(i)}$ of the $i$-th gauge group $SU(N)_{(i)}$. The Coulomb moduli of the $i$-th gauge group are given by $Q_{\alpha\beta}^{(i)}$. Since $SU(N)_{(0)}$ and $SU(N)_{(M)}$ are in fact not  gauge groups but global, flavor symmetries, the K\"ahler parameters $Q_{\alpha\beta}^{(0)}$ and $Q_{\alpha\beta}^{(M)}$ on the edges encode the masses of the (anti-) fundamental hypermultiplets living on the endpoints of the corresponding quiver diagram.}
 \label{fig:large}
\end{figure}

The generic $SU(N)^{M-1}$ linear quiver theories with fundamental and bifundamental hypermultiplets are engineered using CY$_3$ with
 linear toric diagrams  that are obtained by gluing local structures depicted in Figure~\ref{fig:quiv2}.
The partition function for $SU(N)^{M-1}$ quivers is read off from  Figure~\ref{fig:large} and is written in terms of the local structure of the geometry that is illustrated in Figure~\ref{fig:genQuiver}, 
\begin{equation}
\label{genQuivPartFuncmain}
\begin{split}
Z_{\,\textrm{inst}} &=
\sum
\cdots
\sum_{Y_1^{(i)},\cdots,Y_N^{(i)}}
\cdots
\prod_{\alpha}(-Q_{B\alpha}^{(i)})^{|Y_\alpha|}
\cr
&\quad \cdots \tilde{H}_{\,Y_1^{(i+1)} Y_2^{(i+1)} \cdots Y_N^{(i+1)}}%
^{\,Y_1^{(i)} Y_2^{(i)} \cdots Y_N^{(i)}}\,
(\,Q^{(i)}_{m1}, \cdots,Q^{(i)}_{mN},Q^{(i)}_{F1},\cdots,Q^{(i)}_{FN})
\cr
&\quad \times
\tilde{H}^{\,Y_1^{(i-1)} Y_2^{(i-1)} \cdots Y_N^{(i-1)}}%
_{\,Y_1^{(i)} Y_2^{(i)} \cdots Y_N^{(i)}}\,
(\,Q^{(i-1)}_{m1},\cdots,Q^{(i-1)}_{mN},Q^{(i-1)}_{F1},\cdots,Q^{(i-1)}_{FN})
\cdots.
\end{split}
\end{equation}
\normalsize
This expression is written in terms of the string theory parameters. In order to make contact  with Nekrasov's partition function 
 we introduce the following identification for the K\"ahler parameters
\begin{align}
Q^{(i)}_{\alpha\beta}=e^{-\beta(a^{(i)}_\alpha-a^{(i)}_\beta)}
= \frac{\tilde{a}_{\alpha}^{(i)}}{\tilde{a}_{\beta}^{(i)}} \, , \qquad 
Q^{(i)}_{m\alpha}=e^{-\beta(a^{(i)}_\alpha-{a}^{(i+1)}_\alpha-m^{(i,i+1)})}
= \frac{\tilde{a}^{(i)}_{\alpha}}{\tilde{a}^{(i+1)}_{\alpha}} \, ,
\label{Qab_Qm}
\end{align}
where $\tilde{a}_{\alpha}^{(i)}$ are the M-theory parameters from Section \ref{sec:MtheoryDeriv}. Here, we have defined 
\begin{align}
Q_{\alpha \, \alpha+1}^{(i)} \equiv Q_{m \, \alpha}^{(i)} Q_{F\alpha}^{(i)}
\label{QF_Qab}
\end{align}
(see Figure~\ref{fig:genQuiver}), which leads to the identification
\begin{align}
Q_{F\alpha}^{(i)}
{= \frac{\tilde{a}_{\alpha}^{(i+1)}}{\tilde{a}_{\alpha+1}^{(i)}}}
= \exp \left[ - \beta (a_{\alpha}^{(i+1)}-a_{\alpha+1}^{(i)} + m^{(i,i+1)}) \right] \, .
\label{QF_param}
\end{align}
Note that the parameters above satisfy the following relations
\begin{equation}
\begin{split}
Q^{(i)}_{\alpha\beta} 
&= Q^{(i-1)}_{\alpha\beta} \frac{Q^{(i-1)}_{m\beta}}{Q^{(i-1)}_{m\alpha}} \, ,
\\
Q_{F\alpha}^{(i)} 
&= Q_{F \alpha}^{(i-1)} \frac{Q_{m \, \alpha+1}^{(i-1)}}{Q_{m\alpha}^{(i)}} \, .
\label{rec_QF}
\end{split}
\end{equation}

When comparing with the expression (\ref{GluingOfQuiver}) of the Nekrasov partition function \cite{Nekrasov:2002qd, Fucito:2004gi}, we can show that the topological string partition function (\ref{genQuivPartFuncmain}) is almost the same as the partition function for the quiver gauge theory.
\begin{figure}[htbp]
 \centering
  \includegraphics[width=120mm,clip]{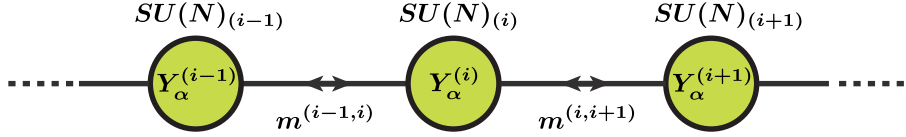}
 \caption{The quiver diagram for the $SU(N)$ quiver gauge theory associated with Figure \protect\ref{fig:genQuiver}. } 
 \label{fig:genQuiverDiag}
\end{figure}
The remaining problem is to find the identification between the base K\"ahler parameters $Q_{B}^{(i)}$ and the gauge coupling constants $q^{(i)}$. For the purpose, let us compute the corresponding part of the partition function (\ref{genQuivPartFuncmain}). The K\"ahler parameters $Q_{B\alpha}$ of the two-cycles $B+m_\alpha F+n_\alpha F'$ are given by
\begin{align}
Q_{B1}^{(i)}=Q_B^{(i)}
\qquad \text{and} \qquad
Q_{B\alpha}^{(i)} 
= Q^{(i)}_{B\,\alpha-1} \frac{Q^{(i)}_{m \, \alpha-1}}{ Q^{(i-1)}_{m\alpha}} \, .
\label{rec_QB}
\end{align}
The part of the partition function (\ref{genQuivPartFuncmain}) that contains these parameters is
\begin{equation}
\begin{split}
\prod_{\alpha}(-Q_{B\alpha}^{(i)})^{|Y_\alpha^{(i)}|}
&=(-Q_B^{(i)})^{\sum{|Y_\alpha^{(i)}|}}
\frac
{
\prod_{1\leq\alpha<\beta\leq N}
(Q^{(i)}_{m\alpha})^{|Y_\beta^{(i)}|}
}
{
\prod_{2\leq\alpha\leq\beta\leq N}
(Q^{(i-1)}_{m\alpha})^{|Y_\beta^{(i)}|}
}\\
&=(-Q_B ^{(i)}Q^{(i-1)}_{m1})^{\sum{|Y_\alpha^{(i)}|}}
\frac
{
\prod_{1\leq\alpha<\beta\leq N}
(Q^{(i)}_{m\alpha})^{|Y_\beta^{(i)}|}
}
{
\prod_{1\leq\alpha\leq\beta\leq N}
(Q^{(i-1)}_{m\alpha})^{|Y_\beta^{(i)}|}
}
\, .
\label{QB_q}
\end{split}
\end{equation}
By comparing (\ref{QB_q}) with (\ref{GaugeCouplingKahlers}), we find the following relation between the gauge coupling constants and the K\"ahler parameters of the base $\mathbb{P}^1$
\begin{equation}
\begin{split}
Q_B^{(i)}
= &
q^{(i)}
\frac{1}{Q^{(i-1)}_{m1}}
\prod_{\alpha=1}^N
\sqrt{\frac{Q^{(i-1)}_{m\alpha}}
{Q^{(i)}_{m\alpha}}}
= 
q^{(i)}
\frac{\tilde{a}_1^{(i)}}{\tilde{a}_1^{(i-1)}}
\prod_{\alpha=1}^N
\frac{\sqrt{\tilde{a}_{\alpha}^{(i-1)} \tilde{a}_{\alpha}^{(i+1)}}}%
{\tilde{a}_{\alpha}^{(i)}}
\cr
= &
q^{(i)}
\exp \left[
- \beta 
\left( 
a_1^{(i)} - a_1^{(i-1)} - \frac{N-2}{2} m^{(i-1,i)} + \frac{N}{2} m^{(i,i+1)}  
\right)
\right] \, .
\end{split}
\label{QB_param}
\end{equation}
Inserting  (\ref{Qab_Qm}), (\ref{QF_param}) and (\ref{QB_param}) into the topological partition function (\ref{genQuivPartFuncmain})
gives precisely the Nekrasov partition function for the quiver theory in Figure~\ref{fig:genQuiverDiag}.

From the relations (\ref{QF_Qab}), (\ref{rec_QF}), and (\ref{rec_QB}) we see that all the $Q_{\alpha \beta}^{(i)}$, $Q_{F\alpha}^{(i)}$ for $1 \le i \le M-1$ and $Q_{B\alpha}^{(i)}$ for $2 \le \alpha \le N$ are not independent. The toric diagram in Figure~\ref{fig:genQuiverDiag} shows that the remaining parameters  $Q_{m\alpha}^{(i)}$, $Q_{B}^{(i)}$ and $Q_{F \alpha}^{(0)}$ are independent; this can also be deduced from the relations (\ref{Qab_Qm}), (\ref{QF_param}) and (\ref{QB_q}). Moreover, the number of parameters add up to $(M+1)(N+1)-3$, which is the same as the number of parameters of the $SU(N)^{M-1}$ gauge theory as discussed in Section \ref{subsec:review_duality}. Therefore, they are one-to-one correspondent with the gauge theory parameters. 

The duality map of the reflection transformation  is given by
\begin{equation}
\begin{split}
(Q_{mi}^{(\alpha-1)})_d = Q_{m \alpha}^{(i-1)} \, , & \quad
(Q_{Fi}^{(\alpha-1)})_d = Q_{B \alpha}^{(i)} \, , \cr
\qquad(Q_{Bi}^{(\alpha)})_d = Q_{F \alpha}^{(i-1)} \, , & \quad
(Q_{\alpha \, \alpha+1}^{(i)})_d = Q_{m \, i+1}^{(\alpha-1)} Q_{B \, i+1}^{(\alpha)} \, .
\label{top_gen_map}
\end{split}
\end{equation}
Again, by taking into account (\ref{QF_Qab}), (\ref{rec_QF}), and (\ref{rec_QB}), we see that in (\ref{top_gen_map}) the second map (in the first line) for $2 \le \alpha \le N$, the third (in the second line) for $2 \le i \le M$ and the fourth can be derived from the remaining maps are redundant. Therefore, the independent ones are the first map, the second with $\alpha=1$ and the third with $i=1$.

Finally, we show that the duality map obtained here is equivalent to the one we found using the M5-brane analysis. To do so, it is enough to show that the independent duality relations in (\ref{top_gen_map}) can also be derived from the relations (\ref{gen_map}). Just like the $SU(2)$ case, we combine this duality map with a {simultaneous} transformation $\tilde{a}_{\alpha}^{(i)} \to \tilde{a}_{\alpha}^{(i)} {}^{-1}$ and $(\tilde{a}_{\alpha}^{(i)})_d \to (\tilde{a}_{\alpha}^{(i)})_d^{-1}$, which is a symmetry of the Nekrasov partition function. Then, the first map, the second map (in the first line) for $\alpha=1$, and the third map (in the second line) for $i=1$ in (\ref{top_gen_map}) respectively become
\small
\begin{equation}
\begin{split}
\left( \frac{\tilde{a}^{(\alpha-1)}_{i}}{\tilde{a}^{(\alpha)}_{i}} \right)_d
&= \frac{\tilde{a}^{(i-1)}_{\alpha}}{\tilde{a}^{(i)}_{\alpha}} \, ,
\\
\left( \frac{\tilde{a}^{(1)}_{i}}{\tilde{a}^{(0)}_{i+1}} \right)_d
&= q^{(i)}
\frac{\tilde{a}_1^{(i-1)}}{\tilde{a}_1^{(i)}}
\prod_{\alpha=1}^N
\frac{\sqrt{\tilde{a}_{\alpha}^{(i-1)} \tilde{a}_{\alpha}^{(i+1)}}}%
{\tilde{a}_{\alpha}^{(i)}} \, ,
\\
\frac{\tilde{a}^{(1)}_{i}}{\tilde{a}^{(0)}_{i+1}} 
&= \left(
q^{(i)}
\frac{\tilde{a}_1^{(i-1)}}{\tilde{a}_1^{(i)}}
\prod_{\alpha=1}^N
\frac{\sqrt{\tilde{a}_{\alpha}^{(i-1)} \tilde{a}_{\alpha}^{(i+1)}}}%
{\tilde{a}_{\alpha}^{(i)}}
\right)_d
\label{top_final_map}
\end{split}
\end{equation}
\normalsize
after inserting  (\ref{Qab_Qm}), (\ref{QF_param}) and (\ref{QB_param}). The first line in (\ref{top_final_map}) is precisely the relation (\ref{aa_aa}), from which the duality map (\ref{gen_map}) is derived. The second line can be derived from (\ref{gen_map}), while the third line in (\ref{top_final_map}) is the same as the second line in (\ref{top_final_map})  with the parameters of the original theory exchanged with the ones of the dual theory. 
Since the role of the original  and the dual theory can be exchanged the third line of (\ref{top_final_map}) is also correct.
We have thus shown that the duality map obtained from the topological string analysis is identical to the one obtained from the M-theory analysis.

\section{From 5D $\mathcal{N}=1$ gauge theory to 2D CFT}
\label{sec:GaugeToCFT}

In this section we discuss the implications of the 5D $SU(N)^{M-1} \leftrightarrow SU(M)^{N-1}$ duality in 2D CFTs and we propose that the DOZZ three-point function of  $q$-deformed Toda theory is obtained from the topological string partition function of $U(1)$ linear quivers.
We rewrite the $U(1)$  gauge theory partition function into the DOZZ three-point function of $q$-deformed Liouville theory that is given in \cite{Kozcaz:2010af}. What is more, we
extend it to $q$-deformed Toda theory and then
 conjecture that  $q$-deformed  Heisenberg free CFT on a multi-punctured sphere is dual to  $q$-deformed  Toda CFT on a three-punctured sphere.
We begin with a short review of the AGTW duality \cite{Alday:2009aq,Wyllard:2009hg} between 4D $\mathcal{N}=2$ $SU(N)$ conformal gauge quivers and 2D $A_{N-1}$ conformal Toda field theories and then turn to its  5D generalization between $\mathcal{N}=1$ gauge theories and $q$-deformed Virasoro and $W_N$ algebra \cite{Awata:2010yy}. The 5D gauge theory duality studied in this article then implies relations between correlation functions (conformal blocks) of the $q$-deformed Virasoro algebra and those of the $q$-deformed $W_N$ algebra. Ultimately, the 4D version of this duality should lead to relations between Liouville and Toda theories.

In \cite{Gaiotto:2009we} Gaiotto was able to obtain a large class of $\mathcal{N}=2$ superconformal field theories in four dimensions by compactifying (a twisted version of) the six-dimentional $(2,0)$ SCFT on a Riemann surface with genus $g$ and $n$ punctures. The parameter space of the exactly marginal gauge couplings of the 4D gauge theory is identified with the complex structure moduli space $\mathcal{C}_{g,n} /\Gamma_{g,n}$ of the Riemann surface. The discrete group $\Gamma_{g,n}$ is the generalized S-duality transformations of the 4D theory.

Soon after, Alday, Gaiotto and Tachikawa conjectured \cite{Alday:2009aq} that the instanton partition function of a $\mathcal{N}=2$ $SU(2)$ quiver gauge theory in $\Omega$ background is equal to the conformal block of the conformal Liouville theory on a certain Riemann surface $\mathcal{C}_{g,n}$. This Riemann surface can be found in a systematic way from the quiver diagram of the 4D gauge theory\footnote{The quiver diagram drawn \`a la Gaiotto \cite{Gaiotto:2009we} looks identical to the diagram associated with the conformal block.}. The two theories are equal under the following identificaton between their parameters
\begin{equation} 
\epsilon_1 = b \, , \qquad \epsilon_2 = \frac{1}{b} \, ,
\end{equation}
with the central charge of the Virasoro algebra being $c=1+6\left(b+\frac{1}{b}\right)^2$. The coupling constants $q$ are identified with the cross-ratios $z$, the hypermultiplet masses $m$ (both flavor and bifundamental) correspond to the external momenta in the Liouville theory and the Coulomb moduli $a$ correspond to the internal momenta in the conformal block. Both external and internal momenta are denoted by $\alpha$ here. 
The AGTW conjecture has been proved for a special case in \cite{Fateev:2009aw, Hadasz:2010xp, Mironov:2009qn, Mironov:2010pi}, and attempts for proof in more generic settings have been made by using a new basis of the Verma module \cite{Alba:2010qc, Fateev:2011hq, Belavin:2011js}.

The one-loop contribution in the partition function precisely reproduces the product of the so called DOZZ three-point function of the Liouville theory
\cite{Dorn:1994xn,Zamolodchikov:1995aa,Teschner:2001rv,Nakayama:2004vk}
\small
\begin{equation}
\begin{split}
\label{dozz}
&C_{\,{}_{\textrm{DOZZ}}}(\alpha_1,\alpha_2,\alpha_3)=
\left[
\pi\, \mu\, \gamma\,\left(b^2\right)\, b^{2-2b^2}
\right]^{\frac{b+1/b-\sum_{i}\alpha_i}{b}}
\\
&\quad \quad \rule{0pt}{5ex}
\times\frac{\Upsilon_0\,\Upsilon_b (2\alpha_1)\Upsilon_b (2\alpha_2)\Upsilon_b (2\alpha_3)}
{\Upsilon_b (\alpha_1+\alpha_2+\alpha_3-b-1/b)
\Upsilon_b (\alpha_1+\alpha_2-\alpha_3)
\Upsilon_b (\alpha_2+\alpha_3-\alpha_1)
\Upsilon_b (\alpha_3+\alpha_1-\alpha_2)} \,,
\end{split}
\end{equation}
\normalsize
where the special function $\Upsilon_b (x)$ is defined by
\begin{align}
&\Upsilon_b (x)=\frac{1}{\Gamma_b(x)\Gamma_b(\epsilon_1+\epsilon_2-x)},
\\
&\Gamma_b(x)
=
\exp \frac{d}{ds}
\frac{1}{\Gamma(s)}
\int_0^\infty
\frac{dt}{t}\,
\frac{t^se^{-tx}}{(1-x^{-\epsilon_1t})(1-e^{-\epsilon_2t})}
\Big|_{s=0}
\propto
\sum_{m,n=0}^\infty
(x+\epsilon_1m+\epsilon_2n)^{-1}.
\end{align}
Finally, the partition function of the 4D SCFT on $S^4$,
\begin{equation} 
\int d a \, a^2 \, |\,Z_{\,\textrm{Nek}}(a)\,|^2
\end{equation}
with $Z_{\textrm{Nek}} =Z_{\textrm{tree}} Z_{\textrm{1-loop}}Z_{\textrm{inst}}$ being the full partition function, is equal to the correlation function of primary fields $V_\alpha =e^{2 \alpha \phi}$ in the Liouville theory with conformal dimension $\Delta =\alpha\left(b+\frac{1}{b} - \alpha \right)$. Take the $SU(2)$ gauge theory with four flavors as an example, this theory corresponds to the Liouville CFT on the Riemann sphere with four punctures $\mathcal{C}_{0,4}$. Quantitatively, the AGTW conjecture states
\begin{equation} 
\int d \mu(\alpha)\,C_{\,\textrm{DOZZ}}C_{\,\textrm{DOZZ}} \, | \,q^{\Delta-\Delta_1-\Delta_2} \mathcal{B}_{0,4}(\alpha)\,|^2
\propto
\int d a \, a^2 \, |\,Z_{\,\textrm{Nek}}(a)\,|^2 \, ,
\end{equation}
where the two DOZZ factors come from the decomposition of the four punctured sphere into two pants. The conformal block $\mathcal{B}$ is then equal to the instanton part of the Nekrasov partition function, and the ``square root'' of the DOZZ part gives the perturbative correction of the partition function.

The 5D extension of the conjecture suggests that the instanton part of the 5D Nekrasov partition function is equal to the conformal block of a $q$-deformed CFT. Schematically this conjectured duality is the following equality
\begin{equation} 
\mathcal{B}^{\,q-\textrm{Liouville}}(\alpha)=
Z^{\,\textrm{5D}}_{\,\textrm{Nek}}(a) \, .
\end{equation}
In \cite{Awata:2010yy} the authors studied the case of $SU(2)$ pure SYM, which is the simplest setup of the AGT duality, and they found that the partition function coincides with the irregular conformal block of the $q$-deformed Virasoro algebra. Although the 5D extension of the instanton counting is established, the theoretical framework of $q$-deformed CFT's is not well developed. Therefore, we cannot establish the duality for the full sector yet.
The $q$-deformation of conformal field theory should first be developed to reveal the scope of the AGTW duality. However, by assuming the 5D AGTW conjecture, we will now illustrate how the gauge theory duality studied in Section \ref{sec:MtheoryDeriv} and \ref{sec:TopStringDeriv} can be used to make predictions about $q$-deformed CFT's. Although we have mostly reviewed the $SU(2)$ quiver case, the ideas can be generalized to $SU(N)$ quivers.

\begin{figure}[t]
\qquad   \qquad  \qquad  \qquad  \quad 
\includegraphics[scale=0.5]{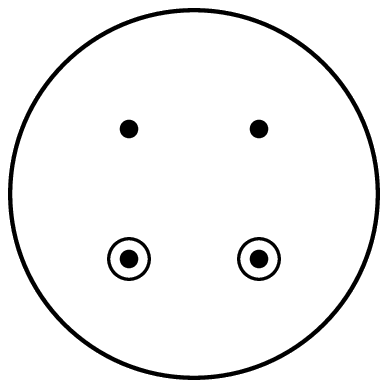} 
\qquad   \qquad  \qquad  \qquad  \qquad 
\includegraphics[scale=0.5]{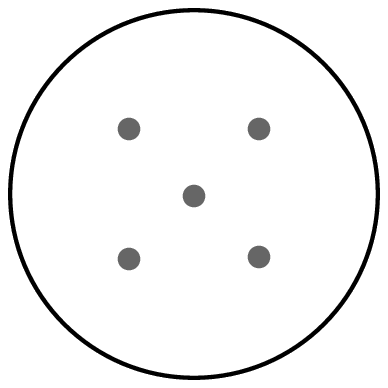}
\caption{The $SU(3) \leftrightarrow SU(2)\times SU(2)$ duality implies that the four-point $W_{3}$ Toda correlator on a sphere (left) should be equal to the five-point Liouville correlator on a sphere (right). The black points denote $U(1)$ punctures and the encircled ones $SU(3)$ punctures in the $W_{3}$ Toda theory respectively, whereas the grey points correspond to $SU(2)$ punctures in the Liouville theory.}
\label{SU(3)example}
\end{figure}

\subsection{5D quiver $U(1)$ gauge theories and $q$-deformation of DOZZ}

In this subsection we give an example involving $U(1)$ gauge theory, whose instanton partition function is given by
\small
\begin{equation} 
Z^{\,\textrm{5D inst}}_{\,U(1)}=
\sum_{Y}\,
{q}^{|Y|}
\frac{\prod_{(i,j)\in Y}\sinh \frac{\beta}{2}(m_1+\hbar(i-j))\sinh \frac{\beta}{2}(-m_2+\hbar(i-j))}
{\prod_{(i,j)\in Y}\sinh \frac{\beta}{2}(\hbar(Y_i+Y^T_j-i-j+1))\sinh \frac{\beta}{2}(-\hbar(Y_i+Y^T_j-i-j+1))}
\, ,
\end{equation}
\normalsize
with one fundamental and one anti-fundamental hypermultiplet.
Moreover, we introduce the perturbative part of the partition function:
\begin{equation} 
Z^{\,\textrm{5D pert}}_{\,U(1)}={[\emptyset,\emptyset]_{e^{-\beta m_1}}[\emptyset,\emptyset]_{e^{-\beta m_2}}} \, ,
\end{equation}
where the bracket is defined in (\ref{bracket}). The full Nekrasov partition function is the product of the two:
$Z^{\,\textrm{5D}}_{\,U(1)}=Z^{\,\textrm{5D pert}}_{\,U(1)}Z^{\,\textrm{5D inst}}_{\,U(1)}$.
By using the techniques from the topological vertex formalism, we can perform the summation inside the full partition function to obtain
\begin{equation}
\label{eq:U1PartFunc}
Z^{\,\textrm{5D}}_{\,U(1)}=
\frac{[\emptyset,\emptyset]_{Q_1}[\emptyset,\emptyset]_{Q_F}[\emptyset,\emptyset]_{Q_2}[\emptyset,\emptyset]_{Q_1Q_FQ_2}}
{[\emptyset,\emptyset]_{Q_1{Q_F}}[\emptyset,\emptyset]_{{Q_F}Q_2}} \, .
\end{equation} 
The right hand side of this equation has appeared already in \cite{Iqbal:2004ne,Kozcaz:2010af}.
The parameters are defined as
\begin{equation} 
Q_i=e^{-\beta m_i}\,\,(i=1,2)
\quad \text{and} \quad
-Q_F{\sqrt{Q_1Q_2}}=q \, .
\end{equation}

\begin{figure}[htbp]
 \begin{center}
  \includegraphics[width=120mm,clip]{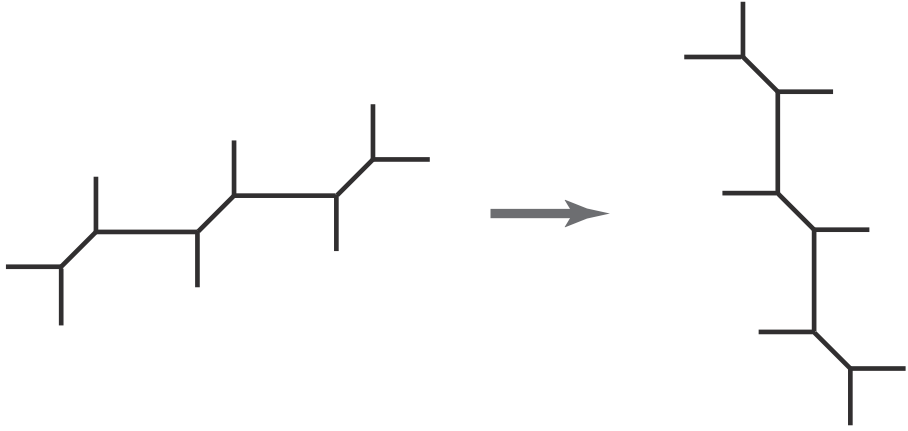}
 \end{center}
 \caption{The toric diagram for $U(1)^2$ linear quiver (left) and the free hypermultiplets (right).}
 \label{fig:abelquiv}
\end{figure}

What is interesting here is that the expression (\ref{eq:U1PartFunc}) corresponds to the $q$-deformed DOZZ function \cite{Kozcaz:2010af}
\begin{equation} 
\mid[\emptyset,\emptyset]_{Q_1}[\emptyset,\emptyset]_{Q_F}[\emptyset,\emptyset]_{Q_2}[\emptyset,\emptyset]_{Q_1Q_FQ_2}\mid^2
\propto
C_{\,\textrm{DOZZ}}^{\,\mathfrak{q}} \, ,
\end{equation} 
where the $q$-deformation parameter is $\mathfrak{q}=e^{-\beta \hbar}$ and the identification of parameters takes the form
\begin{equation} 
Q_1=e^{-\beta(-\alpha_1-\alpha_2+\alpha_3)} \, , \quad
Q_F=e^{-\beta(-\alpha_1+\alpha_2-\alpha_3)} \, , \quad
Q_2=e^{-\beta(\alpha_1-\alpha_2-\alpha_3)} \, .
\end{equation}
By using the rotational duality described in Appendix~\ref{app:90Rotation}, the $q$-deformed DOZZ function is expected to be given by the following replacement of the $\Upsilon$-function in the definition (\ref{dozz}):
\begin{align}
\Upsilon_b (x)=\frac{1}{\Gamma_b(x)\Gamma_b(\epsilon-x)}
\quad \longrightarrow \quad
\Upsilon_b^{\,\mathfrak{q}} (x)
=\frac{1}{\Gamma_b^{\,\mathfrak{q}}(x)\Gamma_b^{\,\mathfrak{q}}(\epsilon-x)} \, ,
\end{align}
where $\Gamma_b^{\,\mathfrak{q}}(x)\propto \prod_{i,j} \left( \sinh \frac{\beta}{2}(x+\epsilon_1 i+\epsilon_2j) \right)^{-1}$. The idea is illustrated using the toric diagrams in Figure~\ref{fig:abelquiv}, where the $U(1)^{N-1}$ quiver gauge theory is on the left\footnote{See \cite{Tai:2010ps} for work related to this idea.}. The rotated diagram on the right depicts the so-called 4D Gaiotto theory for the sphere with two full punctures and one simple puncture.
The AGT dual of this $U(1)$ gauge theory partition function is the DOZZ three-point function of the rank-$N$ Toda field theory. In response to Gaiotto's construction of the 4D gauge theory, the DOZZ function is the three-point function for two full primary fields and one semi-degenerate field. Above we studied $U(1)$ gauge theory with two flavors, which is dual to Liouville theory on the sphere with 3 punctures. Since we consider the 5D uplift of the gauge theory, 2D CFT is replaced by the $q$-analogue of it. It is straightforward to extend this argument to generic $\Omega$ background, in which case the 2D CFT with generic central charge appears.

Using the idea above we can conjecture the $q$-analogue of the Toda DOZZ function. We consider the $U(1)^{N-1}$  linear quiver gauge theory. The toric diagram for this theory is shown on the left in Figure~\ref{fig:abelquiv}. With the formalism of the \textit{refined} topological vertex, we can compute the closed form of the full Nekrasov partition function
\small
\begin{equation} 
 \label{5Dabelianquiver}
Z^{\,\textrm{5D}}_{\,U(1)^{N-1}}=
\prod_{i,j=1}^\infty
\frac{\prod_{1\leq a\leq b\leq N}\left(1-Q_{ab}Q_{m\,b}\mathfrak{t}^{i-\frac{1}{2}}\mathfrak{q}^{j-\frac{1}{2}}  \right)
\prod_{1\leq a\le b\leq N}\left(1-Q_{m\,a}^{-1}Q_{ab}\mathfrak{t}^{i-\frac{1}{2}}\mathfrak{q}^{j-\frac{1}{2}}  \right)
}
{
\prod_{1\leq a\le b\leq N}\left(1-Q_{ab}\mathfrak{t}^{i-\frac{3}{2}}\mathfrak{q}^{j+\frac{1}{2}}  \right)
\left(1-Q_{m\,a}^{-1}Q_{ab}Q_{ab}\mathfrak{t}^{i+\frac{1}{2}}\mathfrak{q}^{j-\frac{3}{2}}  \right)
} \, ,
\end{equation}
\normalsize
where $Q_{ab}=\prod_{i=a}^{b-1}Q_{m\,i}{q}^{(i)}$. In order to relate this expression to the combinatorial form of the instanton part of the partition function, we have to assume the slicing invariance of the refined topological vertex (see \cite{Iqbal:2008ra} for details). Our claim is that the square of (\ref{5Dabelianquiver}) gives the major portion of the DOZZ three-point function of the ``$q$-deformed $sl(N)$ Toda field theory'' on sphere with two full primary fields and one semi-degenerate field. This result would be a powerful guide to formulate a yet-unknown $q$-deformation of the Toda field theory.

We can also recast our proposal as a duality between the $(M+2)$-point function of $W_{N}$-algebra and the $(N+2)$-point function of $W_{M}$-algebra. 
See Figure \ref{SU(3)example} for an example. The $q$-deformed conformal blocks for the Heisenberg algebra are defined in the form of the Dotsenko-Fateev integral representation \cite{Mironov:2011dk,Awata:2010yy}, and we can see that these conformal blocks give the 5D Nekrasov partition functions for $U(1)$ quiver gauge theories\footnote{See \cite{Losev:2003py,Marshakov:2009gs,Alba:2009ya}
for the 4D version of the AGTW ``Heisenberg/$U(1)$'' duality, which implies the equality between the free conformal block for the Heisenberg algebra and the Nekrasov partition function for $U(1)$ quiver gauge theory. This is a simplified toy model for the original AGT duality of Virasoro/$SU(2)$.}.  The simplest situation we have studied in this subsection is thus the equivalence between the $(N+2)$-point function of ``$W_1$'' (Heisenberg) algebra and the three-point function of $W_{N}$ algebra. This conjecture for $W_{N}$ algebras is the direct consequence of combining the duality from Section~\ref{sec:TopStringDeriv} and the AGTW conjecture. It gives a CFT analogue of this duality, which can be valuable in the studies of 2D CFT.

\section{Summary and discussion}
\label{sec:Discussion}

In this paper, we have studied the duality between two 5D ${\cal N}=1$  linear quiver gauge theories compactified on $S^1$ with gauge groups $SU(N)^{M-1}$ and $SU(M)^{N-1}$ respectively. We have found the explicit map between the gauge theory parameters of these two theories, under which they describe the same 
low energy effective theory on the Coulomb branch. We have derived the map both by considering the M5-brane configuration and by calculating the topological string partition function. There are several interesting extensions and applications of this duality.

\smallskip

The implications of this duality in 2D CFT through the 5D extension of the AGTW conjecture have been discussed above.
We conjuctured the three-point function of $q$-deformed Toda theory from the topological string partition function of the U(1) linear quiver. Moreover, the duality between $(M+2)$-point function of $q$-deformed $W_{N}$-algebra and $(N+2)$-point function of $q$-deformed $W_{M}$-algebra is proposed. An interesting future direction is to study in detail the duality we have proposed here between Liouville and Toda correlation functions.

\smallskip

Although it is natural and interesting to consider the 4D limit of this duality, it seems to be subtle. In an upcoming paper \cite{index} we follow a simple path to the 4D version of this duality, where the 4D superconformal index \cite{Kinney:2005ej,Romelsberger:2005eg} is used to study the duality between the 4D conformal $\mathcal{N}=2$ $SU(N)^{M-1}$ and $SU(M)^{N-1}$ line quivers. The superconformal index counts  the multiplets that obey shortening conditions, up to equivalence relations that set to zero all the short multiplets that can recombine into long multiplets. Basically, it knows the complete list of protected operators in a superconformal theory. Together with one-loop computations, the analysis of the chiral ring and representation theory arguments it was used to study the spectrum of $\mathcal{N}=2$ superconformal QCD at large $N$ in \cite{Gadde:2009dj,Gadde:2010zi}. What is more, there is a relation between the 4D superconformal index and topological quantum field theories in 2D \cite{Gadde:2009kb, Gadde:2010te, Gadde:2011ik}, which provides a simpler version of the AGTW  relation between 4D partion functions and 2D CFT correlators. The index is the  partition function on $S^3 \times S^1$ \cite{Festuccia:2011ws}, it is coupling-independent and easier to calculate  than  Pestun's partition function on $S^4$. It is related to a 2D TQFT correlation function  \cite{Gadde:2009kb} as opposed to the full-fledged CFT correlation function that is required in AGTW. The superconformal index has been successfully used to test ${\mathcal N}=1$ Seiberg duality \cite{Romelsberger:2005eg,Romelsberger:2007ec} 
and ${\mathcal N}=1$ toric duality \cite{Gadde:2010en} (as well as AdS/CFT  \cite{Kinney:2005ej}).

\smallskip

Low energy physics of supersymmetric gauge theories can also be captured by matrix models. Different types of matrix models have been studied in this context. First, the (``old'') Dijkgraaf-Vafa matrix model \cite{Dijkgraaf:2002fc,Dijkgraaf:2002vw,Dijkgraaf:2002dh} gives the low energy effective superpotential of 4D ${\cal N}=1$ gauge theory that is obtained by 
deforming ${\cal N}=2$ with the addition of superpotential terms of polynomial type. 
The action of this matrix model is given by its tree-level superpotential.
Another matrix model was later proposed by the same authors in \cite{Dijkgraaf:2009pc}. The ``new''  Dijkgraaf-Vafa matrix model gives Nekrasov's partition function of 4D ${\cal N}=2$ gauge theory, and though the AGTW conjecture, the conformal block of the Liouville/Toda CFT  \cite{Itoyama:2009sc,Awata:2010yy,Eguchi:2009gf,Mironov:2011dk}.
Since the prepotential of the ${\cal N}=2$ gauge theory can be reproduced from the low energy effective superpotential \cite{Cachazo:2002pr}, these two matrix models should be closely related even if they are computing different quantities. Indeed, both of them are introduced in the context of topological string theory in such a way that 
the spectral curves of these matrix models reproduce the Seiberg-Witten curve.

However, at first sight they look quite different in the following way. On the one hand, in the ``old'' Dijkgraaf-Vafa matrix model the matrix corresponds to the zero-modes of the adjoint scalar fields. Therefore, $SU(N)$ theory is studied using a single matrix while $SU(2)^{M-1}$ a quiver matrix (multi-matrix) model. On the other hand,  in the ``new''  Dijkgraaf-Vafa matrix model $SU(2)^{M-1}$ theory corresponds to the single matrix model with multi-Penner type action, while $SU(N)$ theory corresponds to the quiver matrix model with $N-1$ adjoint matrices \cite{Schiappa:2009cc}. As was already pointed out in \cite{Dijkgraaf:2009pc}, the role of the base and the fiber of the Calabi-Yau geometry is inverted in the second matrix model compared to the first one. Since the structure of the base and the fiber 
are related to the numbers $N$ and $M$ of the $SU(N)^{M-1}$ gauge theory, it implies that these matrix models are related by the duality studied in this paper. We expect that it will play an important role to understand the relation between these matrix models.

\smallskip

Several other kinds of extensions of the duality we study here are also possible. In this article we focus on the duality between theories which are 5D uplifts of 4D superconformal field theories. It should be possible to extend the duality to the theories which are uplifts of asymptotically free theories.
In such cases it is known that we can introduce the Chern-Simons term \cite{Seiberg:1996bd} in the action. The configuration of the M5-brane curve depends on the Chern-Simons level \cite{Brandhuber:1997ua} and thus the duality will also act  on it. Considering such an effect would be interesting.

The extension to the elliptic quiver gauge theories, including $\mathcal{N}=2^*$ theory, is another future direction. Such quiver gauge theories are obtained by further compactifying the $x^6$ direction in addition to the $x^5$ direction in Table \ref{config}. Following \cite{Vafa:1997mh,Tachikawa:2011ch}, the S-duality corresponding to the electric-magnetic duality appears by compactifying the $x^6$ direction in the special case where no NS5-branes are placed. The duality studied in this article can be also interpreted as S-duality, but it acts on the gauge theories in a totally different manner than the conventional electric-magnetic duality. The elliptic quiver gauge theories will offer an interesting playground to understand these two different types of S-dualities in a unified manner.

In the present article, we have studied the duality in the self-dual $\Omega$  background. The extension to the generic $\Omega$ background would be an important direction related to the existence of a preferred direction in the refined topological vertex. The conjectured slicing invariance would then be crucial for extending our result to the refined case. The duality maps we have derived are maintained even after switching on the self-dual $\Omega$ background. However, it is non-trivial whether the generic $\Omega$ background modifies the maps.

Considering this duality for generic $\Omega$ background in the context of the integrable system would be also interesting, where the ``quantum Seiberg-Witten curve"
appears as the Hamiltonian of the Schr\"odinger equation. If we manage to find the explicit expression of the 5D Hamiltonian \cite{Aganagic:2011mi}, then it would be straightforward to obtain the duality map by using the same method we employed in this paper. 
The Nekrasov-Schatashivilli \cite{Nekrasov:2009ui,Nekrasov:2009rc,Nekrasov:2009zz} limit is especially interesting because the time-dependent terms\footnote{The time coordinates are interpreted as  gauge couplings.} in the Schr\"odinger equation are expected to vanish there. We then get a simple eigenvalue problem as an alternative way to solve quantum gauge theory.

\section*{Acknowledgements}
It is a pleasure to thank Giulio Bonelli, Nadav Drukker, Tohru Eguchi,  Amihay Hanany, Kazunobu Maruyoshi, Sara Pasquetti, Filippo Passerini, Leonardo Rastelli, Kazuhiro Sakai, Yuji Tachikawa, Alessandro Tanzini, Niclas Wyllard and Konstantinos Zoubos for useful discussions and correspondence. L.B., E.P. and F.Y. would like to thank IHES for providing a stimulating atmosphere during the course of this work. E.P. wishes to thank IHP for its warm hospitality as this work was in progress. The work of E.P. is supported  by the Humboldt Foundation.
The research of M.T. is supported in part by JSPS Grant-in-Aid for Creative Scientific Research No. 19GS0219.
 F.Y. is partially supported by the INFN project TV12. 

\appendix

\section{Physical parameters}
\label{phys-par}

The duality map (\ref{gen_map}), as discussed in Section \ref{sec:MtheoryDeriv}, seems to depend on the choice of coordinates when written in terms of the {\it position parameters}\footnote{In this appendix we set $\ell_{s}=1$.}. However, when rewritten in terms of the {\it physical parameters} it is manifestly independent of the choice of coordinates. In this appendix we define the {physical gauge theory parameters} in terms of the position parameters.
We also introduce the ``traceless'' and ``trace'' parameters that are more natural in the context of the AGTW conjecture, where these two kinds of mass parameters correspond to the $SU(N)$ and the $U(1)$ punctures respectively.
 
The {\it position parameters} $\tilde{a}^{(i)}_\alpha$, introduced in (\ref{tildema}) and depicted in Figure \ref{branesetup}, denote the positions of the D4-branes in the $w$ coordinate. On the other hand, the {\it physical parameters} are defined as distances between the endpoints of open strings and relative distances do not depend on the choice of coordinates. As we will see, the definitions of the physical masses and the physical Coulomb moduli are such that it is difficult to define them in a unified way as we do for the position parameters (\ref{mass-a}) or the ``traceless'' and ``trace'' parameters (\ref{trace_and_traceless}).

\begin{figure}[htb]
\centering
\subfigure
{
\label{fig:AFMass}
\tiny
\setlength{\unitlength}{4cm}
\begin{picture}(1,1)(0,0)
\linethickness{0.25mm}
\put(0.5,0){\line(0,1){1}}
\put(0.45,-0.1){NS5}
\put(0.5,0.15){\line(1,0){0.4}}
\put(0.95,0.15){$a^{(1)}_{1}$}
\put(0.5,0.3){\line(1,0){0.4}}
\put(0.95,0.3){$a^{(1)}_{2}$}
\put(0.5,0.75){\line(1,0){0.4}}
\put(0.95,0.75){$a^{(1)}_{N-1}$}
\put(0.5,0.85){\line(1,0){0.4}}
\put(0.95,0.85){$a^{(1)}_{N}$}
\put(0.1,0.7){\line(1,0){0.4}}
\put(-0.03,0.7){$m_{\alpha}$}
\linethickness{0.4mm}
\multiput(0.5,0.5)(0.04,0){10}{\line(1,0){0.025}}
\linethickness{0.25mm}
\put(0.95,0.5){$a^{(1)}_{\textrm{cm}}$}
\Blue{
\put(0.4,0.7){\vector(0,-1){0.2}}
\put(0.25,0.55){$m^{\textrm{af}}_{\alpha}$}
}
\put(0.7,0.40){\circle*{0.01}}
\put(0.7,0.58){\circle*{0.01}}
\put(0.7,0.66){\circle*{0.01}}
\end{picture}
\normalsize
}
\hspace{2.5cm}
\subfigure
{
\label{fig:BiFMass}
\tiny
\setlength{\unitlength}{4cm}
\begin{picture}(1,1)(0,0)
\linethickness{0.25mm}
\put(0.1,0){\line(0,1){1}}
\put(0.05,-0.1){NS5}
\put(0.5,0){\line(0,1){1}}
\put(0.45,-0.1){NS5}
\put(0.9,0){\line(0,1){1}}
\put(0.85,-0.1){NS5}
\put(0.1,0.2){\line(1,0){0.4}}
\put(-0.12,0.2){$a^{(i-1)}_{1}$}
\put(0.1,0.35){\line(1,0){0.4}}
\put(-0.12,0.35){$a^{(i-1)}_{2}$}
\put(0.1,0.8){\line(1,0){0.4}}
\put(-0.12,0.8){$a^{(i-1)}_{N-1}$}
\put(0.1,0.9){\line(1,0){0.4}}
\put(-0.12,0.9){$a^{(i-1)}_{N}$}
\put(0.5,0.15){\line(1,0){0.4}}
\put(0.95,0.15){$a^{(i)}_{1}$}
\put(0.5,0.3){\line(1,0){0.4}}
\put(0.95,0.3){$a^{(i)}_{2}$}
\put(0.5,0.75){\line(1,0){0.4}}
\put(0.95,0.75){$a^{(i)}_{N-1}$}
\put(0.5,0.85){\line(1,0){0.4}}
\put(0.95,0.85){$a^{(i)}_{N}$}
\linethickness{0.4mm}
\multiput(0.1,0.65)(0.04,0){10}{\line(1,0){0.025}}
\multiput(0.5,0.5)(0.04,0){10}{\line(1,0){0.025}}
\linethickness{0.25mm}
\put(-0.12,0.65){$a^{(i-1)}_{\textrm{cm}}$}
\put(0.95,0.5){$a^{(i)}_{\textrm{cm}}$}
\Blue{
\put(0.4,0.65){\vector(0,-1){0.15}}
\put(0.13,0.53){$m^{(i-1,i)}_{\textrm{bif}}$}
}
\put(0.3,0.4){\circle*{0.01}}
\put(0.3,0.45){\circle*{0.01}}
\put(0.3,0.72){\circle*{0.01}}
\put(0.7,0.40){\circle*{0.01}}
\put(0.7,0.58){\circle*{0.01}}
\put(0.7,0.66){\circle*{0.01}}
\end{picture}
\normalsize
}
\caption{Definitions of an anti-fundamental (left) and a bifundamental (right) mass.}
\label{fig:PhyMass}
\end{figure}
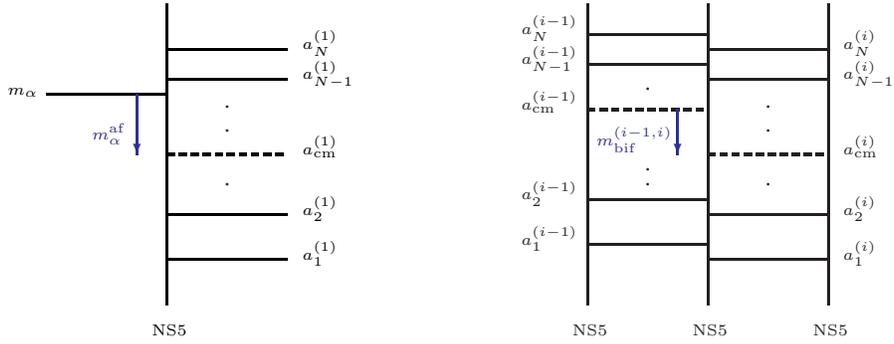

First, we turn to the {\it physical flavor mass} that correspond to the distance (along $v$) between a flavor D4-brane position and the center of mass position of the adjacent color branes. D4-branes attached from the right to an NS5-brane correspond to fundamental masses, whereas D4-branes attached from the left to an NS5-brane correspond to anti-fundamental masses. The first $N$ flavor masses ($m_1,\cdots,m_N$) on the left of the quiver are, thus, anti-fundamental under the first gauge group. Moreover, they are fundamental under the ``$0$-th gauge group'', which is in fact a global symmetry. According to these conventions, we define the {\it anti-fundamental flavor mass}
\begin{align}
m_{\alpha}^{\text{af}} = \frac{1}{N}\sum_{\beta=1}^N a^{(1)}_{\beta} - m_{\alpha} \, ,
\end{align}
where $m_{\alpha}$ is the position of the semi-infinite flavor D4-brane on the left of the quiver, see Figure \ref{fig:PhyMass}.
The anti-fundamental flavor mass can then be exponentiated, following (\ref{tildema}), as
\begin{align}
\tilde{m}_{\alpha}^{\text{af}} 
= \left( \tilde{a}^{(0)}_{\alpha} \right)^{-1}
\prod_{\beta=1}^N \left( \tilde{a}^{(1)}_{\beta} \right)
^{\frac{1}{N}} \, .
\end{align}
We can, moreover, think of the flavor masses on the left of the quiver as  fundamental masses if we define
\begin{align}
\tilde{m}_{\alpha}^{\text{f}} 
= \tilde{a}^{(0)}_{\alpha} \prod_{\beta=1}^N \left( \tilde{a}^{(1)}_{\beta} \right)
^{-\frac{1}{N}}
= \frac{1}{\tilde{m}_{\alpha}^{\text{af}} } \, .
\end{align}
In this paper we use this convention when there is only one gauge group factor. However, we find the anti-fundamental definition more natural for a generic quiver. In addition, the last $N$ masses ($m_{N+1}, \cdots m_{2N}$) on the right of the quiver are fundamental under the $(M-1)$-th gauge group and anti-fundamental under the ``$M$-th gauge group'', with the latter being a global symmetry. Following the  ``right minus left'' convention, we have
\begin{align}
m_{N+\alpha}^{\text{f}} = m_{N+\alpha} - \frac{1}{N}\sum_{\beta=1}^N a^{(M-1)}_{\beta} \, ,
\end{align}
which becomes
\begin{align}
\tilde{m}_{N+\alpha}^{\text{f}} 
= \tilde{a}^{(M)}_{\alpha} \prod_{\beta=1}^N \left( \tilde{a}^{(M-1)}_{\beta} \right)^{-\frac{1}{N}}
\end{align}
after exponentiation (\ref{tildema}).

\begin{figure}[htb]
\centering
\tiny
\setlength{\unitlength}{4cm}
\begin{picture}(1,1)(0,0)
\linethickness{0.25mm}
\put(0.5,0){\line(0,1){1}}
\put(0.45,-0.1){NS5}
\put(0.9,0){\line(0,1){1}}
\put(0.85,-0.1){NS5}
\put(0.5,0.15){\line(1,0){0.4}}
\put(0.95,0.15){$a^{(i)}_{1}$}
\put(0.5,0.3){\line(1,0){0.4}}
\put(0.95,0.3){$a^{(i)}_{2}$}
\put(0.5,0.65){\line(1,0){0.4}}
\put(0.95,0.65){$a^{(i)}_{\alpha}$}
\put(0.5,0.85){\line(1,0){0.4}}
\put(0.95,0.85){$a^{(i)}_{N}$}
\linethickness{0.4mm}
\multiput(0.5,0.5)(0.04,0){10}{\line(1,0){0.025}}
\linethickness{0.25mm}
\put(0.95,0.5){$a^{(i)}_{\textrm{cm}}$}
\Blue{
\put(0.4,0.5){\vector(0,1){0.15}}
\put(0.378,0.5){\line(1,0){0.04}}
\put(0.02,0.54){$- \frac{1}{\beta} \log \hat{a}^{(i)}_{\alpha}$}
}
\put(0.7,0.40){\circle*{0.01}}
\put(0.7,0.575){\circle*{0.01}}
\put(0.7,0.75){\circle*{0.01}}
\end{picture}
\normalsize
\vspace{5mm}
\caption{Definition of a Coulomb modulus.}
\label{fig:CoulombModuli}
\end{figure}
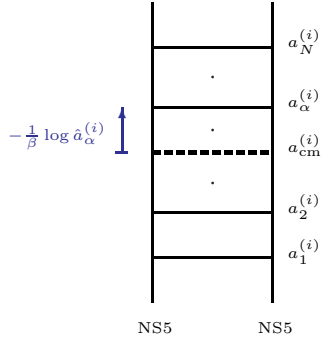

Next, we turn to the definition of the {\it physical Coulomb moduli parameter}.
This should be thought of as the distance between a color D4-brane position and the center of mass position of the color branes within a single gauge group factor, see Figure \ref{fig:CoulombModuli}.
In other words, this is the {\it ``traceless part''} of the Coulomb moduli
\begin{align}
{a}^{(i)}_\alpha - \frac{1}{N}\sum_{\beta=1}^N {a}^{(i)}_\beta \, ,
\end{align}
which in terms of the 5D parameters (\ref{mass-a}) is defined as
\begin{align}
\hat{a}^{(i)}_{\alpha}
= \frac{\tilde{a}_{\alpha}^{(i)}}%
{\prod_{\beta=1}^N \left( \tilde{a}_{\beta}^{(i)} \right)^{\frac{1}{N}}} \, .
\end{align}

One last definition is in order, that of the {\it bi-fundamental masses} ${m}^{(i-1,i)}_{\text{bif}}$. Recall that a conformal quiver gauge theory has not only the overall $U(1)$ factored out\footnote{The overall $U(1)$ corresponds to the center of mass in the $v$ coordinate when all the flavor masses are identically equal to zero.}, but also all the relative $U(1)$s so that each factor in the quiver is $SU(N)$ and not $U(N)$. However, we want to study not just the conformal quiver, but the more general asymptotically conformal quiver with non-zero bi-fundamental masses. The bi-fundamental masses ${m}^{(i-1,i)}$ (for $2 \le i \le M-1$) are related to the relative $U(1)$s between the $(i)$-th and $(i-1)$-th $SU(N)$ gauge factors. They are equal to the distance (along $v$) between the center of mass positions of the color D4-branes that correspond to these two adjacent gauge group factors. As above, we use the notation that the bi-fundamental fields are fundamental under the gauge group on the left ($(i-1)$ -th gauge group) and anti-fundamental under the gauge group on the right ($i$-th gauge group)
\begin{equation}
\label{biff-def}
{m}^{(i-1,i)}_{\text{bif}} = \frac{1}{N}\sum_\alpha a^{(i)}_\alpha  -  \frac{1}{N}\sum_\alpha a^{(i-1)}_\alpha \, .
\end{equation}
This definition (\ref{biff-def}) can be extended to include $1 \le i \le M$ as long as we keep in mind that ${m}_{\text{bif}}^{(0,1)}$ and ${m}_{\text{bif}}^{(M-1,M)}$ are not bifundamental masses, although the subscript suggests otherwise. In terms of the 5D parameters the definition takes the form
\begin{align}
\tilde{m}_{\text{bif}}^{(i-1,i)} 
= \frac{\prod_{\beta=1}^N ( \tilde{a}^{(i)}_{\beta} )^{\frac{1}{N}}}
{\prod_{\alpha=1}^N ( \tilde{a}^{(i-1)}_{\alpha} )^{\frac{1}{N}}} \, .
\end{align}

In the context of the AGTW conjecture mass parameters that correspond to the $U(1)$ and the $SU(N)$ punctures are introduced. To make contact with Gaiotto's quiver diagrams we define
\begin{equation}
\label{trace_and_traceless}
\begin{split}
\hat{a}^{(0)}_{\alpha} 
= \frac{\prod_{\beta=1}^N (\tilde{m}_{\beta}^{\text{af}})^{\frac{1}{N}} }
{\tilde{m}_{\alpha}^{\text{af}} } \, ,&
\qquad
\tilde{m}_{\text{bif}}^{(0,1)} 
= \prod_{\alpha=1}^N \left(
\tilde{m}_{\alpha}^{\text{af}} 
\right)^{\frac{1}{N}} \, ,
\\
\hat{a}^{(M)}_{\alpha} 
= \frac{\tilde{m}_{\alpha}^{\text{f}} }
{\prod_{\beta=1}^N (\tilde{m}_{\beta}^{\text{f}})^{\frac{1}{N}} } \, ,&
\qquad
\tilde{m}_{\text{bif}}^{(M-1,M)} 
= \prod_{\alpha=1}^N \left(
\tilde{m}_{\alpha}^{\text{f}} 
\right)^{\frac{1}{N}} \, ,
\end{split}
\end{equation}
where $\hat{a}^{(0)}_{\alpha}$ and $\hat{a}^{(M)}_{\alpha}$ are the {\it traceless parts} of the flavor masses while $\tilde{m}_{\text{bif}}^{(0,1)}$ and $\tilde{m}_{\text{bif}}^{(M-1,M)}$ the {\it trace parts}. In the AGTW language, the traceless  masses $\hat{a}^{(0)}_{\alpha}$ and $\hat{a}^{(M)}_{\alpha}$ correspond to $SU(N)$ punctures, whereas the trace part $\tilde{m}_{\text{bif}}^{(0,1)}$ and $\tilde{m}_{\text{bif}}^{(M-1,M)}$ together with all the bifindamental masses $\tilde{m}_{\text{bif}}^{(i-1,i)}$ ($2 \le i \le M-1$) correspond to $U(1)$ punctures.

\section{4D limit of the SW curve for $SU(2)$ gauge theory with four flavors}
\label{app:4DLimit}

We consider the 4D limit of the Seiberg-Witten curve (\ref{curveC}) of the $\mathcal{N}=2$ $SU(2)$ gauge theory with four flavors. 
First, by multiplying $\tilde{m}_1^{-1/2} \tilde{m_2}^{-1/2} w^{-1}$ to the curve and imposing $a_1=-a_2=a$ as in (\ref{U(1)}), it can be rewritten as
\small
\begin{equation}
\begin{split}
0 \,\,  = \,\,
&4 \sinh \left( \frac{\beta}{2} (v-m_1) \right) \sinh \left( \frac{\beta}{2} (v-m_2) \right) 
t^2
\\
& + \left(
 - 2C \cosh \left( \beta v \right)
- 2Cq \cosh \left( \frac{\beta}{2} 
\left( 2v - \sum_{i=1}^4 m_i \right) \right)
+ \frac{b}{\tilde{m}_1^{\frac{1}{2}} \tilde{m}_2^{\frac{1}{2}}} 
\right) t
\\
& + 4 C^2 q \sinh \left( \frac{\beta}{2} (v-m_3) \right) 
\sinh \left( \frac{\beta}{2} (v-m_4) \right) \, .
\label{failedC}
\end{split}
\end{equation}
\normalsize
Further, by expanding the coefficients in powers of the  circumference of the 5D $\beta$
\begin{equation}
\begin{split}
C &= C_{(0)} + C_{(1)} \beta + C_{(2)} \beta^2 + \cdots \, ,
\\
\frac{b}{\tilde{m}_1^{\frac{1}{2}}\tilde{m}_2^{\frac{1}{2}}} 
&= b_{(0)} + b_{(1)} \beta + b_{(2)} \beta^2 + \cdots \, ,
\end{split}
\end{equation}
the leading and the next-to-leading order of (\ref{failedC}) lead to the relations
\begin{equation}
\begin{split}
- 2 C_{(0)} (1+q) + b_{(0)} &= 0 \, , \\
- 2 C_{(1)} (1+q) + b_{(1)} &= 0 \, ,
\end{split}
\end{equation}
respectively. In other words, the expansion coefficients are related to each other. The next-to-next-to-leading order gives the following non-trivial result:
\small
\begin{equation}
\begin{split}
0 
= &(v-m_1)(v-m_2) t^2
\\
& + \left(
- C_{(0)} (1+q) v^2
+ C_{(0)}q \sum_{i=1}^4 m_i v
- 2 C_{(2)} (1+q) 
- \frac{C_{(0)} q}{4} \left( \sum_{i=1}^{4} m_i \right)^2 
+ b_{(2)}
\right) t
\\
& + C_{(0)}{}^2 q (v-m_3)(v-m_4) \, .
\end{split}
\end{equation}
\normalsize
By defining the parameter $U$ as
\begin{align}
C_{(0)}U \equiv - 2 C_{(2)} (1+q)
- \frac{C_{(0)}q}{4} \left( \sum_{i=1}^{4} m_i \right)^2 
+ b_{(2)}
\end{align}
and rescaling the coordinate $t$ as $t \to C_{(0)} t$, we obtain
\begin{align}
0
= (v-m_1)(v-m_2) t^2 
+ \left( - (1+q) v^2 + q \sum_{i=1}^{4} m_i \, v + U \right) t 
+ q (v- m_3)(v- m_4) \, ,
\end{align}
which is precisely the SW curve for the 4D superconformal $SU(2)$ gauge theory found in \cite{Eguchi:2009gf}. In conclusion, we find that the 5D SW curve (\ref{curveC}) correctly reproduces the known 4D SW curve in the 4D limit $\beta \to 0$.

\section{Details of Nekrasov partition function and topological strings}
\label{app:NekrasovTop}

In this appendix we provide the details that are needed for the computation of the Nekrasov partition function using the topological vertex formalism. In particular, we give the Nekrasov partition function in a way that is convenient for the comparison with the topological string partition function in Section~\ref{sec:TopStringDeriv}.

\subsection{Young diagrams and combinatorial relations}

Before writing down the Nekrasov partition function itself, we will provide some useful formulas. We start by proving a few combinatorial relations for Young diagrams. Let $Y$ be a Young diagram, which can be viewed as a decreasing sequence of non-negative integers $Y_1\geq Y_2\geq \cdots Y_{d(Y)}>Y_{d(Y)+1}=Y_{d(Y)+2}=\cdots=0$. Taking the summation over the boxes $(i,j)\in Y$ we find
\begin{equation}
\label{eq:SumYoungDiagram}
\begin{split}
\sum_{(i,j)\in Y}i
&=
\sum_{j=1}^{Y_1}
\sum_{i=1}^{Y^T_j}i
=
\frac{\| Y^T\|^2}{2}
+
\frac{| Y|}{2} \, ,
\\
\sum_{(i,j)\in Y}j
&=
\sum_{i=1}^{Y^T_1}
\sum_{j=1}^{Y_i}j
=
\frac{\| Y\|^2}{2}
+
\frac{| Y|}{2} \, ,
\\
\sum_{(i,j)\in Y}Y_i
&=
\| Y\|^2 \, ,
\end{split}
\end{equation}
where $Y^T$ is the transpose of the diagram $Y$, $|Y| \equiv \sum_{i=1}^{Y^T_1} Y_i = \sum_{i=1}^{Y_1} Y^T_i$ is the total number of boxes in the diagram, and $\| Y\|^2 \equiv \sum_{i=1}^{Y^T_1} (Y_i)^2$. By combining the formulas in (\ref{eq:SumYoungDiagram}) we obtain
\begin{align}
\sum_{(i,j)\in Y}\left(
Y_i-i-j+1
\right)
=\frac{\kappa_{Y}}{2} \, ,
\end{align}
where the second Casimir is defined as $\kappa_Y\equiv \|Y\|^2-\|Y^T\|^2$.

Next, we introduce the 5D ``Nekrasov factor"
\begin{align}
N_{Y_1Y_2}(\mathfrak{q},Q)
\equiv
\prod_{(i,j)\in Y_1}
\left(1-Q
\mathfrak{q}^{Y_{1\,i}+Y^T_{2\,j}-i-j+1}
\right)
\prod_{(i,j)\in Y_2}
\left(1-Q
\mathfrak{q}^{-Y^T_{1\,j}-Y_{2\,i}+i+j-1}
\right) \, .
\label{def_Nfac}
\end{align}
Moreover, following \cite{Konishi:2003qq} we also define a function $P$ as
\small
\begin{equation}
\begin{split}
&\frac{1
}{P_{Y_1Y_2}(\mathfrak{q},Q)}
\equiv
\prod_{(i,j)\in Y_1}
\sinh \frac{\beta}{2}
\left(
a+\hbar(Y_{1\,i}+Y^T_{2\,j}-i-j+1)
\right)
\\
&\rule{0pt}{5ex}
\quad\qquad\qquad\quad
\times
\prod_{(i,j)\in Y_2}
\sinh \frac{\beta}{2}
\left(a+\hbar
(-Y^T_{1\,j}-Y_{2\,i}+i+j-1)\right)
\end{split}
\end{equation}
\normalsize
with $\mathfrak{q}=e^{-\beta \hbar}$ and $Q=e^{-\beta a}$. After using the identity $1 - e^{x} = 2 e^{x/2} \sinh (x/2)$ we find an important relation between these two ubiquitous factors:
\begin{align}
N_{Y_1Y_2}(\mathfrak{q},Q)
=\frac{(2\,Q^{\frac{1}{2}})^{|Y_1|+|Y_2|}\,\mathfrak{q}^{\frac{\kappa_{Y_1}}{4}-\frac{\kappa_{Y_2}}{4}}
}{P_{Y_1\,Y_2}(Q)} \, .
\label{NPrelation}
\end{align}
In order to obtain (\ref{NPrelation}) we have used the identity\footnote{See p.12 of \cite{Taki:2007dh}.}
$$\sum_{(i,j)\in Y_2}Y^T_{1\,j}
=
\sum_{(i,j)\in Y_1}Y^T_{2\,j} \, .$$
The combinatorial properties of these functions are essential when we rewrite the topological partition function to the form of Nekrasov's partition function. A basic formula we use frequently is
\begin{align}
P_{R_1R_2}(\mathfrak{q},Q)
=(-1)^{|R_1|+|R_2|}P_{R_2R_1}(\mathfrak{q},Q^{-1}) \, ,
\label{Pinverse}
\end{align}
which follows immediately from the definition of $P$. Moreover, the following infinite product expressions of the Nekrasov factor $N_{Y_1Y_2}$
\begin{equation}
\label{InfinFin}
\prod_{i,j=1}^\infty
\frac{1-Q
\mathfrak{q}^{-Y_{1\,i}-Y^T_{2\,j}+i+j-1}}
{1-Q
\mathfrak{q}^{i+j-1}}
=\prod_{i,j=1}^\infty
\frac{1-Q
\mathfrak{q}^{Y^T_{1\,i}+Y_{2\,j}-i-j+1}}
{1-Q
\mathfrak{q}^{-i-j+1}}
=
N_{Y_1Y_2}(\mathfrak{q},Q) \, 
\end{equation}
taken from (3.9) and (3.10) in \cite{Taki:2007dh}, will also be useful. By defining a bracket $[*,*]_Q$
\begin{align}
\label{bracket}
\left[Y_1, Y_2 \right]_{Q}
\equiv
\prod_{i,j=1}^\infty
(1-Q\mathfrak{q}^{Y_{1i}+Y_{2j}-i-j+1})
=\left[Y_2, Y_1 \right]_{Q} \, ,
\end{align}
we can recast the relation (\ref{InfinFin}) into a simple form
\begin{align}
\label{bracket_N}
\frac{
\left[Y_1^T, Y_2 \right]_{Q}}
{\left[\emptyset, \emptyset \right]_{Q}}
=N_{Y_1Y_2}(\mathfrak{q},Q)
=N_{Y_2^TY_1^T}(\mathfrak{q},Q) \, .
\end{align}
This formula now implies
\begin{align}
P_{Y_1 Y_2}(\mathfrak{q},Q)=P_{Y_2^T Y_1^T}(\mathfrak{q},Q) \, .
\label{Ptranspose}
\end{align}

The Schur functions play a special role in our derivation, since they describe the ``topological vertex decomposition" of Nekrasov's partition function. The specialized Schur function $S_{R}(\mathfrak{q}^\rho)$ takes the form
\begin{align}
S_{R}(\mathfrak{q}^{\rho})
=(-1)^{|R|}S_{R^T}(\mathfrak{q}^{-\rho})
=\mathfrak{q}^{-n(R)-\frac{|R|}{2}}
\prod_{(i,j)\in R}
(1-\mathfrak{q}^{-R_i-R^T_j+i+j-1})^{-1} \, ,
\end{align}
where $n(R)=\sum_{R}(i-1)$ satisfies $n(R^T)-n(R)=\kappa_R/2$. Taking into account the definition (\ref{def_Nfac}) and the relation (\ref{NPrelation}), we obtain the following relation
\begin{equation}
\begin{split}
S_{R}(\mathfrak{q}^{\rho})
S_{R^T}(\mathfrak{q}^{\rho})
=
(-1)^{|R|}
N_{RR}^{-1}(\mathfrak{q},1)
=
(-4)^{-|R|}
P_{RR}(\mathfrak{q},1) \, .
\label{specializedSchur}
\end{split}
\end{equation}

\subsection{Nekrasov partition function}

The Nekrasov partition function for the linear quiver gauge theories that we have investigated in this article is given by \cite{Nekrasov:2002qd,Fucito:2004gi}
\small
\begin{equation}
\begin{split}
Z
= &
\sum_{\vec{Y}^{(1)}} \cdots \sum_{\vec{Y}^{(M-1)}}\,
\left( q^{(1)} \right)^{|\vec{Y}^{(1)}|} \cdots 
\left( q^{(M-1)} \right)^{|\vec{Y}^{(M-1)}|} 
\cr
& \times 
\prod_{i=1}^{M-1} Z_{\,\textrm{vect}}(\vec{a}^{(i)},\vec{Y}^{(i)}, \hbar; \beta)\,
\prod_{i=1}^{M-2} Z_{\,\textrm{bifund}}
(\vec{a}^{(i)}, \vec{Y}^{(i)}, \vec{a}^{(i+1)}, \vec{Y}^{(i+1)}, m_{\text{bif}}^{(i,i+1)}, \hbar; \beta)\,
\cr
& \times 
\prod_{\gamma=1}^N Z_{\,\textrm{antifund}} (\vec{a}^{(1)} ,\vec{Y}^{(1)}, m^{\text{af}}_\gamma, \hbar; \beta) 
\prod_{\delta=1}^N Z_{\,\textrm{fund}} (\vec{a}^{(M-1)} ,\vec{Y}^{(M-1)}, m^{\text{f}}_{N+\delta}, \hbar; \beta) \, ,
\label{Nek_formula}
\end{split}
\end{equation}
\normalsize
where 
$$\vec{a}^{(i)} = (a^{(i)}_1, \cdots, a^{(i)}_N)
\qquad \text{and} \qquad
\vec{Y}^{(i)} = (Y^{(i)}_1, \cdots, Y^{(i)}_N)$$
denote the Coulomb moduli parameters and the Young diagrams for the corresponding gauge group factors, respectively. The Young diagrams describe the fixed points of the localization computation. The explicit forms and the basic properties of the factors $Z_{\textrm{vect}}$, $Z_{\textrm{bifund}}$, $Z_{\textrm{fund}}$ and $Z_{\textrm{antifund}}$ will now be described separately.

\subsubsection*{Vector multiplet contribution}

The contribution from a vector multiplet is the following product of $\sinh$ functions
\small
\begin{equation}
\begin{split}
Z_{\,\textrm{vect}}(\vec{a},\vec{Y},\hbar ; \beta)
&=
\prod_{\alpha,\beta=1}^N
\prod_{(i,j)\in Y_\alpha}
\sinh^{-1} \frac{\beta}{2}
\left(
a_\alpha-a_\beta
-\hbar(
Y_{\alpha i}+Y^T_{\beta j}-i-j+1
)
\right) \\
&\qquad\qquad\qquad
\times
\prod_{(i,j)\in Y_\beta}
\sinh^{-1} \frac{\beta}{2}
\left(
a_\alpha-a_\beta
+\hbar(
Y_{\beta i}+Y^T_{\alpha j}-i-j+1
)
\right) \\
&
=\prod_{\alpha,\beta=1}^N
P_{Y_\beta Y_\alpha}(\mathfrak{q},Q_{\alpha\beta}) \, ,
\label{Nek5Dvect}
\end{split}
\end{equation}
\normalsize
where the argument $Q_{\alpha\beta}$ is defined as
\begin{align}
Q_{\alpha\beta}=e^{-\beta(a_\alpha-a_\beta)} \, .
\end{align}
By separating the products in (\ref{Nek5Dvect}) into three parts ($\alpha=\beta$, $\alpha < \beta$, $\beta < \alpha$)
and applying (\ref{Pinverse}), 
we obtain
\footnotesize
\begin{align}
Z_{\,\textrm{vect}}(\vec{a},\vec{Y},\hbar ; \beta)
=
(-1)^{(N-1)\sum_\alpha |Y_\alpha|}
\prod_{\alpha=1}^N\,
P_{Y_\alpha Y_\alpha}(\mathfrak{q},1)\
\prod_{1\leq\alpha<\beta\leq N}
P_{Y_\beta Y_\alpha}(\mathfrak{q},Q_{\alpha\beta})
P_{Y_\beta Y_\alpha}(\mathfrak{q},Q_{\alpha\beta}) \, .
\label{Nek5Dvect2}
\end{align}
\normalsize
Further, when combining this result with (\ref{NPrelation}) and (\ref{specializedSchur}) it follows that
\begin{align}
Z_{\,\textrm{vect}}(\vec{a},\vec{Y},\hbar ; \beta)
&= C( Q_{\alpha\beta}, \vec{Y} )\,
\prod_{\alpha=1}^N\,
S_{Y_\alpha}(\mathfrak{q}^\rho)\,
S_{Y_\alpha^T}(\mathfrak{q}^\rho)
\prod_{1\leq\alpha<\beta\leq N}
\left( N_{Y_\beta Y_\alpha}(\mathfrak{q},Q_{\alpha\beta})\right)^{-2} \, ,
\end{align}
where the coefficient $C$ is defined as
\begin{align}
C ( Q_{\alpha\beta}, \vec{Y} ) =
(-4)^{N\sum_\alpha |Y_\alpha|}
\prod_{1\leq\alpha<\beta\leq N}
(Q_{\alpha\beta})^{|Y_\alpha|+|Y_\beta|}
\mathfrak{q}^{-\frac{\kappa_{Y_\alpha}}{2}+\frac{\kappa_{Y_\beta}}{2}} \, .
\label{defC}
\end{align}

\subsubsection*{Hypermultiplet contribution}

The hypermultiplets appearing in this article transform as fundamental and bifundamental representations. We start with studying the bifundamental one. Their contribution to the 5D Nekrasov partition function is
\small
\begin{equation}
\begin{split}
Z_{\,\textrm{bifund}}(\vec{a},\vec{R},\vec{\tilde{a}},\vec{Y},m,\hbar ; \beta)
=&
\prod_{\alpha,\beta=1}^N
\prod_{(i,j)\in R_\alpha}
\sinh \frac{\beta}{2}
\left(
a_\alpha-\tilde{a}_\beta-m
-\hbar(
R_{\alpha i}+Y^T_{\beta j}-i-j+1
)
\right)\\
&\qquad
\times
\prod_{(i,j)\in Y_\beta}
\sinh\frac{\beta}{2}
\left(
a_\alpha-\tilde{a}_\beta-m
+\hbar (
Y_{\beta i}+R^T_{\alpha j}-i-j+1
)
\right)\\
=&\prod_{\alpha,\beta=1}^N
P^{-1}_{Y_\beta R_\alpha}(\mathfrak{q},e^{-\beta (a_\alpha-\tilde{a}_\beta-m )}) \, .
\label{Nek5Dbifund}
\end{split}
\end{equation}
\normalsize
We introduce the variable $Q_{m\alpha}$ for later convenience
\begin{align}
Q_{m\alpha}=e^{-\beta(a_\alpha-\tilde{a}_\alpha-m)} \, .
\end{align}
The arguments of $P^{-1}$ in (\ref{Nek5Dbifund}) can then be written as
\begin{align}
e^{-\beta(a_\alpha-\tilde{a}_\beta-m)}
={Q}_{\alpha\beta}Q_{m\beta} \qquad \text{and} \qquad
e^{-\beta(-a_\beta+\tilde{a}_\alpha+m)}=
Q_{m\alpha}^{-1}{Q}_{\alpha\beta} \, .
\end{align}
In terms of these new variables (\ref{Nek5Dbifund}) reads
\begin{align}
Z_{\,\textrm{bifund}} & (\vec{a},\vec{R},\vec{\tilde{a}},\vec{Y},m,\hbar ; \beta)
\cr
= &
\prod_{1\leq\alpha\leq\beta\leq N}
P^{-1}_{Y_\beta R_\alpha}(\mathfrak{q},{Q}_{\alpha\beta}Q_{m\beta})
\prod_{1\leq\alpha<\beta\leq N}
(-1)^{|R_\beta|+|Y_\alpha|}\,
P^{-1}_{R_\beta Y_\alpha }(\mathfrak{q},Q_{m\alpha}^{-1}{Q}_{\alpha\beta}) \, ,
\end{align}
\normalsize
where we have separated the product into $\alpha \le \beta$ and $\beta < \alpha$
and applied (\ref{Pinverse}). After also applying (\ref{NPrelation}) we obtain
\begin{align}
Z_{\,\textrm{bifund}} & (\vec{a},\vec{R},\vec{\tilde{a}},\vec{Y},m,\hbar ; \beta)
\cr
& = D\,
\prod_{1\leq\alpha\leq\beta\leq N}
N_{Y_\beta R_\alpha}(\mathfrak{q},{Q}_{\alpha\beta}Q_{m\beta})
\prod_{1\leq\alpha<\beta\leq N}
N_{R_\beta Y_\alpha }(\mathfrak{q},Q_{m\alpha}^{-1}{Q}_{\alpha\beta}) \, ,
\end{align}
\normalsize
where $D$ is defined as
\footnotesize
\begin{equation}
D =
\prod_{1\leq\alpha\leq\beta\leq N}
\left(-2\sqrt{Q_{\alpha\beta}Q_{m\beta}}\right)^{-|R_\alpha|-|Y_\beta|}
\mathfrak{q}^{-\frac{\kappa_{Y_\beta}}{4}+\frac{\kappa_{R_\alpha}}{4}}
\prod_{1\leq\alpha<\beta\leq N}
\left(2\sqrt{Q_{m\alpha}^{-1}{Q}_{\alpha\beta}}\right)^{-|R_\beta|-|Y_\alpha|}
\mathfrak{q}^{-\frac{\kappa_{R_\beta}}{4}+\frac{\kappa_{Y_\alpha}}{4}} \, .
\end{equation}
\normalsize
By decomposing this expression into the $\vec{R}$ dependent and the $\vec{Y}$ dependent part, and separating out the $\alpha=\beta$ term in the first products, we find
\begin{equation}
D = D_{L}(Q_{\alpha \beta}, Q_{m\alpha}, \vec{R})\,D_{R}(Q_{\alpha \beta}, Q_{m\alpha}, \vec{Y})
\end{equation}
with
\footnotesize
\begin{equation}
\label{defDLR}
\begin{split}
&D_{L}(Q_{\alpha \beta}, Q_{m\alpha}, \vec{R})=
\prod_{\alpha=1}^N
\frac{
(-)^{(N-\alpha+1)|R_\alpha|}
\mathfrak{q}^{\frac{1}{4} \kappa_{R_\alpha}}}
{2^{N |R_\alpha|} \left( Q_{m\alpha} \right)^{\frac{|R_\alpha|}{2}}}
\prod_{1\leq\alpha<\beta\leq N}
(Q_{\alpha\beta}^{-\frac{1}{2}})^{|R_\alpha|+|R_\beta|}
\frac
{\left(Q_{m\alpha}\right)^{\frac{|R_\beta|}{2}}}
{\left(Q_{m\beta}\right)^{\frac{|R_\alpha|}{2}}}
\mathfrak{q}^{
\frac{\kappa_{R_\alpha}}{4}-\frac{\kappa_{R_\beta}}{4}} \, ,
\\
&D_{R}(Q_{\alpha \beta}, Q_{m\alpha}, \vec{Y})=
\prod_{\alpha=1}^N
\frac{
(-)^{\alpha|Y_\alpha|}
\mathfrak{q}^{ - \frac{1}{4} \kappa_{Y_\alpha}}}
{ 2^{N |Y_\alpha|}\, \left( Q_{m\alpha} \right)^{\frac{|Y_\alpha|}{2}} }
\prod_{1\leq\alpha<\beta\leq N}
(Q_{\alpha\beta}^{-\frac{1}{2}})^{|Y_\alpha|+|Y_\beta|}
\frac{\left(Q_{m\alpha}\right)^{\frac{|Y_\alpha|}{2}}}
{\left(Q_{m\beta}\right)^{\frac{|Y_\beta|}{2}} }
\mathfrak{q}^{\frac{\kappa_{Y_\alpha}}{4}-\frac{\kappa_{Y_\beta}}{4}} \, .
\end{split}
\end{equation}
\normalsize

The contribution from a fundamental hypermultiplet
\begin{equation}
\begin{split}
Z_{\,\textrm{fund}}(\vec{a},\vec{Y}, m, \hbar ; \beta)
&= \prod_{\alpha=1}^N \prod_{(i,j) \in Y_{\alpha}}
\sinh \frac{\beta}{2} (a_{\alpha} - m + \hbar (i-j ) )
\cr
&= \prod_{\alpha=1}^N P^{-1}_{Y_{\alpha \emptyset}} \left( \mathfrak{q}, e^{-\beta(a_{\alpha}-m)} \right)
\label{Nek5Dfund}
\end{split}
\end{equation}
is related to that from an anti-fundamental hypermultiplet as
\begin{align}
Z_{\textrm{antifund}}(\vec{a},\vec{R},m)=Z_{\textrm{fund}}(\vec{a},\vec{R},-m) \, .
\label{Nek5Dantifund}
\end{align}
Moreover, the product of contributions from all the $N$ fundamental hypermultiplets obeys
\begin{align}
\prod_{\alpha=1}^N
Z_{\,\textrm{fund}}(\vec{a},\vec{R},{m}_\alpha,\hbar ; \beta)
=
Z_{\,\textrm{bifund}}(\vec{a},\vec{R},\vec{m}-m\vec{1},\vec{\emptyset},m,\hbar ; \beta) \, ,
\label{fund_bif}
\end{align}
where $m \equiv \frac{1}{N}\sum_\alpha m_\alpha$. A similar relation exists for the anti-fundamental contribution:
\begin{align}
\prod_{\alpha=1}^N
Z_{\,\textrm{antifund}}(\vec{a},\vec{R},{m}_\alpha,\hbar ; \beta)
=
(-1)^{N\sum |R_\alpha|}
Z_{\,\textrm{bifund}}(-\vec{m}+m\vec{1},\vec{\emptyset},\vec{a},\vec{R},-m,\hbar ; \beta) \, .
\label{antifund_bif}
\end{align}

\subsubsection*{Gluing the multiplets}

We will now go back to the full Nekrasov partition function in (\ref{Nek_formula}) and rewrite it into a form which is convenient for the comparison with the topological string partition function. To do this, we focus on the $i$-th vector multiplet corresponding to $\vec{Y}^{(i)}$ together with the bifundamental hypermultiplets which are charged under the associated vector field. The local structure of the Nekrasov partition function for $\vec{Y}^{(i)}$
\begin{equation}
\begin{split}
\cdots
\sum_{\vec{Y}^{(i)}}
& \left( q^{(i)} \right) ^{\sum_{\alpha} |Y^{(i)}_\alpha|}\,
Z_{\,\textrm{bifund}}(\vec{a}^{(i-1)},\vec{Y}^{(i-1)},\vec{a}^{(i)}, \vec{Y}^{(i)},{m}_{\text{bif}}^{(i-1,i)},\hbar ; \beta)\,
\cr
& \times Z_{\,\textrm{vect}}(\vec{a}^{(i)},\vec{Y}^{(i)},\hbar ; \beta)\,
Z_{\,\textrm{bifund}}(\vec{a}^{(i)},\vec{Y}^{(i)},\vec{a}^{(i+1)}, \vec{Y}^{(i+1)}, {m}_{\text{bif}}^{(i,i+1)}, \hbar ; \beta)
\cdots
\end{split}
\end{equation}
dictates the contribution from these multiplets. By employing the relations (\ref{fund_bif}) and (\ref{antifund_bif}), this expression becomes valid also for $i=1$ and $i=M-1$. After collecting the factors which depend on $\vec{Y}^{(i)}$ we find the following contribution in Nekrasov's formula
\small
\begin{equation}
\begin{split}
&
D_{R}({Q}_{\alpha\beta}^{(i-1)}, {Q}_{m\alpha}^{(i-1)}, \vec{Y}^{(i)})\,
C(Q_{\alpha\beta}^{(i)}, \vec{Y}^{(i)})\,
D_{L}(Q_{\alpha\beta}^{(i)}, {Q}_{m\alpha}^{(i)}, \vec{Y}^{(i)})\,
q^{\sum |Y^{(i)}_\alpha|}
\cr
&\times
\prod_{\alpha=1}^N\,
S_{Y^{(i)}_\alpha}(\mathfrak{q}^\rho)\,
S_{Y^{(i) \, T}_\alpha}(\mathfrak{q}^\rho)
\prod_{1\leq\alpha<\beta\leq N}
\left( N_{Y^{(i)}_\beta, Y^{(i)}_\alpha} (\mathfrak{q},Q_{\alpha\beta}^{(i)}) \right)^{-2}
\cr
& \times
\prod_{1\leq\alpha\leq\beta\leq N}
N_{Y^{(i)}_\beta Y^{(i-1)}_\alpha}
\left( \mathfrak{q},{Q}_{\alpha\beta}^{(i-1)} {Q}_{m\beta}^{(i-1)} \right)
\prod_{1\leq\alpha<\beta\leq N}
N_{Y_\beta^{(i-1)} Y^{(i)}_\alpha }
\left( \mathfrak{q}, ({Q}_{m\alpha}^{(i-1)})^{-1} {Q}_{\alpha\beta}^{(i-1)} \right)
\cr
& \times
\prod_{1\leq\alpha\leq\beta\leq N}
N_{Y^{(i+1)}_\beta Y^{(i)}_\alpha}
\left( \mathfrak{q},{Q}_{\alpha\beta}^{(i)} Q_{m\beta}^{(i)} \right)
\prod_{1\leq\alpha<\beta\leq N}
N_{Y^{(i)}_\beta Y^{(i+1)}_\alpha }
\left( \mathfrak{q},(Q_{m\alpha}^{(i)}) ^{-1}{Q}_{\alpha\beta}^{(i)} \right)\, .
\label{GluingOfQuiver}
\end{split}
\end{equation}
\normalsize
The instanton factor $q^{(i)}$ will absorb most of the coefficients $C$, $D_L$ and $D_R$ in (\ref{GluingOfQuiver})
\small
\begin{equation}
\begin{split}
&D_{R}({Q}^{(i-1)}_{\alpha\beta}, {Q}^{(i-1)}_{m\alpha}, \vec{Y}^{(i)})\,
C(Q^{(i)}_{\alpha\beta}, \vec{Y}^{(i)})\,
D_{L}(Q^{(i)}_{\alpha\beta}, {Q}^{(i)}_{m\alpha}, \vec{Y}^{(i)})\,
\left( q^{(i)} \right) ^{\sum |\vec{Y}^{(i)}_\alpha|}
\\
&=
\prod_\alpha
\left(\frac{-q^{(i)}}{\sqrt{Q^{(i)}_{m\alpha}Q^{(i-1)}_{m\alpha}}}\right)^{|Y^{(i)}_\alpha|}
\prod_{1\leq\alpha<\beta\leq N}
\sqrt{\frac{(Q^{(i)}_{m\alpha})^{|Y^{(i)}_\beta|}}{(Q^{(i)}_{m\beta})^{|Y^{(i)}_\alpha|}}}
\sqrt{
\frac
{
(Q^{(i-1)}_{m\beta})^{|Y^{(i)}_\alpha|}
}
{
(Q^{(i-1)}_{m\alpha})^{|Y^{(i)}_\beta|}
}
}\\
&=
\prod_\alpha
\label{GaugeCouplingKahlers}
\left(
-q^{(i)} \prod_\beta
\sqrt{\frac
{
Q^{(i-1)}_{m\beta}
}
{
Q^{(i)}_{m\beta}
}}
\right)^{|Y^{(i)}_\alpha|}
\frac
{
\prod_{1\leq\alpha<\beta\leq N}
(Q^{(i)}_{m\alpha})^{|Y^{(i)}_\beta|}
}
{
\prod_{1\leq\alpha\leq\beta\leq N}
(Q^{(i-1)}_{m\alpha})^{|Y^{(i)}_\beta|}
} \, ,
\end{split}
\end{equation}
\normalsize
where in the first equality we have used the definitions (\ref{defC}) and (\ref{defDLR}) together with the relation (\ref{rec_QF}). This result plays a key role in identifying the relation between the gauge coupling $q$ and the K\"ahler parameter $Q_B$ of the base $\mathbb{P}^1$ of the corresponding CY$_3$. The last line of (\ref{GaugeCouplingKahlers}) is thus related to the product $\prod (-Q_{B \alpha}^{(i)})^{|Y^{(i)}_\alpha|}$.

\subsection{Comparison with the topological string partition function}

In the topological vertex computation in Section \ref{sec:TopStringDeriv}, first we calculated a sub-diagram corresponding to a decomposed toric diagram. To get the full string partition function, we had to glue together the sub-diagrams. That procedure is similar to the gluing construction of the Nekrasov partition functions in the previous subsection. Here, we demonstrate the gluing construction for the $SU(N)^{M-1}$ linear quiver theory. Since the gluing of the toric sub-diagrams is done by taking the summation over the Young diagrams, the full partition function corresponding to Figure~\ref{fig:large} is given by
\begin{equation}
\begin{split}
Z_{\,\textrm{inst}}&=
\label{genQuivPartFunc}
\sum\cdots
\sum_{Y^{(i)}_1,\cdots,Y^{(i)}_N}
\prod_{\alpha}(-Q^{(i)}_{B\alpha})^{|Y^{(i)}_\alpha|}\\
& \times \tilde{H}_{\,Y^{(i+1)}_1Y^{(i+1)}_2\cdots Y^{(i+1)}_N}^{\,Y^{(i)}_1Y^{(i)}_2\cdots Y^{(i)}_N}\,(\,Q^{(i)}_{m1},\cdots,Q^{(i)}_{mN},Q^{(i)}_{F1},\cdots,Q^{(i)}_{FN})\\
& \times \tilde{H}^{\,Y^{(i-1)}_1Y^{(i-1)}_2\cdots Y^{(i-1)}_N}_{\,Y^{(i)}_1Y^{(i)}_2\cdots Y^{(i)}_N}\,(\,Q^{(i-1)}_{m1},
\cdots,Q^{(i-1)}_{mN},Q^{(i-1)}_{F1},\cdots,Q^{(i-1)}_{FN})
\cdots.
\end{split}
\end{equation}
Using the explicit expression for the sub-diagram $\tilde{H}$ in (\ref{Htilde}), which resembles the Nekrasov factor very much, we can rewrite (\ref{genQuivPartFunc}) as the Nekrasov partition function for a linear quiver gauge theory.

We will again focus on the contribution from the $i$-th gauge group in the full partition function (\ref{genQuivPartFunc}). In the topological vertex computation, the $i$-th color D4-branes are assigned with the Young diagrams $Y^{(i)}$ and correspond to the chopped lines of the web-diagram in Figure~\ref{fig:genQuiver}. The contribution from the $i$-th gauge group depends therefore on $Y^{(i)}$. By collecting such factors, we obtain
\small
\begin{equation}
\label{eq:GluingOfVertex}
\begin{split}
&\prod_{\alpha}(-Q^{(i)}_{B\alpha})^{|Y^{(i)}_\alpha|}\\
&\times\frac{\prod_{\alpha=1}^N\,
S_{Y^{(i)}_\alpha}(\mathfrak{q}^{\rho})\,
S_{{Y^{(i)\,T}_\alpha}}(\mathfrak{q}^{\rho})}
{\prod_{1\leq \alpha<\beta\leq N}
N_{Y^{(i)}_\beta Y^{(i)}_\alpha}\left(Q_{\alpha\beta}^{(i)}\right)\,
N_{Y^{(i)}_\beta Y^{(i)}_\alpha}\left((Q_{m\alpha}^{(i-1)})^{-1}Q_{\alpha\beta}^{(i-1)}Q_{m\beta}^{(i-1)}\right)
}\\
&\times
\rule{0pt}{4ex}
\prod_{1\leq \alpha<\beta\leq N}
N_{Y^{(i)}_\beta Y^{(i+1)}_\alpha}\left((Q^{(i)}_{m\alpha})^{-1}Q^{(i)}_{\alpha\beta}\right)
N_{Y^{(i-1)}_\beta Y^{(i)}_\alpha}\left((Q^{(i-1)}_{m\alpha})^{-1}Q^{(i-1)}_{\alpha\beta}\right)
\\
&\times
\rule{0pt}{4ex}
\prod_{1\leq \alpha\leq\beta\leq N}
N_{Y^{(i+1)}_\beta Y^{(i)}_\alpha}\left(Q^{(i)}_{\alpha\beta}Q^{(i)}_{m\beta}\right)
N_{Y^{(i)}_\beta Y^{(i-1)}_\alpha}\left(Q^{(i-1)}_{\alpha\beta}Q^{(i-1)}_{m\beta}\right) \, .
\end{split}
\end{equation}
\normalsize
From the web-diagram in Figure~\ref{fig:genQuiver} we see that the arguments of the Nekrasov factors in denominator of the second line satisfy
\begin{align}
Q_{\alpha\beta}^{(i)}=(Q_{m\alpha}^{(i-1)})^{-1} Q_{\alpha\beta}^{(i-1)}Q_{m\beta}^{(i-1)} \, .
\end{align}
After inserting the known factor $\prod_{\alpha}(Q^{(i)}_{B\alpha})^{|Y^{(i)}_\alpha|}$ from (\ref{QB_q}) together with the gauge theory parametrization (\ref{Qab_Qm}) of the string parameters, we are now ready to compare the topological string partition function (\ref{eq:GluingOfVertex}) with the Nekrasov partition function (\ref{GluingOfQuiver}). By inspection these two expressions are identical.

For the full Nekrasov partition function, the above argument is applied successively for each index $i$. We find that $Z_{\,\textrm{inst}}$ exactly matches the instanton partition function for the linear quiver gauge theory, provided that $Q^{(i)}_B$ and ${q}^{(i)}$ are related as
\begin{align}
Q_B^{(i)}
=
{q}^{(i)}
\frac{1}{Q^{(i-1)}_{m1}}
\prod_{\alpha=1}^N
\sqrt{\frac{Q^{(i-1)}_{m\alpha}}
{Q^{(i)}_{m\alpha}}} \, .
\end{align}
This is the relation between the K\"ahler parameters of the base $\mathbb{P}^1$ and the gauge couplings. Together with (\ref{Qab_Qm}), they provide the complete identification rules between the gauge theory parameters and the K\"ahler parameters of CY$_3$. Using this identification, the topological vertex computation gives precisely the Nekrasov partition function, which proves the geometric engineering for the quiver gauge theories.

\subsection{Symmetry of the Nekrasov partition function}

We end this section by commenting on a specific symmetry of the Nekrasov partition function. By using the identities of $P_{Y_1Y_2}(Q)$ in (\ref{Pinverse}) and (\ref{Ptranspose}), the contributions from the vector multiplet (\ref{Nek5Dvect})
and the hypermultiplets (\ref{Nek5Dbifund}) (\ref{Nek5Dfund}) (\ref{Nek5Dantifund})
respectively satisfy the following properties
\small
\begin{equation}
\begin{split}
&Z_{\,\textrm{vect}}(\vec{a},\vec{Y},\hbar ; \beta)
=Z_{\,\textrm{vect}}(-\vec{a},\vec{Y^T},\hbar ; \beta) \, ,
\\
&Z_{\,\textrm{bifund}}(\vec{a},\vec{R},\vec{\tilde{a}},\vec{Y},m,\hbar ; \beta)
=(-1)^{N\sum_{\alpha}|R_\alpha|+|Y_\alpha|}
Z_{\,\textrm{bifund}}(-\vec{a},\vec{R^T},-\vec{\tilde{a}},\vec{Y^T},-m,\hbar ; \beta) \, ,
\\
&Z_{\,\textrm{fund}}(\vec{a},\vec{R},{m},\hbar ; \beta)
= (-1)^{N\sum_{\alpha}|R_\alpha|} Z_{\,\textrm{fund}}(-\vec{a},\vec{R}^T,-{m},\hbar ; \beta) \, ,
\\
&Z_{\,\textrm{antifund}}(\vec{a},\vec{R},{m},\hbar ; \beta)
= (-1)^{N\sum_{\alpha}|R_\alpha|} Z_{\,\textrm{antifund}}(-\vec{a},\vec{R}^T,-{m},\hbar ; \beta) \, .
\end{split}
\end{equation}
\normalsize
All the signs will cancel out if we substitute them into the Nekrasov partition function. When we sum over all the Young diagrams to obtain the expression (\ref{Nek_formula}), we find a symmetry with the signs of all the Coulomb moduli and the masses being inverted:
\begin{equation}
Z(\vec{a}, \vec{m}, q, \hbar, \beta) = Z(-\vec{a}, -\vec{m}, q, \hbar, \beta) \, .
\end{equation}

\section{Rotation of 90 degrees}
\label{app:90Rotation}

In Section \ref{subsec:Msu2} and \ref{subsec:topsu2} we analyze the symmetry coming from the reflection in a diagonal axis of the toric diagram for $SU(2)$  SQCD. The invariance of the topological string partition function reproduces the duality transformation found in the M-theory setup. However, both the M5-brane configuration and the toric web-diagram are also invariant under the rotation by 90 degrees. Rigorously, it is this 90 degree rotation which corresponds to part of the $SL(2,Z)$ S-duality transformations, and not the reflection as mentioned in Section~\ref{subsec:review_duality}. We will now give the duality map of the 90 degree rotation.

In the M-theory setup, this rotation leads to the coordinate transformation
\begin{align}
w_d = t, \qquad t_d^{-1} = w
\label{coord_rot}
\end{align}
of the SW curve. Contrary to the reflection transformation in Section \ref{subsec:Msu2}, the dual of the SW one-form is given by
\begin{equation}
(\lambda_{SW})_d = \lambda_{SW} \, .
\end{equation}
However, due to the convention of the direction of the cycle, we have $(A_1)_d = - A'_1$. The second relation in (\ref{aint}) and thus also (\ref{dual_a}) are the same in this case. Taking these into account, we obtain the following duality map for the rotation
\begin{equation}
\begin{split}
(\tilde{m}_1)_d
= \tilde{m}_1^{-\frac{1}{4}} \tilde{m}_2^{-\frac{3}{4}}
\tilde{m}_3^{-\frac{1}{4}} \tilde{m}_4^{\frac{1}{4}} q^{-\frac{1}{2}} \, ,&
\qquad
(\tilde{m}_2)_d
= \tilde{m}_1^{\frac{1}{4}} \tilde{m}_2^{-\frac{1}{4}}
\tilde{m}_3^{\frac{1}{4}} \tilde{m}_4^{\frac{3}{4}} q^{\frac{1}{2}} \, ,
\cr
(\tilde{m}_3)_d
= \tilde{m}_1^{\frac{3}{4}} \tilde{m}_2^{\frac{1}{4}}
\tilde{m}_3^{-\frac{1}{4}} \tilde{m}_4^{\frac{1}{4}} q^{-\frac{1}{2}} \, ,&
\qquad
(\tilde{m}_4)_d
= \tilde{m}_1^{\frac{1}{4}} \tilde{m}_2^{-\frac{1}{4}}
\tilde{m}_3^{-\frac{3}{4}} \tilde{m}_4^{-\frac{1}{4}} q^{\frac{1}{2}} \, ,
\cr
\tilde{a}_d = \left( \frac{\tilde{m}_2 \tilde{m}_4}%
{\tilde{m}_1\tilde{m}_3} \right)^{\frac{1}{4}} q^{-\frac{1}{2}} \tilde{a} \, ,&
\qquad
q_d = \left( \frac{\tilde{m}_2\tilde{m}_4}{\tilde{m}_1 \tilde{m}_3} \right)^{\frac{1}{2}} \, .
\label{map_rotation}
\end{split}
\end{equation}

On the other hand, in the topological string setup we the full partition function to be invariant under the transformation
\begin{align}
(Q_{m1})_d= Q_{m2} \, ,\quad
(Q_{m2})_d =Q_{m4} \, ,&\quad
(Q_{m4})_d =Q_{m3} \, ,\quad
(Q_{m3})_d =Q_{m1} \, ,\quad
\cr
(Q_{B})_d = Q_F \, ,&\quad
(Q_{F})_d = Q_B \, ,
\end{align}
where all the parameters are given by Figure~\ref{fig:4flavSQCD}. Moreover, the relation between the K\"ahler parameters and the gauge theory parameters are found in (\ref{Q_Def}) and (\ref{q_Def}). After combining with the symmetry transformation $\tilde{m}_{i}\to\tilde{m}^{-1}_{i}$, $(\tilde{m}_{i})_d\to(\tilde{m}_{i})_d^{-1}$, $\tilde{a}\to\tilde{a}^{-1}$ and $(\tilde{a})_d\to(\tilde{a})_d^{-1}$ of the Nekrasov partition function, we confirm that it exactly reproduces the duality map (\ref{map_rotation}). We have thus proved that the $\Omega$-background does not break this duality either.

The difference between the reflection and the rotation symmetry can be understood as a simple parity transformation, which corresponds to the coordinate transformation
\begin{align}
w_d = w^{-1} \, , \qquad t_d = t \, .
\label{w_inv}
\end{align}
The corresponding duality map of the gauge theory parameters is
\begin{equation}
\begin{split}
(\tilde{m}_1)_d = \tilde{m}_2^{-1} \, , \qquad
(\tilde{m}_2)_d = \tilde{m}_1^{-1} \, ,& \qquad
(\tilde{m}_3)_d = \tilde{m}_4^{-1} \, , \qquad
(\tilde{m}_4)_d = \tilde{m}_3^{-1} \, ,
\cr
\tilde{a}_d = \tilde{a}^{-1} \, ,& \qquad
q_d = q \, .
\label{map_parity}
\end{split}
\end{equation}
It is straightforward to show that the duality map (\ref{map_reflection_su2}) for the reflection can be obtained by sequentially acting with the transformations (\ref{map_parity}) and (\ref{map_rotation}).

\end{document}